\documentclass[aps,prd,twocolumn,superscriptaddress,amsfont,graphicx,nofootinbib,preprintnumbers]{revtex4}
\usepackage{color,graphicx,epsfig}
\usepackage{ifpdf}
\usepackage{amsmath}
\usepackage{bm}
\usepackage{color}
\usepackage[english]{babel}
\usepackage{graphicx}
\usepackage{amsfonts}
\usepackage{amssymb}
\usepackage[bookmarks=false]{hyperref}
\usepackage{enumerate}
\usepackage{bbold}

\definecolor{nicered}{rgb}{0.7,0.1,0.1}
\definecolor{nicegreen}{rgb}{0.1,0.5,0.1}
\hypersetup{colorlinks,citecolor= nicegreen,linkcolor= nicered}

\newcommand{\beq}{\begin{equation}}
\newcommand{\eeq}{\end{equation}}
\newcommand{\bea}{\begin{eqnarray}}
\newcommand{\eea}{\end{eqnarray}}

\definecolor{Red}{rgb}{1.,0.,0.}

\def\gsim{{~\raise.15em\hbox{$>$}\kern-.85em
          \lower.35em\hbox{$\sim$}~}}
\def\lsim{{~\raise.15em\hbox{$<$}\kern-.85em
          \lower.35em\hbox{$\sim$}~}}

\def\mysection#1{{{\bf #1}.~}}
\def\OMIT#1{}

\def\bea{\begin{eqnarray}}
\def\eea{\end{eqnarray}}
\def\ba{\begin{eqnarray}}
\def\ea{\end{eqnarray}}
\def\be{\begin{equation}}
\def\ee{\end{equation}}
\def\beq{\begin{equation}}
\def\eeq{\end{equation}}
\def\beq{\begin{equation}}
\def\eeq{\end{equation}}
\def\bea{\begin{eqnarray}}
\def\eea{\end{eqnarray}}

\def\D0bar{\overline D{}^0}
\def\K0bar{\overline K{}^0}

\def\DDbar{D{}^0-\overline D{}^0}

\def\lsim{\lesssim}
\def\gsim{\gtrsim}

\newcommand{\leftrightarrowraised}{\mathrel{\rlap{\lower-0pt\hbox{\hskip1pt$\partial$}}
    \raise6 pt\hbox{$\leftrightarrow$}}}

\begin{document}

\title{
Dispersive and Absorptive CP Violation in $D^0- \D0bar$ Mixing }
\def\Cincy{Department of Physics, University of Cincinnati, Cincinnati, Ohio 45221,USA}
\author{Alexander L. Kagan}
\email[]{kaganalexander@ucmail.uc.edu}
\affiliation{\Cincy}
\author{Luca Silvestrini}
\email[]{luca.silvestrini@roma1.infn.it}
\affiliation{CERN, 1211 Geneva 23, Switzerland}
\affiliation{INFN, Sezione di Roma, Piazzale A. Moro 2, I-00185 Roma, Italy}
\begin{abstract}
CP violation (CPV) in $D{}^0-\overline D{}^0$ mixing is described in terms
of the dispersive and absorptive `weak phases' $\phi_f^M$
and $\phi_{f}^\Gamma$.  They parametrize CPV originating
from the interference of $D^0$ decays with and without dispersive mixing,
and with and without absorptive mixing, respectively, for CP conjugate hadronic final states $f$,
$\bar f$. These are distinct and separately measurable effects.
For CP eigenstate final states,
indirect CPV only depends on $\phi_f^M$ (dispersive CPV), whereas
$\phi_f^\Gamma$ (absorptive CPV) can only be probed with non-CP
eigenstate final states.  Measurements of the final state dependent
phases $\phi_f^{M}$, $\phi_{f}^\Gamma$ determine the intrinsic
dispersive and absorptive mixing phases $\phi_2^{M}$ and
$\phi_2^\Gamma$. The latter are the arguments of the
dispersive and absorptive mixing amplitudes $M_{12}$ and
$\Gamma_{12}$, relative to their dominant ($\Delta U=2$) $U$-spin
components.  The intrinsic phases are experimentally accessible due
to {\it approximate universality}: in the SM, and in extensions with
negligible new CPV phases in Cabibbo favored/doubly Cabibbo suppressed
(CF/DCS) decays, the deviation of $\phi_f^{M,\Gamma}$ from
$\phi_2^{M,\Gamma}$ is negligible in CF/DCS decays
$D^0 \to K^\pm X$, and below
$10\% $ in CF/DCS decays $D^0 \to K_{S,L} X$ (up to precisely known $O(\epsilon_K )$
corrections).
In Singly Cabibbo Suppressed (SCS) decays, QCD pollution enters at $O(\epsilon)$ in
$U$-spin breaking and can be significant, but is $O(\epsilon^2)$ in
the average over $f=K^+K^-$, $\pi^+\pi^-$. SM %$U$-spin based
estimates yield $\phi_2^M, \phi_2^\Gamma = O(0.2\%)$.
A fit to current data allows $O(10)$ larger phases at $2\sigma$, from new physics.
A
fit based on naively extrapolated experimental precision suggests that sensitivity to $\phi_2^{M}$ and $\phi_2^{\Gamma}$ in the SM
may be achieved at the LHCb
Phase II upgrade.
\end{abstract}
\maketitle
%\newpage
%%%%%%%%%%%%%%%%%%%%%%%
%\section{Introduction}
%%%%%%%%%%%%%%%%%%%%%%%
%%%%%%%%%%%%%%%%%%

\section{Introduction}

In the Standard Model (SM), CP violation (CPV) enters $\DDbar$ mixing and $D$ decays at { $O(V_{cb} V_{ub} / V_{cs} V_{us} ) \sim 10^{-3}$},
due to the weak phase $\gamma $.
Consequently, all three types of CPV \cite{Nir:1992uv} are realized: (i) direct CPV, (ii) CPV in pure mixing (CPVMIX), which is due to interference of the dispersive and absorptive mixing amplitudes, and (iii) CPV due to the interference of decay amplitudes with and without mixing (CPVINT).
In this work, we are particularly interested in the latter two, which result from $D^0 -\D0bar$ mixing, and which we collectively refer to as ``indirect CPV''.
We would like to answer the following questions: How large are the indirect CPV asymmetries in the SM? What is the minimal parametrization appropriate for the LHCb/Belle-II precision era? How large is the current window for new physics (NP)? Can this window be closed
by LHCb and Belle-II?

In order to address these questions we first develop the description of indirect CPV in terms of
the CP violating (CP-odd) and final state dependent dispersive and absorptive ``weak phases''.
These phases, which we denote as $\phi_f^M$ and $\phi_{f}^\Gamma$, respectively, for CP conjugate final states $f$ and $\bar f$, parametrize CPVINT contributions originating from the interference of $D^0$ decays with and without dispersive (absorptive) mixing, respectively.  These are distinct
measurable effects, as we will see below.  Their difference equals the CPVMIX weak phase.

An immediate consequence of our approach is that it yields simplified expressions for the indirect CP asymmetries, which have a transparent physical interpretation (unlike the more familiar description in terms of the mixing parameter $|q/p|$, and the weak phase
$\phi_{\lambda_f}$).  In particular, the requirement that the underlying interfering amplitudes possess non-trivial CP-even ``strong-phase'' differences is manifest,
and accounts for the differences between the $\phi_f^M$ and $\phi_f^\Gamma$
dependence of the CP asymmetries. For example, we will see that the time-dependent CPVINT asymmetries in decays to CP eigenstate final states are purely dispersive, i.e. they only depend on $\phi_f^M$ (apart from subleading direct CPV effects).

In the SM, the dispersive and absorptive $\DDbar$ mixing amplitudes are due to the long distance exchanges of all off-shell and on-shell intermediate states, respectively (short distance dispersive mixing is negligible).
The CPVINT asymmetries are due to the
CP-odd contributions of the subleading $\Delta C=1$ transitions to the mixing amplitudes (via intermediate states) and the decay amplitudes (via final states).
The combined effects of these two CPV contributions
can be expressed in terms of the underlying final state dependent phases $\phi_f^{M,\Gamma}$, as noted above.
Unfortunately, due to their non-perturbative nature,  these phases can not currently be calculated from first principles QCD.
However, we will be able to make meaningful statements using $SU(3)_F$ flavor symmetry arguments.

In order to estimate the magnitudes and final state
dependence of $\phi_f^{M,\Gamma}$ in the different classes of decays,
we compare them to a theoretical pair of dispersive and
absorptive phases. The latter are intrinsic to the mixing amplitudes, and follow from their $U$-spin decomposition. In general, they are defined
as the arguments of the total dispersive and absorptive amplitudes, respectively, relative to a basis choice for the real axis in the complex
mixing plane, given by the common direction of the dominant $\Delta U=2$  mixing amplitudes.  Hence, we denote them as $\phi_2^M$ and $\phi_2^\Gamma$, respectively.
Note that these phases are quark (or meson) phase convention independent and physical, like the phases $\phi_f^{M,\Gamma}$ directly measured in the decays.  $U$-spin based estimates yield
$\phi_2^M, \phi_2^\Gamma = O(0.2\%)$ in the SM.  In principle, they could be measured on the lattice in the future.
Their difference yields the CPVMIX phase, like the final state dependent phases.

In the SM, and for the Cabibbo favored and doubly
Cabibbo suppressed decays (CF/DCS), the differences
between $\phi_f^M$ and $\phi_2^M$, or $\phi_f^\Gamma$ and $\phi_2^\Gamma$ are essentially known.
This allows for precise experimental determinations of the theoretical phases, and their comparison with $U$-spin based estimates and future lattice measurements. 
A single pair of intrinsic dispersive and absorptive mixing phases suffices to parametrize all indirect CPV effects in CF/DCS decays, whereas for SCS decays this may cease to be the case as SM sensitivity is approached.  We refer to this fortunate state of affairs as {\it approximate universality}. In particular, the approximate universality phases are identified with the intrinsic mixing phases, $\phi^M_2$ and $\phi_2^\Gamma$.
Once non-universality is hinted at in the SCS phases, the SCS observables could be dropped from the global fits.  Instead,
one could compare the CF/DCS based fit results for $\phi_2^{M,\Gamma}$
with measurements of $\phi_f^{M,\Gamma}$ and direct CPV in the SCS decays, to learn about the anatomy of the (subleading) SCS QCD penguin amplitudes.
For example, in the SM one could separately determine their relative magnitudes, and strong phases.

Approximate universality generalizes beyond the SM under the following conservative assumptions regarding subleading decay amplitudes containing new weak phases:  (i) they can be neglected in Cabibbo favored and doubly Cabibbo suppressed (CF/DCS) decays,
given that an exotic NP flavor structure would otherwise be required in order to evade
the $\epsilon_K$ constraint \cite{Bergmann:1999pm}; (ii) in singly Cabibbo suppressed (SCS) decays, their magnitudes are similar to, or smaller than the SM QCD penguin amplitudes,
as already hinted at by current bounds on direct CPV in $D^0 \to K^+ K^- , \pi^+ \pi^- $ decays.  These assumptions can ultimately be tested by future direct CPV measurements at LHCb and Belle-II.

The most stringent experimental bounds on indirect CPV phases have  been obtained  %via fits to the indirect CP asymmetries
in the superweak limit \cite{Ciuchini:2007cw,Grossman:2009mn,Kagan:2009gb},
in which the SM weak phase $\gamma$ and potential NP weak phases in the decay amplitudes
are set to zero in the indirect CPV observables.
In this limit, the dispersive and absorptive mixing phases satisfy $\phi_{f}^M = \phi^M_2$ and $\phi_f^\Gamma = \phi_2^\Gamma  = 0$. Thus, indirect CPV is entirely due to short-distance NP.
The superweak fits are highly
constrained, given that only one CPV phase controls all indirect CPV.  Comparison of  superweak fit results
with our estimate, $\phi_2^M,\,\phi_2^\Gamma = O(0.2\%)$ suggests that there is currently an O(10) window for NP in indirect CPV.

Moving forward, the increased precision at LHCb and Belle-II will require fits to the indirect CPV data to be carried out for both $\phi_2^M$ and $\phi_2^\Gamma$, in the approximate universality framework.  The addition of $\phi_2^\Gamma$ yields a less constrained fit. However, this should ultimately be overcome by a large increase in statistics.  %d statistics at LHCb and Belle-II.

Throughout this work we develop, in parallel, the description of indirect CPV for the three relevant classes of decays: (i) SCS (both CP eigenstate and non-CP eigenstate final states), (ii) CF/DCS decays to $K^\pm X$, and (iii) CF/DCS decays to
$K^0 X$, $\K0bar X$.  The last one requires special care due to the intervention of CPV in $K^0 - \K0bar$ mixing.
In Section~\ref{sec:formalism}, the formalism for mixing and indirect CPV is presented,
based on the final state dependent dispersive and absorptive CPVINT observables.
A translation between the dispersive and absorptive
CPV phases, $\phi_f^{M}$, $\phi_f^{\Gamma}$, and more widely used CPV parameters is also provided.    In Section~\ref{sec:indirectCPasym}, we apply this formalism to the derivation of general expressions for the time dependent decay widths and indirect CP
asymmetries in terms of $\phi_f^M$, $\phi_f^\Gamma$.   In CF/DCS decays to
$K^0 X$, $\K0bar X$, the widths depend on two elapsed time intervals:  the time at which the $D$ decays, and the time at which the $K$ decays, following their respective production.
Approximate universality is discussed in Section~\ref{sec:appuniv}.  We begin with the $U$-spin decomposition of the mixing amplitudes in the SM, introduce the intrinsic mixing phases $\phi_2^M$, $\phi_2^\Gamma$, estimate their magnitudes, and derive their deviations from the final state dependent phases.
In Section~\ref{sec:implement} we explain how to convert the expressions for the time dependent decay widths and indirect CP asymmetries, collected in Section~\ref{sec:indirectCPasym}, to the approximate universality framework.  In the case of CF/DCS decays to $K^0X$, $\K0bar X$, the effects of $\epsilon_K$ on the  $K$ decay time scales of relevance for LHCb and Belle-II are compared.
Superweak and approximate universality fits to the current data are presented in Section~\ref{sec:fits}, together with future projections.
We conclude with a summary of our results in Section~\ref{sec:conclusion}.  Appendix~\ref{sec:AppendixA} contains expressions for a selection of time-integrated CP asymmetries, demonstrating that they can also be used to separately measure $\phi_2^{M}$ and $\phi_2^\Gamma$.

\[\]
%\vspace{-0.6cm}
\section{Formalism}
\label{sec:formalism}

\subsection{Mixing and time evolution}
The time evolution of an arbitrary linear combination of the neutral $D^0$ and $\D0bar$ mesons,
\beq a |D^0 \rangle + b | \D0bar\rangle \,\eeq
follows from the time-dependent Schr{\"o}dinger equation (see e.g. \cite{Nir:1992uv}),
\beq  \label{SchEq} i {d~ \over dt } \begin{pmatrix}a \cr b \end{pmatrix} = H \begin{pmatrix}a \cr b \end{pmatrix}\equiv ({\large M} -{\textstyle \frac{i}{2}}\,\Gamma ) \begin{pmatrix}a \cr b \end{pmatrix}\,.\eeq
The $2 \times 2$ matrices $M$ and $\Gamma$ are Hermitian. The former is referred to as the mass matrix,
and the latter yields exponential decays of the neutral mesons. CPT invariance implies $H_{11} = H_{22}$.
The transition amplitudes for $\DDbar$ mixing are given by the off-diagonal entries
\beq\begin{split} \langle  D^0 | H | \D0bar\rangle &= M_{12} - {\textstyle \frac{i}{2}} \Gamma_{12}\,,\cr
 \langle  \D0bar | H | D^0\rangle &= M^*_{12} - {\textstyle \frac{i}{2}} \Gamma^*_{12}.
 \end{split}\eeq
$M_{12}$ is the dispersive mixing amplitude.  In the SM it is dominated by the long-distance contributions of off-shell intermediate states. A significant short distance effect would be due to NP.
$\Gamma_{12} $ is the absorptive mixing amplitude, and is due
to the long distance contributions of on-shell intermediate states, i.e. decays.

The $D$ meson mass eigenstates are
\beq |D_{1,2}\rangle = p |D^0 \rangle \pm q |\overline D^0\rangle, \label{qp}\eeq
where
\beq \label{qovp}
\left({q \over p}\right)^{\!\!2} = {M_{12}^* -{\textstyle \frac{i}{2}}\Gamma_{12}^* \over M_{12} -{\textstyle \frac{i}{2}}\Gamma_{12}}
\eeq
The differences between the masses and widths of the mass eigenstates, $\Delta M_D = m_2 - m_1$ and $\Delta \Gamma_D = \Gamma_2 - \Gamma_1 $, are expressed in terms of the observables
\beq\label{xy}
x= {\Delta M_D  \over \Gamma_D}, \qquad
y= {\Delta \Gamma_D \over 2 \Gamma_D}, %\qquad \Gamma_D \,= {\Gamma_1+ \Gamma_2 \over 2},
\eeq
where the averaged $D^0$ lifetime and mass are denoted by
$\Gamma_D$ and $M_D$.
We can define three ``theoretical" physical mixing parameters: two CP conserving ones,
\beq \label{x12y12}
x_{12}\, \equiv\, {2 |M_{12} |/ \Gamma_D},~y_{12}\, \equiv\, {|\Gamma_{12}|/\Gamma_D}, \eeq
and a CP violating pure mixing (CPVMIX) phase
\beq \label{phi12}
\phi_{12} \,\equiv \, {\rm arg}\left({M_{12} \over \Gamma_{12} }\right)= \phi^M - \phi^\Gamma\,.
\eeq
The CP-odd phases
\beq  \label{phiMunphys}\phi^M = {\rm arg}\left({M_{12}}\right),~~~~\phi^\Gamma = {\rm arg}\left({\Gamma_{12}}\right)\,,\eeq
are separately meson and quark phase convention dependent and unphysical.
The CP conserving parameters in \eqref{xy} and \eqref{x12y12} are related as
\beq (x- i y)^2 = x_{12}^2 - y_{12}^2  - 2 \,i\, x_{12} y_{12} \cos\phi_{12}\,,\label{xyx12y12}\eeq
yielding
\beq \begin{split} |x|=x_{12} ,~~|y| = y_{12}\,,\label{x12vsx}\end{split}\eeq
up to negligible corrections quadratic in $\sin\phi_{12}$.
Two other useful relations are
$$ \left(\left|{q\over p}\right|^2,\left|{p\over q}\right|^2\right) \times (x^2 + y^2) = x_{12}^2 + y_{12}^2 \pm 2  x_{12} y_{12} \sin\phi_{12}\,.$$

Measurements of the $D^0$ meson mass and lifetime differences and CPV asymmetries imply that $x_{12}, y_{12} \sim  0.5 \%$,
while $\sin\phi_{12} \lesssim 0.1$, % at 95\% CL  \cite{UTfittriangle}, \cite{Amhis:2016xyh},
cf. Section~\ref{sec:fits}.
One is free to identify $D_2$ or $D_1$ with either the short-lived meson, % ($D_S$),
or
the heavier meson, %($D_H $),
by redefining $q \to -q$.  This is equivalent to choosing a sign-convention for $y$, which in turn fixes the sign of
$x$, or vice-versa, via the imaginary part of \eqref{xyx12y12}.
In the HFLAV \cite{Amhis:2016xyh} convention, $D_2$ is identified with the would be CP-even state in the limit of no CPV.
Given that the short-lived meson % $D_S$
is approximately CP-even, this is equivalent to the choice $y>0$.

The time-evolved mesons $D^0 (t) $ and $\D0bar (t)$ denote the mesons which start out as a $D^0$ and $\D0bar$ at $t=0$, respectively.
Solving \eqref{SchEq} for their time-dependent components yields,
\begin{eqnarray}
\langle \D0bar | D^0 (t) \rangle &=& - e^{-i \left(M_D - i{\Gamma_D \over 2}\right)t} \left(e^{i \pi/2}  M_{12}^* +{\textstyle \frac12 }\Gamma_{12}^* \right) t \nonumber\\
                                 &&~~~~~~~~~~~~\times ~{\sin\left[{\textstyle \frac12 } \left(\Delta M_D  - i {\textstyle \frac12 }\, \Delta \Gamma_D \right) t \right ] \over {\textstyle \frac12 } \left(\Delta M_D  - i {\textstyle \frac12 }\, \Delta \Gamma_D \right)\, t },\nonumber\\
\langle D^0 | D^0 (t) \rangle  &=&\langle \D0bar | \D0bar (t)
                                   \rangle \label{timeevln} \\
&=&e^{-i \left(M_D - i{\Gamma_D \over 2}\right)t}  \cos\left[{\textstyle \frac12 } \left(\Delta M_D  - i {\textstyle \frac12 }\, \Delta \Gamma_D \right) t \right ] \,,\nonumber
\end{eqnarray}
with $\langle D^0 | \D0bar (t) \rangle$ obtained from $\langle \D0bar | D^0 (t) \rangle$ by substituting $M_{12}^* \to M_{12}$ and
$\Gamma_{12}^* \to \Gamma_{12}$.
The phase $\pi/2$ in the first relation of \eqref{timeevln} originates from the time derivative in \eqref{SchEq}, and is
a dispersive CP-even ``strong phase".
We will keep track of its role in the derivation of the indirect CP asymmetries in Section~\ref{sec:indirectCPasym}.
For the time intervals relevant to experiment, i.e. $t \lsim 1/\Gamma_D $, \eqref{timeevln} reduces to
\begin{eqnarray}
\langle \D0bar | D^0 (t) \rangle &=& e^{-i \left(M_D - i{\Gamma_D \over 2}\right)t} \left(e^{-i \pi/2}  M_{12}^* -{\textstyle \frac12 }\Gamma_{12}^* \right) t \nonumber \\
\langle D^0 | \D0bar (t) \rangle &=& e^{-i \left(M_D - i{\Gamma_D \over 2}\right)t} \left(e^{-i \pi/2}  M_{12} -{\textstyle \frac12 }\Gamma_{12} \right) t \nonumber \\
\langle D^0 | D^0 (t) \rangle  &=&\langle \D0bar | \D0bar (t) \rangle =  e^{-i \left(M_D - i{\Gamma_D \over 2}\right)t} \times \label{timeevlnapp} \\
\Bigl(1&-&{\textstyle \frac18 }\, [x_{12}^2 - y_{12}^2 - 2 i  x_{12} y_{12} \cos\phi_{12} ]\, \Gamma_D^2 \, t^2  \,\Bigr)
\,, \nonumber
\end{eqnarray}
up to negligible corrections entering at $O(t^3)$ and beyond,   %$M_{12}/\Gamma_D$, $\Gamma_{12} /\Gamma_{12}$.
where use has been made of \eqref{xyx12y12} in the last relation.

\subsection{The decay amplitudes}

The amplitudes
for $D^0$ and $\D0bar$ decays to CP conjugate final states $f$ and $\bar f$ are denoted as
\beq \label{eq:amplitudes}
\begin{split}
A_f &= \langle f | {\cal H} | D^0 \rangle\,, \ \ \bar A_f =  \langle f | {\cal H} | \D0bar \rangle \,,\cr A_{\bar f}& = \langle \bar f | {\cal H} | D^0 \rangle\,, \ \ \bar A_{\bar f} =  \langle \bar f | {\cal H} | \D0bar \rangle,
\end{split}
\eeq
where ${\cal H}$ is the $\vert \Delta C\vert=1$ weak interaction effective Hamiltonian.
The tree-level dominated decay amplitudes can, in general, be written as
 \beq
\begin{split}
 \label{eq:fouramp}
A_f&=A^0_{f} e^{+i\phi^0_{f}}[1+r_fe^{i(\delta_f+\phi_f)}],\cr
A_{\overline f}&=A^0_{\overline f} e^{i(\Delta^0_{f}+\phi^0_{\overline f})} [1+r_{\overline f} e^{i(\delta_{\overline
f}+\phi_{\overline f})}] ,\cr
\overline{A}_{\overline f}&=A^0_{f}
e^{-i\phi^0_{f}}[1+r_fe^{i(\delta_f-\phi_f)}],\cr
\overline{A}_f&=A^0_{\overline f} e^{i(\Delta^0_{f}-\phi^0_{\overline f})}
[1+r_{\overline f }e^{i(\delta_{\overline f}-\phi_{\overline f })}],
\end{split}
\eeq
where $A^0_{f} $ and $A^0_{\overline f} $
are the magnitudes of the dominant SM contributions,
the ratios $r_f$ and $r_{\overline f} $ are the relative magnitudes of the subleading amplitudes (which are CKM suppressed in the SM, and potentially contain NP contributions),
$\phi_f^0$, $\phi_{\overline f}^0$, $\phi_f$,
and $\phi_{\overline f}$ are CP-odd weak phases %which appear with opposite signs in CP conjugate
and
$\Delta^0_f$, $\delta_f$,  and $\delta_{\overline f}$ are CP-even strong phases. % which appear with the same signs in CP conjugate amplitudes.
With the exception of the weak phases, the quantities entering \eqref{eq:fouramp} are understood to be phase space dependent for three-body and higher multiplicity decays.
Note that $\phi^0_f$ and $\phi^0_{\overline f}$ are quark and meson phase convention dependent.
However, this dependence cancels in physical observables.

In the case of decays to CP eigenstates, $\Delta^0_f = 0\, (\pi)$ for
CP even (odd) final states.
Eq. \eqref{eq:fouramp} therefore reduces to
\beq\label{eq:twoamp}
\begin{split}
A_f&=A^0_{f} e^{+i\phi^0_{f}}[1+r_fe^{i(\delta_f +  \phi_f)}],\cr
 \overline{A}_{f}&= \eta_{f}^{CP} A^0_{f} e^{-i\phi^0_{f}} [1+r_{ f} e^{i(\delta_{f}-\phi_{ f})}] ,
 \end{split}
 \eeq
where $\eta_f^{CP} = + (-)$ for CP even (odd) final states.

For SCS decays, the choice of the dominant and subleading SM amplitudes in \eqref{eq:fouramp} and \eqref{eq:twoamp} is convention dependent. For example,
using CKM unitarity, the leading SCS $D^0$ decay amplitudes could be chosen to be proportional to
$V_{cs}^* V_{us}$, $V_{cd}^* V_{ud}$, or their difference $V_{cs}^* V_{us}-V_{cd}^* V_{ud}$. The last choice is a particularly convenient one that is motivated by $U$-spin flavor symmetry, cf. Section \ref{sec:Uspindecomp}. In all cases, the subleading SM amplitudes are $\propto V_{cb}^* V_{ub}$, and are included in the second term on the RHS of each relation in \eqref{eq:fouramp}, \eqref{eq:twoamp}.  However, the physical observables must   be convention independent.

We divide the CF/DCS decays into two categories:  (i) decays to $K^\pm X$,
where indirect CPV requires interference between a CF and a DCS decay chain, e.g. $D^0 \to K^- \pi^+$ and
$D^0  \to \bar D^0 \to K^- \pi^+$, respectively;
(ii) decays to
$K^0 X$, $\K0bar X$, %denoted as CF$\!K_s$,
where indirect CPV is dominated by interference between two CF decay chains, e.g.
$D^0 \to \overline  K^0 \pi^+ \pi^-  $ and $D^0 \to \bar D^0 \to  K^0  \pi^+ \pi^- $, with subsequent decays $K^0 / \overline K^0 \to \pi^+  \pi^-$. % decay given primarily by $K_s \to \pi\pi$).
In the SM, the CF and DCS $D^0$ decay amplitudes are proportional to $V^*_{cs} V_{ud}$ and $V_{cd}^* V_{us}$, respectively.
Thus,
only the first terms in \eqref{eq:fouramp} are present.
We choose the CF and DCS amplitudes to be $A_f \,,\bar A_{\bar f}$ and
$A_{\bar f} \,,\bar A_{f}$, respectively.
For the computation of the indirect CP asymmetries in case (i),
all four amplitudes in \eqref{eq:fouramp} must be included, whereas in case (ii) we will see that the contributions of the two DCS amplitudes can
be neglected to good approximation.

\subsection{The CPVINT observables}
\label{sec:CPVINTobs}

The time dependent hadronic decay amplitudes sum over contributions with and without mixing, e.g. for CP conjugate decay modes,
\beq \begin{split} \label{hadampsgen}
%A (D^0 (t)  \to f) &=\bar A_f  \langle \D0bar | D^0 (t) \rangle +  A_f  \langle D^0 | D^0 (t)  \rangle , \cr
A (\D0bar (t)  \to f) &=A_f  \langle D^0 |  \D0bar (t) \rangle +  \bar A_f  \langle \D0bar | \D0bar (t) \rangle\,, \cr
A (D^0 (t)  \to \bar f) &=\bar A_{\bar f}  \langle \D0bar | D^0 (t) \rangle +  A_{\bar f}  \langle D^0 | D^0 (t)  \rangle\,.\end{split}
\eeq
Factoring out the unmixed contributions, the time dependent CP asymmetries are seen to depend on the ratios
$ A_f \langle D^0 |  \D0bar (t) \rangle\,/\, \bar A_f  \langle \D0bar | \D0bar (t) \rangle $, and their CP conjugates.
%(and similarly for $D^0 (t)  \to f $).
In turn, \eqref{timeevlnapp} implies that the CP asymmetries are determined by the
quantities $M_{12} \,A_f / \bar A_f $ and $\Gamma_{12} \,A_f / \bar A_f $, as well as their CP conjugates.
%, for interference between decays with and without dispersive mixing, and with and without absorptive mixing, respectively.
Keeping this in mind, we are now ready to define the CPV phases $\phi^M_f$ and $\phi^\Gamma_f$, responsible for dispersive and absorptive CPVINT, respectively.\footnote{In \cite{Branco:1999fs} it was noted that a non-zero value for ${\rm arg}[M_{12}^2 A_f \overline A_f^* A_{\bar f } \overline A_{\bar f}^*]$
or ${\rm arg}[\Gamma_{12}^2 A_f \overline A_f^* A_{\bar f } \overline A_{\bar f}^*]$, equivalent to $2 \phi_f^M$ and $2 \phi_f^\Gamma$, respectively, cf. \eqref{lambdaMf}, \eqref{lambdaMfnonCP}, \eqref{lambdaMfbarnonCP}, implies CP violation.  However, the phenomenology of these phases was not discussed.}

\subsubsection{SCS decays to CP eigenstates}
For SCS decays to CP eigenstate final states, $\phi^M_f$ and $\phi^\Gamma_f$ are the arguments of the CPVINT observables
\beq \begin{split}
\lambda^M_{f} &\equiv \,{M_{12} \over |M_{12} |}\,
{A_f \over \overline A_f} = \,\eta_{f}^{CP}\,\left|{A_f \over \overline A_f}\right| \, e^{ i  \phi_f^M }\,,\cr
\lambda^\Gamma_{f} &\equiv \,{\Gamma_{12} \over |\Gamma_{12} |}\,
{A_f \over \overline A_f} = \,\eta_{f}^{CP}\,\left|{A_f \over \overline A_f}\right| \, e^{ i  \phi_f^\Gamma }
   \,.\label{lambdaMf}\end{split}\eeq
They are given by
\beq\begin{split} \label{phiMfCP}   \phi^{M\,(\Gamma)}_{f} &=  \phi^{M\,(\Gamma)}  +2 \phi_f^0 + 2 r_f \cos\delta_f \sin\phi_f   \,,\cr
  %  \phi^\Gamma_{f} &=  \phi^\Gamma  +\phi_f^0 + 2 r_f \cos\delta_f \sin\phi_f  \,.
    \end{split}
   \eeq
to first order in $r_f$, cf. \eqref{phiMunphys}, \eqref{eq:twoamp}.
We will see that
$\phi^M_f , \phi_f^\Gamma \approx 0$ (rather than $\pi$), given the
sign of the CP conserving observable $y^f_{CP}$, $f=\pi^+\pi^-$, $K^+ K^-$, cf.~\eqref{yCPapp},~\eqref{phiM0}.

\subsubsection{SCS decays to non-CP eigenstates}

For SCS decays to non-CP eigenstate final states, e.g. $D^0 \to K^{*+} K^-$, two pairs of observables are introduced, % pair of observables is introduced,
\beq \begin{split}
\lambda^M_{f} &\equiv \,{M_{12} \over |M_{12} |}\,
{A_f \over \overline A_f} = \,\left|{A_f \over \overline A_f}\right| \, e^{ i  (\phi_f^M -  \Delta_f  )  }\,,\cr
\lambda^\Gamma_{f} &\equiv \,{\Gamma_{12} \over |\Gamma_{12} |}\,
{A_f \over \overline A_f} = \,\left|{A_f \over \overline A_f}\right| \, e^{ i  (\phi_f^\Gamma -  \Delta_f ) }
   \,,\label{lambdaMfnonCP}\end{split}\eeq
and
\beq \begin{split}
\lambda^M_{\bar f} &\equiv \,{M_{12} \over |M_{12} |}\,
{A_{\bar f} \over \overline A_{\bar f}} = \,\left|{A_{\bar f } \over \overline A_{\bar f}}\right| \, e^{ i  (\phi_f^M + \Delta_f  )  }\,,\cr
\lambda^\Gamma_{\bar f} &\equiv \,{\Gamma_{12} \over |\Gamma_{12} |}\,
{A_{\bar f} \over \overline A_{\bar f} }= \,\left|{A_{\bar f} \over \overline A_{\bar f} }\right| \, e^{ i ( \phi_f^\Gamma+ \Delta_f ) }
   \,.\label{lambdaMfbarnonCP}\end{split}\eeq
The dispersive and absorptive CPV phases now satisfy, cf. \eqref{phiMunphys}, \eqref{eq:fouramp},
\beq\begin{split} \label{phiMfnonCP}   \phi^{M\,(\Gamma)}_{f} &=  \phi^{M\,(\Gamma)} +\phi_f^0  +\phi_{\bar f}^0  \cr
&~~~~~+ r_f \cos\delta_f \sin\phi_f  + r_{\bar f} \cos\delta_{\bar f} \sin\phi_{\bar f}   \,,\cr
\end{split}
   \eeq
while the overall strong phase difference in the decay amplitude ratios is given by
\beq
\begin{split}
\Delta_f  & = \Delta_f^0 - r_f \sin \delta_f \cos\phi_f + r_{\bar f} \sin\delta_{\bar f} \cos\phi_{\bar f} \,,
\end{split}
\label{DeltafnonCP}
\eeq
to first order in $r_f$ and $r_{\bar f}$.

\subsubsection{CF/DCS decays to $K^\pm X$}

For CF/DCS decays to $K^\pm X$, e.g. $D^0 \to K^\pm \pi^\mp $, the definitions in \eqref{lambdaMfnonCP}, \eqref{lambdaMfbarnonCP} apply (recall that $A_f$ is the CF amplitude), however we introduce overall minus signs in the equalities, i.e.
\beq \begin{split}
\lambda^M_{f}  &= \,-\left|{A_f \over \overline A_f}\right| \, e^{ i  (\phi_f^M -  \Delta_f  )  },~~~~\lambda^\Gamma_{f}  = \,-\left|{A_f \over \overline A_f}\right| \, e^{ i  (\phi_f^\Gamma -  \Delta_f  )  }\,\cr
\lambda^M_{\bar f}  &= \,-\left|{A_{\bar f} \over \overline A_{\bar f}}\right| \, e^{ i  (\phi_f^M +  \Delta_f  )  },~~~~\lambda^\Gamma_{\bar f}  = \,-\left|{A_{\bar f} \over \overline A_{\bar f}}\right| \, e^{ i  (\phi_f^\Gamma +  \Delta_f  )  }
\label{lambdaCFDCS}\,.\end{split}\eeq
Thus,
the dispersive and absorptive CPV phases
satisfy
\beq\begin{split} \label{phiMfnonCPCFDCS}   \phi^{M\,(\Gamma)}_{f} &=  \phi^{M\,(\Gamma)} +\phi_f^0  +\phi_{\bar f}^0 +\pi  \cr
&~~~~~+ r_f \cos\delta_f \sin\phi_f  + r_{\bar f} \cos\delta_{\bar f} \sin\phi_{\bar f}   \,,\cr
\end{split}
   \eeq
and the expression for the strong phase in \eqref{DeltafnonCP} is not modified.  The sign convention in \eqref{lambdaCFDCS} yields $\phi_f^M$, $\phi_f^\Gamma \approx 0$ (rather than $\pi$), %after taking into account the phases of the CF/DCS CKM factors,
as in SCS decays.
In the SM and, more generally,
in models with negligible new weak phases in CF/DCS decays, the second line in \eqref{phiMfnonCPCFDCS} is absent, and the
dispersive and absorptive phases are separately equal for all decays in this class. Moreover, the absence of direct CPV yields
the relation $ |A_{\bar f} /\overline{A}_{\bar f}|=|A_f /\overline A_{f} |^{-1}$.

\subsubsection{CF/DCS decays to $K^0 X ,\,\K0bar X$}
\label{sec:K0observables}
Next, we define the CPVINT observables for $D^0 $/$\D0bar$ decays to final states $f=[\pi^+\pi^-] X$,
where the square brackets indicate that the pion pair originates from decays of a $K_S$ or $K_L$,
i.e. two step transitions of the form
$D^0 \to  [ K_{S,L} \to  \pi^+ \pi^-  ] +X $.
In order to achieve SM sensitivity to CPVINT,
the contributions of CPV in the $K$ system must be taken into account.
The neutral $K$ mass eigenkets are written as,
\beq\begin{split}\label{KaonEig}
 | K_S \rangle &= p_K | K^0 \rangle  + q_K|  \K0bar \rangle ,\cr
| K_L \rangle  &= p_K  |  K^0 \rangle- q_K |  \K0bar \rangle  \,. \end{split}
\eeq
The corresponding eigenbras are given in the ``reciprocal basis''  \cite{Branco:1999fs,Silva:2000db},
\beq\begin{split}\label{recipKaonEig}
 \langle \tilde K_S | &= \frac12 \left(p^{-1}_K \langle K^0 |+ q^{-1}_K \langle \K0bar |  \right),\cr
\langle  \tilde K_L |  &= \frac12 \left( p^{-1}_K  \langle  K^0 | - q^{-1}_K  \langle \K0bar | \right)\,,\end{split}
\eeq
where CPT invariance has been assumed.
To excellent approximation (see, e.g. \cite{Nir:1992uv}),
\beq \left|{p_K \over  q_K }\right| = 1+ 2  {\rm Re}[\epsilon_K]   \, .\label{absepsK} \eeq
The experimental values of the real and imaginary parts of the kaon CPV parameter
$\epsilon_K $  are \cite{PDG},
\beq\label{epsKexp} \begin{split}
\epsilon_R \equiv {\rm Re}[\epsilon_K]  &=(1.62 \pm  0.01)  \times 10^{-3}\,,\cr
\epsilon_I   \equiv {\rm Im}[\epsilon_K]  &=(1.53 \pm  0.01)  \times 10^{-3} \,.     \end{split}\eeq
We have obtained them from the quoted measurements of
$\eta_{00}$ and $\eta_{+-}$, ignoring correlations in their errors.

In general, due to the presence of the two intermediate states $K_S  X$ and $K_L X$, there are four pairs of
CPVINT observables, % to consider for final state $f$, as well as for final state $\bar f$,
\begin{eqnarray}\label{genlambdaK0}
\lambda^M_{K_a X} &\equiv&  \,{M_{12} \over |M_{12} |}{A_{K_a X} \over \overline A_{K_a X }}\,,~~\lambda^\Gamma_{K_a X} \equiv  \,{\Gamma_{12} \over |\Gamma_{12} |}{A_{K_a X} \over \overline A_{K_a X }} \,,\\
\lambda^M_{{\overline{K_a \!X}}} &\equiv&  \,{M_{12} \over |M_{12} |}{A_{\overline{K_a X}} \over \overline A_{\overline{K_a \!X} }}\,,~~\lambda^\Gamma_{{\overline{K_a \!X}}} \equiv  \,{\Gamma_{12} \over |\Gamma_{12} |}{A_{\overline{K_a X}} \over \overline A_{\overline{K_a \!X} }} \,,~a=S,L, \nonumber
\end{eqnarray}
where the first and second lines correspond to the CP conjugate final states $f=[\pi^+\pi^-] X$ and $\bar f = \overline{[\pi^+\pi^-] X}$, respectively.  Note that for the important case of
$X=\pi^+ \pi^-$, $\bar{f}$ corresponds to interchange of the Dalitz plot variables
$(p_K + p_{\pi^+})^2 \leftrightarrow (p_K + p_{\pi^-})^2$ in $f$.
We can express the CPVINT observables \eqref{genlambdaK0} in the form
\begin{eqnarray}\label{genlambdaK0form}
&&\!\!\!\!\!\!\!\!\!\!\lambda^{M,\Gamma}_{K_{S/L} X} =\pm  \left|{A_{K_{S/L} X}\over\overline A_{K_{S/L} X}}\right| \,e^{i (\phi^{M,\Gamma} [K_{S/L} X]\, -\,\Delta [K_{S/L} X])}\,,\\
&&\!\!\!\!\!\!\!\!\!\!\lambda^{M,\Gamma}_{{\overline{K_{S/L} \!X}}} = \pm\left|{\overline A_{\overline{ K_{S/L} X}} \over A_{\overline{ K_{S/L} X }}}\right|\,
e^{i (\phi^{M,\Gamma} [K_{S/L} X]\,+\,\Delta [K_{S/L} X])}\,,\nonumber
\end{eqnarray}
where the overall plus and minus signs refer to the $K_S $ and $K_L $, respectively.
The four CPVINT phases and two strong phases in \eqref{genlambdaK0form} are $\phi^{M,\Gamma} [K_{S/L} X]$ %, $\phi^{M,\Gamma} [K_{L} X ]$
and $\Delta [K_{S/L} X]$, respectively.

The $D$ decay amplitudes in \eqref{genlambdaK0} satisfy,
\beq
\begin{split}  \label{DtoKaXamps}
 A_{K_{S/L} X} &= \frac12 (\pm  q_K^{-1} A_{\K0bar X} + p_K^{-1} A_{K^0 X} ) \,,\cr
 \overline A_{K_{S/L} X} &= \frac12 (p_K^{-1} \overline A_{K^0 X}  \pm  q_K^{-1} \overline A_{\K0bar X} )\,,\cr
 A_{\overline{K_{S/L} \!X}} &= \frac12 (q_K^{-1} A_{\overline{K^0 \!X}} \pm p_K^{-1} A_{\overline{\K0bar\! X}} ) \,,\cr
\overline A_{\overline{K_{S/L} \!X}} &= \frac12 (\pm p_K^{-1} \overline A_{\overline{\K0bar \! X}} +q_K^{-1}\overline A_{\overline{K^0 \!X}} )\,,
\end{split}
\eeq
where we have used the reciprocal basis \eqref{recipKaonEig}, and the first and second terms on the RHS in each relation are the dominant CF and subleading DCS contributions, respectively.

In the SM and, more generally, in models with negligible new CPV phases in CF/DCS decays,
the DCS decay amplitudes introduce relative corrections of $O(\theta_{C}^2)$ to the weak phases, strong phases, and magnitudes of $\lambda^{M,\Gamma}_{K_{S/L} X}$,
$\lambda^{M,\Gamma}_{\overline{K_{S/L} \!X}}$, making it a good approximation to neglect them.  % in \eqref{genlambdaK0}.
(We assess the impact of the DCS amplitudes on approximate universality in Section \ref{finstatedepK0}.)
In this limit,  \eqref{genlambdaK0} reduces to
\beq  \begin{split} \label{lamKSvsKL}
\lambda_f^M \equiv  \lambda^{M}_{K_S X } &= -\lambda^{M}_{K_L X} ={M_{12} \over |M_{12} |}\,{p_K\over q_K }\,
{A_{\K0bar X} \over \overline A_{K^0 X }}\,,\cr
\lambda_f^\Gamma \equiv \lambda^{\Gamma}_{K_S X } &= -\lambda^{\Gamma}_{K_L X} ={\Gamma_{12} \over |\Gamma_{12} |}\,{p_K\over q_K }\,
{A_{\K0bar X} \over \overline A_{K^0 X }}\,,\cr
\lambda_{\bar f}^M\equiv  \lambda^{M}_{\overline{K_S X } } &= -\lambda^{M}_{\overline{K_L X}} ={M_{12} \over |M_{12} |}\,{p_K\over q_K }\,
{A_{\overline{K^0 \!X}}\over  \overline A_{\overline{\overline K^0\! X}} }\,,\cr
\lambda_{\bar f}^\Gamma \equiv \lambda^{\Gamma}_{\overline{K_S X }} &= -\lambda^{\Gamma}_{\overline{K_L X}} ={\Gamma_{12} \over |\Gamma_{12} |}\,{p_K\over q_K }\,
{A_{\overline{K^0 \!X}} \over  \overline A_{\overline{\overline K^0\! X}} }\,. \cr
\end{split}\eeq
Thus, in the limit of negligible new CPV phases in CF/DCS decays, it is a good approximation to consider a single pair of CPVINT observables for final state $f=[\pi^+ \pi^- ] X$, and a single pair for $\bar f=\overline{[\pi^+ \pi^- ] X}$, which we have denoted in \eqref{lamKSvsKL} as $\lambda_f^M $, $\lambda_f^\Gamma $ and  $\lambda_{\bar f}^M $, $\lambda_{\bar f}^\Gamma $, respectively.
They can be expressed in terms of dispersive and absorptive CPVINT phases as
\beq \begin{split}
\lambda^{M\,(\Gamma)}_f &=
\left|{ p_K\, A_{\K0bar X} \over q_K\,\overline A_{K^0 X }}\right| \, e^{ i  (\phi_{f}^{M\,(\Gamma)} -  \Delta_{f } )  }\,
,\cr
\lambda^{M\,(\Gamma)}_{\bar f }&=
\left|{p_K\, \overline A_{K^0 X } \over q_K\,A_{\K0bar X}}\right| \, e^{i  (\phi_{f}^{M\,(\Gamma)} + \Delta_{f } )  },
   \label{lambdaMfCFK0}\end{split}\eeq
where the amplitude relations,
\beq | \overline A_{\overline{\K0bar \! X}}/A_{{\K0bar \! X}}|=|A_{\overline{K^0 \!X}}/ \overline A_{K^0 X} |=1\,,\eeq
valid in the limit of vanishing direct CPV, have been employed in the second relation.
Note that the weak phases $\phi^{M\,,\Gamma} [K_{S/L} X]$ and strong phases $\Delta [K_{S/L} X]$, defined in general in
\eqref{genlambdaK0form}, reduce to $\phi_f^{M,\Gamma}$ and $\Delta_f$, respectively.

The strong phase difference $\Delta_{f}$ (between $\overline A_{K^0 X }$ and $A_{\K0bar X}$) is generally non-vanishing and phase space dependent for multi-body intermediate states, e.g.
$X= \pi^+ \pi^- $.
The weak phases %in \eqref{lambdaMfCFK0}, \eqref{lambdaMfCFK0twobody}
satisfy
\beq\begin{split} \label{phiMfCFK0}   \phi^{M \,(\Gamma)}_{f} &=  \phi^{M\,(\Gamma)}  +2 \phi_{\K0bar X}^0 + {\rm arg}(p_K /q_K )\,,
  %  \phi^\Gamma_{f} &=  \phi^\Gamma  +\phi_f^0 + 2 r_f \cos\delta_f \sin\phi_f  \,.
    \end{split}
   \eeq
where $\phi_{\K0bar X}^0$ is the weak phase of the CF amplitudes $A_{\K0bar X}$, $A_{\overline{K^0 X}}$, cf. \eqref{eq:fouramp}, while ${\rm arg}(p_K /q_K )$ introduces a dependence on CPV in the $K$ system, cf. Section \ref{finstatedepK0}.
Note that $\phi_f^M$ and $\phi_f^\Gamma$ are separately equal for all %decays in this class
final states in this class.

In the case of two-body (and quasi two-body) intermediate states, the CPVINT observables in \eqref{lambdaMfCFK0}    %, \eqref{lambdaMfCFK0CP}
reduce to
\beq
\lambda^{M\,(\Gamma)}_{f} = \eta_{f}^{CP} \,\left|{ p_K \over q_K }\right| \, e^{ i  \phi_{f}^{M\,(\Gamma )}}\,,
\label{lambdaMfCFK0twobody}
\eeq
where
\beq \label{etaKsX} \eta_{f}^{CP} \equiv  (-)^L \times CP[X]\,,\eeq
$L$ is the orbital angular momentum of the intermediate states $K_{S/L} X$, and $CP[X] = +\,(-)$
for CP even (odd) $X$.
For example, $\eta_f^{CP} =-1$ for $f =K_S\, \omega$, $K_S \pi^0$, and $\eta_f^{CP} =+1$ for
$f=K_S f_0$. (Equivalently, $\eta_{CP}^f = +1(-1)$ for CP even (odd) intermediate state $K_S X$.)

Finally, we point out that in all three classes of $D^0$ decays discussed in this Section, the quark (CKM) phase convention dependence
cancels in $\phi^M_f$ and $\phi_f^\Gamma$, i.e.
between the first two terms on the RHS of \eqref{phiMfCP}, the first three terms on the RHS of \eqref{phiMfnonCP}, and  between all three terms in \eqref{phiMfCFK0}, cf. Section \ref{sec:finstatedep}. %, as is required for physical observables.
Moreover, they are always related to the pure mixing phase $\phi_{12}$ %, see \eqref{phi12}, \eqref{ASL},
as
\beq \phi_{12} = \phi_{f}^M - \phi_{f}^\Gamma \,,\label{phiMfmphiGf}\eeq
i.e. the final state dependent effects are common to the dispersive and absorptive phases.

\subsubsection{Relation to other parametrizations of CPVINT}

It is instructive to relate the parametrization of indirect CPV effects in terms of absorptive and dispersive phases
to the more familiar one currently in use.  The latter consists
of the CPVMIX parameter,
\beq |q/p|-1 \,,\eeq
and the final state dependent phenomenological CPVINT phases $\phi_{\lambda_f}$, which appear in the arguments of the observables $\lambda_f$,
see e.g. \cite{Nir:1992uv}.  We begin with the definitions of the $\lambda_f$, corresponding to the absorptive and dispersive observables $\lambda_f^{M,\Gamma}$, in the different classes of decays.
For SCS decays to CP eigenstate final states, they correspond to the observables in  \eqref{lambdaMf}, and are given by\footnote{In our convention for $\lambda_f^M$, $\lambda_f^\Gamma$, the numerators
correspond to the transitions $\D0bar \to D^0 \to f$, whereas in $\lambda_f$ they correspond to $D^0 \to \D0bar \to f$.}
\beq{\lambda_f} \equiv \frac{q}{p} \frac{ \bar A_f }{A_f} = - \eta_f^{ CP} \left|\lambda_f \right| e^{i \,\phi_{\lambda_f}}\,.\label{lambdaf}\eeq
For SCS decays to non-CP eigenstate final states, and CF/DCS decays to $K^\pm X$, the $\lambda_f$
corresponding to the observables in \eqref{lambdaMfnonCP}, \eqref{lambdaMfbarnonCP}, and \eqref{lambdaCFDCS}
are given by,
\beq
\begin{split}
 {\lambda_f}& \equiv\frac{q}{p} \frac{ \bar A_{f} }{A_{f}}=\mp\left|\lambda_f\right| e^{i \,(\phi_{\lambda_f}+ \Delta_f )}, \cr
 {\lambda_{\bar f}} &\equiv  \frac{q}{p} \frac{ \bar A_{\bar f} }{A_{\bar f}}=\mp\left|\lambda_{\bar f}\right| e^{i \,(\phi_{\lambda_{f}}- \Delta_f ) } \,,\label{lambdaffbar}
 \end{split}
 \eeq
where the $\mp$ sign conventions in the right-most relations apply to the SCS and CF/DCS cases,
respectively.

Finally, for CF/DCS decays to $K^0 X$, $\K0bar X$ (given negligible new CPV phases in the decay amplitudes, and neglecting the  DCS contributions)
the $\lambda_f$ correspond to
the absorptive and dispersive observables in \eqref{lamKSvsKL}, \eqref{lambdaMfCFK0}, and are given by
\beq  \begin{split} \label{lamfKSvsKL}
\lambda_f \equiv  {q \over p}\,{q_K\over p_K }\,
{\overline A_{K^0 X } \over A_{\K0bar X}}  = - |\lambda_f |\, e^{i \,(\phi_{\lambda_f}+ \Delta_f )}    \,,\cr
\lambda_{\bar f} \equiv  {q \over p}\,{q_K\over p_K }\,
{ \overline A_{\overline{\overline K^0\! X}} \over A_{\overline{K^0 \!X}} }  = - |\lambda_f | \,e^{i \,(\phi_{\lambda_f}- \Delta_f )}, \cr
\end{split}\eeq
for final states $f=[\pi^+ \pi^- ] X$ and $\bar f=\overline{[\pi^+ \pi^- ] X}$.
In the case of two-body or quasi two-body intermediate states, corresponding to the observables in \eqref{lambdaMfCFK0twobody},
these expressions reduce to,
\beq
\lambda^{M\,(\Gamma)}_{f} = \eta_{f}^{CP} \,\left| {q \over p} { q_K \over p_K }\right| \, e^{ i  \phi_{\lambda_f}}\,.
\label{lambdafCFK0twobody}
\eeq
The sign conventions in the right-most relations of \eqref{lambdaf}--\eqref{lambdafCFK0twobody} yield all $\phi_{\lambda_f }\approx 0$ (HFLAV convention for $D_2$), or all $\approx \pi$,
for the three classes of decays.

The CPV parameters $|q/p|-1$ and $\phi_{\lambda_f}$ are expressed in terms of the absorptive and dispersive
CPV phases as
\beq \label{qovpphi12}
\left|{q\over p}\right| -1  ={  x_{12} \,y_{12} \sin\phi_{12}\over x_{12}^2 + y_{12}^2 } \,  \left[1+ O(\sin\phi_{12}) \right]\,,\eeq
where $\phi_{12} =  \phi_f^M - \phi^\Gamma_f $, cf. \eqref{phiMfmphiGf},
and
\beq
\tan 2 \phi_{\lambda_f}  = - \left(\,{x_{12}^2 \sin 2 \phi_{f}^M + y_{12}^2 \sin 2 \phi_{f}^\Gamma \over
x_{12}^2 \cos 2 \phi_{f}^M + y_{12}^2 \cos 2 \phi_{f}^\Gamma }\,\right).\label{tan2phifnew}\eeq
Eq. \eqref{tan2phifnew} is obtained by multiplying both sides of \eqref{qovp} by $(\bar A_f/ A_f  )^2$ and $(\bar A_f \bar A_{\bar f} / A_f   A_{\bar f} ) $ for CP eigenstate and non-CP eigenstate final states, respectively, and holds for all classes of decays.
To lowest order in the CPV phases, % and, correspondingly, to $O(x_{12} , y_{12})$ in the mixed component probability amplitudes, cf.
it equates %yields
the phenomenological CPVINT phase $\phi_{\lambda_f}$ to a sum over the dispersive and absorptive
CPVINT phases, $\phi_f^M$ and $\phi_f^\Gamma$, weighted by the
ratios $x_{12}^2 / (x_{12}^2 + y_{12}^2)$ and $y_{12}^2 / (x_{12}^2 + y_{12}^2)$, respectively.  These weights are, respectively, the leading dispersive and absorptive contributions to the CP averaged mixing probability, $ |\langle \D0bar | D^0 (t) \rangle |^2 + |\langle D^0 | \D0bar (t) \rangle |^2$, cf. \eqref{timeevlnapp}.

Indirect CPV can be equivalently described in terms of the
parameters emphasized in this work, i.e. $\phi_{f}^M$, $\phi_{f}^\Gamma$, $x_{12}$, $y_{12}$, or the more familiar ones $|q/p|$, $\phi_{\lambda_f}$, $x$, $y$, cf.  \eqref{x12vsx}, \eqref{phiMfmphiGf}, \eqref{qovpphi12}, \eqref{tan2phifnew}.  %cf. Eqs. \eqref{qovpphi12}, \eqref{tan2phifnew}, and \eqref{x12vsx}.
Indeed, \eqref{phiMfmphiGf} implies that the same number of independent parameters is employed in each case.

Finally, we remark on the CPV observables $\Delta x_f $
\cite{DiCanto:2018tsd} and
$\Delta y_f$, which have been measured in
tandem by the LHCb collaboration \cite{Aaij:2019jot} in
$D^0 \to K_S \pi^+ \pi^-$ decays.  They are defined in terms of
$\phi_{\lambda_f}$ and $|q/p|$ as\footnote{To be fully general, we
  have replaced $\phi$ with $\phi_{\lambda_f} $, and added a subscript
  $f$ to $\Delta x$ and $\Delta y$ in the definitions of
  \cite{DiCanto:2018tsd}.}
\begin{equation*}
\begin{split}
2 \,\Delta x_f &= x \cos\phi_{\lambda_f} \left( \left|{q\over p} \right| - \left|{p\over q} \right|\right) + y \sin\phi_{\lambda_f } \left( \left|{q\over p} \right| + \left|{p\over q} \right|\right)\,,\cr
2 \,\Delta y_f & =  y \cos\phi_{\lambda_f} \left( \left|{q\over p} \right|- \left|{p\over q} \right|\right)- x \sin\phi_{\lambda_f } \left( \left|{q\over p} \right| + \left|{p\over q} \right|\right)\,.
\end{split}
\end{equation*}
The observable $-\Delta y_f$ is equivalent to the familiar CPVINT asymmetry $\Delta Y_f$ for
SCS decays to CP eigenstate final states, cf. \eqref{eq:yCPdefExp}.
Translating to the dispersive/absorptive parametrization via \eqref{qovpphi12}, \eqref{tan2phifnew}, we obtain\footnote{We have used the relations
$y \cos\phi_{\lambda_f}= y_{12} \cos\phi_f^\Gamma$, and $ x \cos\phi_{\lambda_f}=x_{12} \cos\phi_f^M $, which hold up to negligible relative corrections quadratic in the CPV phases.}
\beq \label{xytransln}
\Delta x_f  = -y_{12} \sin\phi_f^\Gamma,~~~\Delta y_f  = x_{12} \sin\phi_f^M\,,\eeq
to leading order in $\sin \phi_f^{M,\Gamma}$.
Thus, the use of the parameters $\Delta x_f $ and $\Delta y_f $ is equivalent to the CPVINT parametrization
in terms of $\phi_f^M$ and $\phi_f^\Gamma $, respectively, modulo the corresponding dispersive and absorptive mixing factors. (It is amusing that interchange of the $\Delta x$ and $\Delta y$ labels turns out to be appropriate).
Interestingly, we will see that experimental sensitivity to $\phi_f^\Gamma$ (or $\Delta x_f$) requires a non-trivial strong phase difference between decay amplitudes, i.e. non-CP eigenstate final states, e.g. $f= K_S \pi^+ \pi^-  , \,K^+ \pi^-$.

\section{The indirect CP asymmetries}

\label{sec:indirectCPasym}

We can now derive expressions for the time-dependent decay widths and CP asymmetries in terms of the absorptive and dispersive CPV phases.
(A discussion of CPV in certain time-integrated decays is deferred to  Appendix A.)
\subsection{Semileptonic decays}
We begin with the CPVMIX ``wrong sign'' semileptonic CP asymmetry,
\beq
\begin{split}
a_{\rm SL} &\equiv {\Gamma (D^0 (t) \to \ell^- X) -\Gamma (\overline{D^0} (t) \to \ell^+  X) \over
       \Gamma (D^0 (t) \to \ell^-  X) +\Gamma (\overline{D^0} (t) \to \ell^+  X)  }\,,
\cr
       &={  |\langle \D0bar | D^0 (t) \rangle |^2 -  |\langle D^0 | \overline{D^0} (t) \rangle |^2  \over  |\langle \D0bar | D^0 (t) \rangle |^2 + |\langle D^0 | \overline{D^0}  (t) \rangle |^2} \,.
       \end{split}\label{semileptonic}
       \eeq
In the second line the semileptonic decay amplitude factors have been cancelled, given negligible direct CPV in these decays, i.e. $|\bar{A}_{\ell^- X}|=|{A}_{\ell^+ X}|$.
In turn, the expressions for the mixed amplitudes in \eqref{timeevln} or \eqref{timeevlnapp} yield the semileptonic asymmetry,
\beq
\label{aslphi12}
a_{\rm SL}= { 2 x_{12} \,y_{12}\over x_{12}^2 + y_{12}^2  }  \,\sin\phi_{12} \,.
\eeq
Note that the CP-even phase difference between the interfering
dispersive and absorptive mixing amplitudes, required to obtain CPVMIX, is provided by the dispersive mixing phase $\pi/2$
in the first line of \eqref{timeevln}.

\subsection{Hadronic decays}
\label{sec:hadronicdecays}
The hadronic decay amplitudes sum over contributions with and without mixing, cf. \eqref{hadampsgen} (substitute $f \leftrightarrow \bar f$ for the CP conjugate final states).
%\beq \begin{split}
%A (D^0 (t)  \to f) &=\bar A_f  \langle \D0bar | D^0 (t) \rangle +  A_f  \langle D^0 | D^0 (t)  \rangle , \cr
%A (\D0bar (t)  \to f) &=A_f  \langle D^0 |  \D0bar (t) \rangle +  \bar A_f  \langle \D0bar | \D0bar (t) \rangle\,.\end{split}
%\eeq
The corresponding time-dependent decay rates are identified with their magnitudes squared.  They are expressed in terms of the CPVINT observables $\lambda_{f,\bar f}^M$, $\lambda_{f,\bar f}^\Gamma$,  cf. \eqref{lambdaMf}, \eqref{lambdaMfnonCP}, \eqref{lambdaMfbarnonCP}, as ($\tau \equiv \Gamma_D t$),
\begin{eqnarray}
  \label{timedepwidth}
\! \Gamma (\D0bar (t) \to f )&=&  e^{-\tau} |\bar{A}_f |^2 \bigg\{ 1 -
\tau   \,{{\rm Re}}\big[ i \,\lambda_f^M  x_{12}  +
\lambda_f^\Gamma y_{12}  \big] \nonumber \\
&+&{\tau^2\over 4} \bigg(  \big(\,   | \lambda_f^M|^2 -1\, \big)
   x^2_{12}+\big(\,|\lambda_f^\Gamma|^2 +1\,\big) y^2_{12} \nonumber
  \\
&+& 2\, x_{12}\,y_{12} \,{{\rm Im}}\big[{\lambda_f^M}^* \lambda_f^\Gamma  \big] \bigg)   \bigg\},\\
\! \Gamma (D^0 (t) \to f )&
  =&  e^{-\tau} |{A}_f |^2 \bigg\{ 1 -
\tau   \,{{\rm Re}}\big[ i\, { x_{12}/ \lambda_f^M }  +
{ y_{12}/ \lambda_f^\Gamma} \big] \nonumber\\
&&\mkern-40mu +{\tau^2\over 4} \bigg(   \left({1/| \lambda_f^M|^{2} }
   -1\right)x^2_{12} +\left({1/|\lambda_f^\Gamma|^{2} } +1
   \right)y^2_{12}\nonumber\\ &&\mkern-40mu +   {2 \,x_{12}\,y_{12}\,\ {\rm Im}\big[1/({\lambda_f^M}^* \lambda_f^\Gamma) \big] }  \bigg)   \bigg\},\nonumber
\end{eqnarray}
with the expressions for $\Gamma(\D0bar (t) \to \bar f)$ and $\Gamma(D^0 (t) \to \bar f)$ obtained via the substitutions $f \to \bar f$  in \eqref{timedepwidth}.
Note that throughout this work appropriate normalization factors are implicit in all decay width formulae, including \eqref{timedepwidth}.  The expressions in \eqref{timedepwidth} are applied to the following cases: SCS decays to CP eigenstates, SCS decays to non-CP eigenstates, and CF/DCS decays to $K^\pm X$.  The description of CF/DCS decays to $K^0 X$, $\K0bar X$ requires a separate treatment, cf Section~\ref{sec:K0Xtimedep}.

\subsubsection{SCS decays to CP eigenstates}
\label{sec:SCSCPwidths}
This category includes, for example, the decays $D^0\! \to\! K^+ K^- \! / \pi^+ \pi^-$.
(We comment on the decay $D^0 \to K^0 \overline K^0$ at the end of Section~ \ref{sec:SCSfinstatedep} ).
The
time-dependent decay widths $D^0 (t) \to f$ and $\D0bar (t) \to f$, expressed in terms of $\phi_f^M$ , $\phi_f^\Gamma$, cf. \eqref{phiMfCP}, and the direct CP asymmetry,
\beq a_f^d \equiv  1 - \left|\bar A_f / A_f \right| = - 2 r_f \sin\delta_f\,\sin\phi_f \,,\label{afdCP}\eeq
cf. \eqref{eq:twoamp}, are given by
\beq\begin{split} \label{timedepSCSCP}
\Gamma (D^0 (t) \to  f ) &=e^{-\tau} |A_f |^2  \left( 1   + c_f^{+ } \,\tau +  c_f^{\prime +  }\,\tau^2 \right)\,,\cr
\Gamma (\D0bar (t) \to  f ) &=e^{-\tau} |\bar A_{f} |^2  \left( 1   +  c_f^{- }\, \tau+ c_f^{\prime-} \,\tau^2\right)\,,\cr
\end{split}\eeq
where the coefficients $c_f^\pm$, $c_f^{\prime \pm}$ satisfy
\beq \label{cpmSCSCP}
\begin{split}
c_f^\pm\, = \,& \eta^f_{CP}  \left[\, \mp x_{12} \sin\phi_f^M   -y_{12}\cos\phi_f^\Gamma   \,(1 \mp a_f^d )\, \right],\cr
 c_f^{\prime \pm}\, = \,&   {\textstyle \frac{1}{2}} y_{12}^2   \pm {\textstyle \frac{1}{4}} (x_{12}^2 +y_{12}^2 ) \,\left(a_{\rm SL}  - 2 a_f^d \right)\,.
 \end{split}\eeq
Terms involving $a_f^d$ have been expanded to first order in CPV quantities, and
the semileptonic CP asymmetry, expressed in terms of $\phi_{12}$,
is given in \eqref{aslphi12}.

The $O(\tau^2 )$ terms in the SCS widths are usually neglected, due to an
$O(x_{12}, y_{12} )$ suppression relative
to the $O(\tau )$ term.   Thus, it has been traditional to express the SCS widths in the approximate exponential forms,
\begin{eqnarray} \label{time-dep}
\Gamma({D}^0(t)\to f)&=&|A_f |^2 \exp[-\hat\Gamma_{D^0\to f}\ \tau],\nonumber\\
\Gamma(\overline{D^0}(t)\to f)&=&|\bar A_f |^2 \exp[-\hat\Gamma_{\overline{D^0}\to
  f}\ \tau],
\end{eqnarray}
where the decay rate parameters satisfy
\beq \label{eq:Gammahatabsdisp}
\hat\Gamma_{D^0 /\D0bar \to f}=
1-c^\pm ,\eeq
cf. \eqref{cpmSCSCP}.
As the goal of SM sensitivity comes into view, i.e. $\phi^M_f  , \phi^\Gamma_f  = O({\rm few})\!\times\! 10^{-2}$,
this will not necessarily be a good approximation, as can be seen by comparing the CP-odd terms in $c_f^\pm$, and the CP-even term in $c_f^{\prime \pm }$.  However, the CP-odd terms in $c_f^{\prime \pm }$ are
further suppressed by CPV parameters, and can be neglected. Thus, to good approximation,
\beq c_f^{\prime \pm}\, = {\textstyle \frac{1}{2}} y_{12}^2  \,.\eeq

Measurements of the time-dependent decay rates at linear order in $\tau$ yield the known
CP conserving observables,
\beq\begin{split}
\label{eq:yCPdef} y^f_{CP}&\equiv  -{(c_f^+ + c_f^-  )\over 2} \,
,\end{split}
\eeq
and the CPVINT asymmetries,
\beq \begin{split} \label{dYfdef}
\Delta Y_f & \equiv   {(c_f^+ - c_f^- )\over 2}\,.\end{split}
\eeq
The average of $\Delta Y_f$ over $f=K^+ K^- , \pi^+\pi^-$ is denoted by $A_\Gamma$.
In the exponential approximation, the corresponding definitions are,
\beq\begin{split}
\label{eq:yCPdefExp}   y^f_{CP}&\equiv{\hat\Gamma_{D^0\to f_{CP}} +\hat\Gamma_{\overline{D^0}\to f_{CP}}  \over 2  } -1, \cr
\Delta Y_f &\equiv   {\hat \Gamma_{\overline D^0 \to f}- \hat \Gamma_{D^0 \to f} \over 2 }.\end{split}
\eeq

Applying \eqref{cpmSCSCP}, and neglecting contributions quadratic in CPV, we obtain
\beq
y^f_{CP}= \eta^{CP}_f   y_{12} \cos\phi_f^\Gamma \,.\label{yCPapp}\eeq
The experimental average over $f=K^+ K^-, \pi^+ \pi^-$ \cite{Amhis:2016xyh}
yields $y^f_{CP} /\eta^{CP}_f  >0$, or
\beq y^f_{CP} = \eta^{CP}_f   y_{12} =  \eta^{CP}_f   |y |\,,\eeq
to excellent approximation.
Furthermore, fits to the data \cite{Amhis:2016xyh,UTfit} yield $x y  >0$ at $3\sigma$, or $\phi_{12} \approx 0$ (rather than $\pi$), cf. \eqref{xyx12y12}.
Thus,  % with the fit result $\phi_{12} \approx 0$ \cite{Amhis:2016xyh,UTfit} (rather than $\pi$), which follows from
we learn that both
\beq \phi_{f}^M \approx 0,~~~~~\phi_f^\Gamma \approx 0\,.\label{phiM0}\eeq

At first order in CPV, \eqref{cpmSCSCP} yields the relation (already noted in \eqref{xytransln} for the CPVINT part),
\beq \begin{split} \label{DYfresult} \Delta Y_f &= \eta_{CP}^f \,(- x_{12} \sin\phi_f^M  + a_f^d \,y_{12}  )\,.
\end{split}
\eeq
The direct CPV contribution in \eqref{DYfresult} is formally subleading, cf. Section~\ref{sec:SCSfinstatedep}.  In general, it can be disentangled experimentally from the dispersive CPV contribution
with the help of time integrated CPV measurements, in which $a_f^d$ enters without mixing suppression, cf. Appendix~\ref{sec:AppendixA}.

It is noteworthy that $\Delta Y_f$ depends on $\phi_f^M$, but not on $\phi_f^\Gamma$.  This is because CP asymmetries require a non-trivial CP-even phase difference $\delta$ between the interfering amplitudes, i.e., they are proportional to $\sin\delta$.  In general, for CP eigenstate final states there is a CP-even phase difference between decays with and without dispersive mixing, namely the $\pi/2$ dispersive phase in \eqref{timeevln}.  However, there is none between decays with and without absorptive mixing (the strong phase between $A_f$ and $\overline A_f$ is trivial).  Therefore, in general, $\phi_f^\Gamma$ can only be measured %via CPVINT
in decays to non-CP eigenstate final states, where the requisite CP-even phase is provided by
the strong phase difference $\Delta_f$ between $A_f$ and $\overline A_f$, as we will see explicitly below.
Finally, in the case of CP averaged decay rates, interference terms are in general proportional to $\cos\delta$, rather than $\sin\delta$.  Therefore,
in the CP averaged time dependent decay rates for CP eigenstate final states, the interference between decays with and without dispersive mixing will vanish at leading order in the mixing, i.e. $O(\tau)$, only leaving a dependence on $y_{12}$. This is borne out by the expression for $y^f_{CP}$ in \eqref{yCPapp}. %, and the CP-even contribution to the coefficients $c_f^{\prime \pm}$, cf. \eqref{cpmSCSCP}.

\subsubsection{SCS decays to non-CP eigenstates}
\label{sec:SCSnonCPwidths}
This category includes, for example, the decays $D^0 \to \rho \pi$, $K^{*+} K^- $.
The time dependent decay widths are of the form
\beq\begin{split} \label{timedepSCSCPf}
\Gamma (D^0 (t) \to  f ) &=e^{-\tau} |A_f |^2  \left( 1   +  \sqrt{R_f } c_f^{+ } \,\tau + R_f  c_f^{\prime +  }\,\tau^2 \right)\,,\cr
\Gamma (\D0bar (t) \to  { f} ) &=e^{-\tau} |\bar A_{ f} |^2  \left( 1   + {1\over    \sqrt{R_{f} } } c_f^{- }\, \tau+ {1\over    R_f  }c_f^{\prime-} \,\tau^2\right)\,,\cr
\end{split}\eeq
for final state $f$, and
\beq\begin{split} \label{timedepSCSCPfbar}
\Gamma (D^0 (t) \to  \bar f ) &=e^{-\tau} |A_{\bar f} |^2  \left( 1   +     \sqrt{R_{\bar f} }  \,c_{\bar f}^{+ } \,\tau + R_{\bar f}  c_{\bar f}^{\prime +  }\,\tau^2 \right)\,,\cr
\Gamma (\D0bar (t) \to  \bar f ) &=e^{-\tau} |\bar A_{\bar f} |^2  \left( 1   + {1\over     \sqrt{R_{\bar f} } } \,c_{\bar f}^{- }\, \tau+ {1\over    R_{\bar f}  }c_{\bar f}^{\prime-} \,\tau^2\right)\,,\cr
\end{split}\eeq
for final state $\bar f$,
where
\beq R_f \equiv |\bar A_f / A_f |^2 \,,~~~~~R_{\bar f} \equiv | \bar A_{\bar f}/ A_{\bar f}  |^2\,. \label{RfRfbar}\eeq
In general, the ratios satisfy $R_f , R_{\bar f} = O(1)$ for SCS decays.
The coefficients $c_{f}^\pm$ and $c_{\bar f}^\pm$ in \eqref{timedepSCSCPf}, \eqref{timedepSCSCPfbar}, expressed in terms of $\phi_f^M$, $\phi_f^\Gamma$, and $\Delta _f $, cf. \eqref{lambdaMfnonCP}--\eqref{DeltafnonCP},
are given by
\beq \label{eq:coeffsSCSnonCP}
\begin{split}
c_f^\pm  &=     \mp x_{12} \sin(\phi_f^M  -\Delta_f )-  y_{12}\cos(\phi_f^\Gamma -\Delta_f) ,\cr
c_{\bar f}^\pm &=\mp x_{12} \sin(\phi_f^M  +\Delta_f ) -y_{12}\cos(\phi_f^\Gamma +\Delta_f) .
\end{split}
\eeq
The coefficients in the $O(\tau^2)$ terms %are final state independent and
satisfy
\beq
\label{coeffsSCSnonCPsub}
   \begin{split}
c_f^{\prime \pm}\, = \,&  {\textstyle \frac{1}{4}} \left[ {R^{\mp 1}_f }\,( y_{12}^2 - x_{12}^2 ) +  (x_{12}^2 +y_{12}^2 ) \,\left(1 \pm a_{\rm SL}  \right)\,\right]\,,\cr
c_{\bar f}^{\prime \pm}\, = \,&  {\textstyle \frac{1}{4}} \left[ {R^{\mp 1}_{\bar f }}\,( y_{12}^2 - x_{12}^2 ) +  (x_{12}^2 +y_{12}^2 ) \,\left(1 \pm a_{\rm SL}  \right)\,\right]\,.
\end{split}
\eeq
As in the prior case of decays to CP eigenstates, the CP-even terms in $c_{f ,\bar f}^{\prime \pm}$ should be kept, with future sensitivity at the level of SM indirect CPV in mind. However, the CP-odd terms ($\propto a_{\rm SL}$) can be neglected.

The time dependent measurements yield pairs of CPVINT asymmetries (normalized rate differences for $D^0 (t) \to  f $ vs. $\D0bar (t) \to {\bar  f }$, and $D^0 (t) \to \bar  f $ vs. $\D0bar (t) \to { f }$) at linear order in $\tau$,
\beq  \begin{split}\label{DYfnonCP}
\Delta Y_f   &\equiv {  \sqrt{R_f }  \,c_f^+ - c_{\bar f}^{- } /  \sqrt{R_{\bar f} } \over 2 }\,,\cr
\Delta Y_{\bar f} & \equiv { { \sqrt{R_{\bar f} }\, c_{\bar f}^{+} - c_{ f}^{- } /  \sqrt{R_{ f} } } \over 2 }
\,.\end{split}\eeq
To first order in CPV parameters, \eqref{eq:coeffsSCSnonCP} yields the expressions,
\begin{eqnarray}
{\Delta Y_f } &=&  \sqrt{R_f }\bigg[-x_{12} \sin\phi^M_f \cos\Delta_f -
y_{12} \sin\phi^\Gamma_f \sin\Delta_f \nonumber \\
              && -{\textstyle \frac{1}{2}}{(a_f^d + a_{\bar f}^d )}
              (x_{12} \sin\Delta_f - y_{12} \cos\Delta_f
              )\bigg]\,,\nonumber \\
 \Delta Y_{\bar f}  &=&  {1\over \sqrt{R_{f} } } \bigg[-x_{12}
 \sin\phi^M_f \cos\Delta_f + y_{12} \sin\phi^\Gamma_f \sin\Delta_f
 \nonumber \\
              && +{\textstyle \frac{1}{2}}{(a_f^d + a_{\bar f}^d )} (x_{12} \sin\Delta_f + y_{12} \cos\Delta_f )\bigg] ,
\label{DYfnonCPdetail}
\end{eqnarray}
where the direct CP asymmetries,
\beq \begin{split}
a_f^d &= 1 - \left|\bar A_{\bar f} / A_f \right|=-2 r_f \,\sin\phi_f\,\sin\delta_f , \cr
a_{\bar f}^d & = 1 - \left|\bar A_{ f} / A_{\bar f} \right|   = -2 r_{\bar f} \,\sin\phi_{\bar f}\,\sin\delta_{\bar f}\,,\label{dirCPnonCP}
\end{split}
\eeq
cf. \eqref{eq:fouramp}, enter via the deviation of $\sqrt{R_f R_{\bar f} }$ from unity.
In \eqref{DYfnonCPdetail}, replacing the numerator and denominator in the ratio $R_f$, cf. \eqref{RfRfbar}, with their CP averaged counterparts would
introduce a negligible higher order correction in the CPV parameters.

Note that the CP-even phase differences for dispersive and absorptive CPVINT are given by
$\Delta_f -\pi/2$ and $\Delta_f$, respectively, where $\pi/2$ is the ``dispersive" phase in the first line of \eqref{timeevln}, thus accounting for the factors
$\cos\Delta_f $ and $ \sin\Delta_f$ in the first
two terms of $\Delta Y_f$ and $\Delta Y_{\bar f}$ in \eqref{DYfnonCPdetail}.
In particular, Eq. \eqref{DYfnonCPdetail} confirms that sensitivity to the absorptive phase $\phi_f^\Gamma$ requires a strong phase difference between decay amplitudes, i.e. non-CP eigenstate final states, as argued at the end of Section \ref{sec:SCSCPwidths}.

\subsubsection{CF/DCS decays to $K^\pm  X$ }
\label{sec:CFDCSnonCP}

This category consists of the CF/DCS decays $D^0 \to K^\pm X $, with a single $K$ in the final state.
As noted previously, we choose the DCS decay amplitudes in \eqref{eq:fouramp}, \eqref{lambdaMfnonCP}, \eqref{lambdaMfbarnonCP}, and \eqref{lambdaCFDCS},
to be $A_{\overline f}$ and $\bar A_f$, e.g.  $\bar f = K^+ \pi^-$.
Thus, we denote the time dependent CF/DCS decays to ``wrong-sign'' (WS) final states as
$D^0 (t ) \to \bar f$ and  $\D0bar (t ) \to f$, while the ``right-sign'' (RS) decays are $D^0 (t ) \to f$ and  $\D0bar (t ) \to \bar f$.
The $O(\tau^2 )$ terms in \eqref{timedepwidth} and its CP conjugate can not be neglected, given that the decay amplitude ratios
entering $\lambda^{M,\Gamma}_{f , \bar f}$ are now of $O(1/\theta_{C}^2 )$. %, cf. \eqref{lambdaMfnonCP}, \eqref{lambdaMfbarnonCP}.
The RS and WS decay widths following from \eqref{timedepwidth} and \eqref{phiM0} can be expressed as
\beq\begin{split} \label{timedepRS}
\Gamma (D^0 (t) \to f ) &=e^{-\tau} |A_f |^2  \left(1   + \sqrt{R_f} c_{\scriptscriptstyle{\mathrm{RS}},f}^{+} \,\tau + R_f c_{\scriptscriptstyle{\mathrm{RS}},f}^{\prime +  }\,\tau^2 \right)\,,\cr
\Gamma (\D0bar (t) \to  \bar f ) &=e^{-\tau} |\bar A_{\bar f} |^2  \left(1   +
  \frac{1}{\sqrt{R_{\bar f}}} c_{\scriptscriptstyle{\mathrm{RS}},f}^{- }\, \tau+ \frac{1}{R_{\bar f}} c_{\scriptscriptstyle{\mathrm{RS}}, f}^{\prime-} \,\tau^2\right)\,
\end{split}\eeq
and
\beq\begin{split} \label{timedepWS}
\Gamma (D^0 (t) \to \bar f ) &=e^{-\tau} |A_f |^2  \left( R_f^+   + \sqrt{R_f^+ } c_{\scriptscriptstyle{\mathrm{WS}}, f}^{+ } \,\tau + c_{\scriptscriptstyle{\mathrm{WS}}, f}^{\prime +  }\,\tau^2 \right)\,,\cr
\Gamma (\D0bar (t) \to  f ) &=e^{-\tau} |\bar A_{\bar f} |^2  \left( R_f^-   + \sqrt{R_f^- } c_{\scriptscriptstyle{\mathrm{WS}},f}^{- }\, \tau+ c_{\scriptscriptstyle{\mathrm{WS}},f}^{\prime-} \,\tau^2\right)\,\cr
\end{split}\eeq
where $R_f^\pm$ are the DCS to CF ratios
\beq R_f^+ = |A_{\bar f} / A_f |^2,~~~R_f^- = |\bar A_{f} /\bar A_{\bar  f} |^2,\label{Rfpm}\eeq
the ratios $R_f$ , $R_{\bar f}$ are defined in \eqref{RfRfbar}, and the coefficients $c_{\mathrm{RS(WS)}, f}^\pm$, $c_{\mathrm{WS(WS)},f}^{\prime \pm}$, to first order in CPV parameters,  are given by
\begin{eqnarray} \label{cpm}
c_{\scriptscriptstyle{\mathrm{RS}},f}^\pm &=&-x_{12} \sin\Delta_f  + y_{12} \cos\Delta_f\\
&&\pm \left( x_{12} \sin\phi_f^M \cos\Delta_f  + y_{12} \sin\phi^\Gamma_f
   \sin\Delta_f \right),\nonumber \\
c_{\scriptscriptstyle{\mathrm{WS}},f}^\pm &=&(1\mp a_f^d )\, \left[x_{12} \sin\Delta_f  + y_{12} \cos\Delta_f \right]\nonumber \\
&&\pm x_{12} \sin\phi_f^M \cos\Delta_f  \mp y_{12} \sin\phi^\Gamma_f
   \sin\Delta_f ,\nonumber \\
 c_{\scriptscriptstyle{\mathrm{RS}},f}^{\prime \pm} &=& {\textstyle \frac{1}{4}} \left[ (x_{12}^2 +y_{12}^2 ) \,( 1 \pm a_{\rm SL} ) + \xi^\pm (y_{12}^2 -x_{12}^2 )\right]  ,\nonumber \\
 c_{\scriptscriptstyle{\mathrm{WS}}, f}^{\prime \pm} &=& {\textstyle \frac{1}{4}} (x_{12}^2 +y_{12}^2 ) \,\left[ 1 \pm a_{\rm SL} \mp 2 \, a_f^d  \right]+ {\textstyle \frac{1}{4}} R_f^\pm (y_{12}^2 -x_{12}^2 )  ,\nonumber
\end{eqnarray}
with $\xi^+=R_f^{-1}$, $\xi^-=R_{\bar f}$.
The (CF) direct CP asymmetry, $a_f^d$, appearing in \eqref{cpm} is given by
 \beq \begin{split}
a_f^d &= 1 - \left|\bar A_{\bar f} / A_f \right|=-2 r_f \,\sin\phi_f\,\sin\delta_f \,,\label{dirCPCF}
\end{split}
\eeq
and vanishes in the SM.
In the SM, the $O(\tau^2)$ coefficients are well approximated as
\beq  c_{\scriptscriptstyle{\mathrm{RS,WS}},f}^{\prime \pm}\, =  {\textstyle \frac{1}{4}} (x_{12}^2 +y_{12}^2 )\,.\eeq
The prefactors in \eqref{timedepWS} are, to excellent approximation, equal to the RS time dependent decay widths,
\beq\begin{split} \label{timedepRSsimpl}
\Gamma (D^0 (t) \to  f ) &\sim e^{-\tau} |A_f |^2 \,,\cr
\Gamma (\D0bar (t) \to  \bar f ) &\sim e^{-\tau} |\bar A_{\bar f} |^2 \,,\cr
\end{split}\eeq
where the subleading DCS contributions in \eqref{timedepRS} have been neglected.

A fit to the time-dependence in \eqref{timedepWS}, \eqref{timedepRSsimpl} yields measurements of $R_f^\pm$, $c_{\scriptscriptstyle{\mathrm{WS}}, f}^\pm$, $c_{\scriptscriptstyle{\mathrm{WS}}, f}^{\prime \pm}$, and the indirect CP asymmetries,
\begin{eqnarray}
\delta c_{\scriptscriptstyle{\mathrm{WS}}, f}  &\equiv& {\textstyle \frac{1}{2}}(c_{\scriptscriptstyle{\mathrm{WS}}, f}^+ - c_{\scriptscriptstyle{\mathrm{WS}}, f}^- )=x_{12} \sin\phi^M_f \cos\Delta_f \nonumber\\
&&-y_{12} \sin\phi^\Gamma_f \sin\Delta_f
      -a_f^d (x_{12} \sin\Delta_f + y_{12} \cos\Delta_f ) \,,\nonumber\\
\delta c^\prime_{\scriptscriptstyle{\mathrm{WS}}, f}  &\equiv& {c_{\scriptscriptstyle{\mathrm{WS}}, f}^{\prime +} - c_{\scriptscriptstyle{\mathrm{WS}}, f}^{\prime -}\over c_{\scriptscriptstyle{\mathrm{WS}}, f}^{\prime +} +c_{\scriptscriptstyle{\mathrm{WS}}, f}^{\prime -} }= a_{SL} - 2 a_f^d .\label{CFDCSCP}
\end{eqnarray}
Note that the last terms in \eqref{CFDCSCP} for $\delta c_{\scriptscriptstyle{\mathrm{WS}}, f} $ and $\delta c_{\scriptscriptstyle{\mathrm{WS}}, f}^\prime$ are absent in the SM and, more generally, in models with negligible CP violating NP in CF/DCS decays.
As in \eqref{DYfnonCPdetail},
the $\cos\Delta_f $ and $\sin\Delta_f $ dependence in the first two terms of $\delta c_{\scriptscriptstyle{\mathrm{WS}}, f} $ originates from the
total CP-even phase differences $\Delta_f -\pi/2$ and $\Delta_f$, between decays with and without dispersive mixing and decays with and without absorptive mixing, respectively.  This again confirms that strong phase differences are required in order to measure the absorptive CPV phases,
$\phi_f^\Gamma$.

\subsection{CF/DCS decays to $K^0 X\,, \K0bar X$}
\label{sec:K0Xtimedep}

We derive expressions for the time-dependent $D^0 $ and $ \D0bar$ decay rates
for two step CF/DCS decays of the form
\beq \label{twostagedecay} D^0 (t) \to  [ K_{S,L}(t^\prime ) \to  \pi^+ \pi^-  ] +X\,,\eeq
to final states $f=[\pi^+\pi^-] X$.
These decays depend on two elapsed time intervals, $t$ and $t^\prime$, at which the $D$ and $K$ decay following their respective production.

The $D^0 (t) $ and $\D0bar (t) $ decay amplitudes now sum over contributions with and without $D^0-\D0bar$ mixing, and
with and without $K^0  -\K0bar$ mixing.  The kaon time evolution is conveniently described in the mass basis,
\beq \begin{split}\label{Ktimeevln} | K_S (t ) \rangle  &= e^{-i M_S t } e^{-\Gamma_S t / 2 } | K_S \rangle\,,\cr
| K_L (t ) \rangle  &= e^{-i M_L t } e^{-\Gamma_L  t / 2 } | K_L \rangle \,,\end{split}\eeq
where %the $K_S $ and $K_L $ are the neutral $K$ mass eigenstates, cf. \eqref{KaonEig},
$M_{S,L}$, $\Gamma_{S,L}$, and $\tau_{S,L}$ are the corresponding masses, widths, and lifetimes.
The time-dependent amplitudes for the decay of an initial $D^0$ to final state $f=[\pi^+ \pi^- ] X$, and for the CP conjugate decay of an initial $\D0bar$ to final state $\bar f=\overline{[\pi^+ \pi^- ] X}$, %, as in \eqref{twostagedecay},
are given by
\begin{eqnarray}\label{twotimeamps}
&&\mkern-18mu A_{f} (t,t^\prime )= \sum_{a=S,L}
   A ( K_a \to \pi^+ \pi^- )\,\, \times \\
   &&e^{-(i M_a +\frac12 \Gamma_a ) t^\prime }
  (A_{K_a X}  \langle D^0 | D^0 (t)  \rangle +
\overline A_{K_a X}  \langle \D0bar | D^0 (t)  \rangle\, )\,,\nonumber\\
&&\mkern-18mu \overline A_{\bar f} (t,t^\prime )= \sum_{a=S,L}
   A ( \overline K_a \to \pi^+ \pi^- )\,\, \times \nonumber\\
   &&e^{-(i M_a +\frac12 \Gamma_a ) t^\prime } (A_{\overline{K_a \!X}}  \langle D^0 | \D0bar (t)  \rangle +
\overline A_{\overline{K_a \! X}}  \langle \D0bar |\D0bar (t)  \rangle\, )\,,\nonumber
\end{eqnarray}
where expressions for the $D$ decay amplitudes $A_{K_a X}$, etc. appear in \eqref{DtoKaXamps}.
The $K_{S,L} \to \pi\pi$ decay amplitudes satisfy,
\beq \begin{split}
A ( K_S \to \pi^+ \pi^- ) &=  p_K A_{+-} + q_K \bar A_{+-} \,,\cr
A ( K_L \to \pi^+ \pi^- ) &=  p_K A_{+-} - q_K \bar A_{+-} \,,
\end{split}
\eeq
with
\beq \begin{split} A_{+-}  \equiv \langle \pi^+ \pi^- | H |  K^0 \rangle\,, ~~~
\overline A_{+-} & \equiv  \langle\pi^+ \pi^- | H |  \K0bar\rangle\,.
\end{split}
\eeq
The amplitudes ${\overline A }_{f} (t,t^\prime ) $ and ${A }_{\bar f} (t,t^\prime ) $
are obtained by substituting $|D^0 (t) \rangle \to| \D0bar (t) \rangle$ and vice versa in the first and second relations of \eqref{twotimeamps}, respectively.
Expressing the amplitudes in terms of the CPVINT observables in
\eqref{genlambdaK0} yields the general expressions, valid to linear
order in $\tau$:
\begin{eqnarray}\label{twotimeamps2}
&&\mkern-18mu A_{f} (t,t^\prime )= e^{-(i M_D + \frac12 \Gamma_D) t}\sum_{a=S,L}
   A ( K_a \to \pi^+ \pi^- ) \,  \\
&&  \times \,A_{K_a X}\,\,  e^{-(i M_a +\frac12 \Gamma_a ) t^\prime }
  \bigg( 1 -\frac12 \tau \bigg[ i {x_{12}\over \lambda_{K_a X}^M } +
  {y_{12} \over  \lambda_{K_a X}^\Gamma } \bigg]\bigg)\,,
  \nonumber\\
&&\mkern-18mu  \overline A_{\bar f} (t,t^\prime )= e^{-(i M_D + \frac12 \Gamma_D) t}\sum_{a=S,L}
   A ( \overline K_a \to \pi^+ \pi^- ) \,  \nonumber\\
&&\times \,\overline A_{\overline {K_a X}}\,\,  e^{-(i M_a +\frac12 \Gamma_a ) t^\prime }
  \bigg( 1 -\frac12 \tau  \bigg[ i {x_{12}\lambda_{{\overline{K_a
          \!X}}}^M } + {y_{12}  \lambda_{{\overline{K_a \!X}}}^\Gamma
  } \bigg] \bigg)\,,\nonumber
\end{eqnarray}
where $ \overline A_{f} (t,t^\prime )$ is obtained by substituting $A_{K_a X}\to \overline A_{K_a X}$ and $\lambda_{{K_a X}}^{M\,(\Gamma)} \to 1/ \lambda_{{K_a X}}^{M\,(\Gamma)} $ in the first relation, and $A_{\bar f} (t,t^\prime )$ is obtained by substituting
$\overline A_{\overline{K_a X}}\to A_{\overline{K_a X}}$ and $\lambda_{\overline{K_a X}}^{M\,(\Gamma)} \to 1/ \lambda_{\overline{K_a X}}^{M\,(\Gamma)} $ in the second relation.

The time-dependent decay rates are obtained by squaring the magnitudes of the amplitudes in \eqref{twotimeamps2}, e.g.
$\Gamma_f (t , t^\prime) = |A_f (t,t^\prime)|^2$ etc., and assuming that
CP violating NP is negligible in CF/DCS decays.
Therefore, as in the SM, we assume vanishing direct CPV in the CF decays, neglect the DCS amplitudes (their impact is discussed in Section \ref{finstatedepK0}),
and employ the expressions for the CPVINT observables given in \eqref{lambdaMfCFK0}.  We work to first order in CPV quantities, and also employ
the relations (see e.g. \cite{Nir:1992uv})
\begin{eqnarray} \label{AmpKSsqeps}
&&|A(K_S \to \pi^+ \pi^- )|^2  = 4 |p_K A_{+ -}|^2 \,(1 -
   2\,\epsilon_R\, ) \,\\
&&~~~~~~~~~~~~~~~~~~~~~~~~\,=4 |q_K \overline A_{+ -}|^2 \,(1+ 2\, \epsilon_R\, )\,,\nonumber \\
&&A(K_S \to \pi^+ \pi^- )\,A(K_L \to \pi^+ \pi^- )^* = 4  |p_K A_{+ -}|^2  \epsilon_K^* \nonumber \\
&&~~~~~~~~~~~~~~~~~~~~~~~~\,=4  |q_K \overline A_{+ -}|^2  \epsilon_K^*  \,,\nonumber \\
&&|A(K_L \to \pi^+ \pi^- )|^2  = O(\epsilon_K^2 )\,.\nonumber
\end{eqnarray}
In particular, the last relation in \eqref{AmpKSsqeps} implies that we can neglect the purely $K_L $ contributions to the widths.
The expressions for the time-dependent decay rates are then of the form,
\beq\begin{split} \label{timedepratesK0f}
\Gamma_f (t,t^\prime) =e^{-\tau}&   |\overline A_{+ -}|^2  |A_{\K0bar X} |^2   \,\bigg\{\cr
e^{-\Gamma_S t^\prime }\big[c^+
+&\,\sqrt{R_{f}}\, c_f^{+ } \,\tau + R_f  \,c_f^{\prime +  }\,\tau^2 \big]\,+\cr
\,e^{-\Gamma_K t^\prime}  \big[\, (b^+ &+ \sqrt{R_{f}} \,b_f^+ \tau\, ) \, \cos(\Delta M_K t^\prime)\cr
+ \,(d^+ &+ \sqrt{R_{f}} \,d_f^+ \tau \,) \, \sin(\Delta M_K t^\prime) \,\big]\bigg\}\,,\cr
\overline \Gamma_f (t,t^\prime) =e^{-\tau}&  %  |A_{+ -}|^2
|\overline A_{+ -}|^2  |\overline A_{K^0 X} |^2     \,\bigg\{\cr
e^{-\Gamma_S t^\prime }\bigg[c^-
+&\,{1\over \sqrt{R_{f}}}\, c_f^{- } \,\tau + {1\over R_f }  \,c_f^{\prime-  }\,\tau^2 \bigg]\,+\cr
\,e^{-\Gamma_K t^\prime}  \bigg[\, \bigg(b^- &+ {1\over \sqrt{R_{f}}} \,b_f^- \tau  \,\bigg) \, \cos(\Delta M_K t^\prime)\cr
+ \,\bigg(d^- &+ {1\over \sqrt{R_{f}} }\,d_f^- \tau \,\bigg) \, \sin(\Delta M_K t^\prime) \,\bigg]\bigg\}\,,
\end{split}\eeq
for final state $f$, and
\beq\begin{split} \label{timedepratesK0fbar2} \nonumber
\Gamma_{\bar f} (t,t^\prime) =e^{-\tau}&   |\overline A_{+ -}|^2    |\overline A_{K^0 X} |^2  \,\bigg\{\cr
e^{-\Gamma_S t^\prime }\bigg[c^+
+&\,{1\over \sqrt{R_{f}}}\, c_{\bar f}^{+ } \,\tau + {1\over R_f }  \,c_{\bar f}^{\prime +} \,\tau^2 \bigg]\,+\cr
\,e^{-\Gamma_K t^\prime}  \bigg[\, \bigg(b^+ &+ {1\over \sqrt{R_{f}} }\,b_{\bar f}^+ \tau \,\bigg) \, \cos(\Delta M_K t^\prime)\cr
+ \,\bigg(d^+ &+ {1\over \sqrt{R_{f}}} \,d_{\bar f}^+ \tau  \,\bigg) \, \sin(\Delta M_K t^\prime) \,\bigg]\bigg\}\,,\cr
\end{split}\eeq
\beq\begin{split} \label{timedepratesK0fbar}
  \overline \Gamma_{\bar f} (t,t^\prime) =e^{-\tau}& % |A_{+ -}|^2
 |\overline A_{+ -}|^2 |A_{\K0bar X} |^2      \,\bigg\{\cr
e^{-\Gamma_S t^\prime }\big[c^-
+&\sqrt{R_{f}}\, c_{\bar f}^{- } \,\tau +R_f  \,c_{\bar f}^{\prime-  }\,\tau^2 \big]\,+\cr
\,e^{-\Gamma_K t^\prime}  \big[\, (b^- &+ {\sqrt{R_{f} }}\,b_{\bar f}^- \tau \,) \, \cos(\Delta M_K t^\prime)\cr
+ \,(d^- &+ \sqrt{R_{f}} \,d_{\bar f}^- \tau \,) \, \sin(\Delta M_K t^\prime) \,\big]\bigg\}\,,\end{split}\eeq
for final state $\bar f$, where
\beq \label{Rfdef} R_{f}\equiv \left|{\overline A_{K^0 X}/ A_{\K0bar X}}\right|^2 \,,\eeq
$\Delta M_K\equiv M_L - M_S $, and $\Gamma_K \equiv (\Gamma_L + \Gamma_S)/2$.
We have taken  $|A_{+-} |=|\overline A_{+-}|$, given that the two magnitudes differ by negligible corrections of $O(\epsilon_K^2\,,\epsilon_K^\prime )$.
The coefficients in \eqref{timedepratesK0f}, \eqref{timedepratesK0fbar} depend on the quantities $\phi_f^M$, $\phi_f^\Gamma$,
$\Delta_f$, cf. \eqref{lambdaMfCFK0}--\eqref{phiMfCFK0}, and $\epsilon_K$.
For the purely $K_S X$ contributions ($e^{-\Gamma_S t^\prime}$ dependence), they are given by
\beq\label{KSci}
\begin{split}
c^\pm &=1\,\pm\,2 \epsilon_R,\,\cr
c_f^\pm &=(\pm x_{12} -y_{12} \sin \phi_f^\Gamma )\,\sin\Delta_f \cr
&~~~~~~~~~ - (y_{12} \pm x_{12} \sin\phi_f^M )\cos\Delta_f\,,\cr
c_{\bar f}^\pm &=(\mp x_{12}+y_{12} \sin \phi_f^\Gamma )\,\sin\Delta_f \cr&~~~~~~~~~ - (y_{12} \pm x_{12}  \sin\phi_f^M )\cos\Delta_f\,,\cr
c_f^{\prime \pm}&=   \frac14 \left(x_{12}^2 + y_{12}^2 + [y_{12}^2 - x_{12}^2]\,R_f^{\mp 1}\right)\,,\cr
c_{\bar f}^{\prime \pm}&=   \frac14 \left(x_{12}^2 + y_{12}^2 + [y_{12}^2 - x_{12}^2]\,R_f^{\pm 1}\right).\cr
\end{split}
\eeq
CP-odd contributions to the coefficients $c_f^{\prime \pm}$, $c_{\bar f}^{\prime \pm}$
are of $O[(x_{12}^2, y_{12}^2 ) \times (\epsilon_K, \phi_{12}) ]$ and have been neglected, i.e. they are $O(x_{12} ,y_{12})$ suppressed
relative to the CP-odd terms arising at $O(\tau )$. %, and have therefore been neglected.
Interference between the amplitudes containing intermediate $K_S X$ and $K_L X$ ($e^{-\Gamma_K t^\prime}$ dependence) yields,
\beq\label{KSKLintbidi}
\begin{split}
b^\pm &=\mp \,2 \epsilon_R\,,~~~~d^\pm = \mp\, 2 \epsilon_I \,,\cr
b_f^\pm & =  2\big( \pm x_{12} \cos\Delta_f  + y_{12} \sin\Delta_f \big)\,\epsilon_I  \,,\cr
b_{\bar f}^\pm & =  2\big( \pm x_{12} \cos\Delta_f  - y_{12} \sin\Delta_f \big)\,\epsilon_I \,,\cr
d_f^\pm & =  2\big( \mp x_{12} \cos\Delta_f -  y_{12} \sin\Delta_f \big)\,\epsilon_R \,,\cr
d_{\bar f}^\pm & =  2\big( \mp x_{12} \cos\Delta_f + y_{12} \sin\Delta_f \big)\,\epsilon_R \,.
\end{split}
\eeq
We have neglected interference contributions of $O(x_{12}^2 \,\epsilon_K , y_{12}^2 \,\epsilon_K)$ arising at $O(\tau^2)$ in \eqref{timedepratesK0f}, \eqref{timedepratesK0fbar}. Again, they are $O(x_{12} ,y_{12})$ suppressed
relative to the CP-odd terms arising at $O(\tau )$.

The indirect CP asymmetries are obtained by taking normalized rate differences between $\Gamma_f $ and $\overline \Gamma_{\bar f}$,  and between $\Gamma_{\bar f} $ and $\overline \Gamma_f $.
To first order in CPV quantities, the phases $\phi_f^M , \phi_f^\Gamma$ only enter the CP asymmetries of the purely $K_S$ contributions,
while the CP asymmetries induced by $K_S-K_L$ interference only probe $\epsilon_K$.
The first set of CP asymmetries, between the coefficients in  \eqref{KSci}, are given by ($\delta c^\prime $ is negligible),
\beq\label{K0asymms1}  \begin{split}
\delta c& \equiv \frac12 (c^+ - c^- ) =2 \epsilon_R\,,\cr
\delta c_f &\equiv \frac12 (c_f^+ - c_{\bar f}^- ) \,\cr
           &= - \, (\,y_{12} \sin\phi_f^\Gamma \,\sin\Delta_f + x_{12} \sin\phi_f^M \cos\Delta_f )\,,\cr
\delta c_{\bar f} &\equiv \frac12 (c_{\bar f}^+ - c_{f}^- ) \,\cr
           &= \, (\,y_{12} \sin\phi_f^\Gamma \,\sin\Delta_f - x_{12} \sin\phi_f^M \cos\Delta_f )\,.
\end{split}
\eeq

Again, $\Delta_f \ne 0,\pi$ is required in order to measure $\phi_f^\Gamma$, due to the lack of a non-trivial CP-even phase in the absorptive mixing amplitude.
The six CP asymmetries in the second set of coefficients, cf. \eqref{KSKLintbidi}, are
\beq\label{K0asymms2} \begin{split}
\delta b &\equiv \frac12 (b^+ - b^- ) = - 2 \epsilon_R\,,\cr
\delta d &\equiv \frac12 (d^+ -d^- ) = - 2 \epsilon_I\,,\cr
 \delta b&_f \equiv \frac12 (b_f^+ - b_{\bar f}^- ) \,\cr
&= 2\, (x_{12}  \cos\Delta_f +\,y_{12} \sin\Delta_f )\,\epsilon_I\,,\cr
               \delta b&_{\bar f} \equiv \frac12 (b_{\bar f}^+ - b_{f}^- ) \,\cr
&= 2\, (x_{12}  \cos\Delta_f -\,y_{12} \sin\Delta_f )\,\epsilon_I\,,\cr
               \delta d&_f \equiv \frac12 (d_f^+ - d_{\bar f}^- ) \,\cr
&=- 2\, (x_{12}  \cos\Delta_f +\,y_{12} \sin\Delta_f )\,\epsilon_R\,,\cr
               \delta d&_{\bar f} \equiv \frac12 (d_{\bar f}^+ - d_{f}^- ) \,\cr
&= 2\, (-x_{12}  \cos\Delta_f +\,y_{12} \sin\Delta_f )\,\epsilon_R\,.
              \end{split}
              \eeq
In principle, each of the CP asymmetries in \eqref{K0asymms1} , \eqref{K0asymms2} can be measured by fitting to the dependence of the decay rates on $t$ and $t^\prime$.

In Section \ref{sec:theorphases} we will see that in the SM, $\phi_f^M $ and $\phi_f^\Gamma$ are expected to be of same order as  $\epsilon_K$, implying that the CPVINT asymmetries in \eqref{K0asymms1} and \eqref{K0asymms2} are also of same order.
Thus, the impact of $\epsilon_K$, particularly at linear order in $\tau$, on the asymmetry measurements
needs to be considered.  We will address this point in Section~\ref{sec:implement}, taking into account the typical decay times $t^\prime$ for the intermediate $K^0$'s detected at LHCb and Belle-II.

In the case of two body (and quasi two body) intermediate states, e.g. $X =\pi^0 , \omega, f_0$, expressions for the time dependent decay rates and CP asymmetries are obtained by setting $R_f=1$ [and $|\overline A_{K^0 X}|= |A_{\K0bar X}|$ in
\eqref{timedepratesK0f}], %disregarding \eqref{timedepratesK0fbar},
and $\sin\Delta_f =0$, $\cos\Delta_f = \eta_f^{CP}$ in \eqref{KSci}--\eqref{K0asymms2},
where $\eta_f^{CP}$ is defined in \eqref{lambdaMfCFK0twobody}.
The resulting decay widths are
\beq\begin{split} \label{timedepratesK0twobody1}
\Gamma_f (t,t^\prime)& =e^{-\tau}   |\overline A_{+ -}|^2  |A_{\K0bar X} |^2   \,\bigg\{\cr
&e^{-\Gamma_S t^\prime }\big[c^+
+\,c_f^{+ } \,\tau + \,c^{\prime }\,\tau^2 \big]\,+\cr
&e^{-\Gamma_K t^\prime}  \big[\, (b^+ + \,b_f^+ \tau \,) \, \cos(\Delta M_K t^\prime)\cr
&+ \,(d^+ + \,d_f^+ \tau \,) \, \sin(\Delta M_K t^\prime) \,\big]\bigg\}\,,
\end{split}
\eeq
\beq \label{timedepratesK0twobody2}
\begin{split}
\overline \Gamma_f (t,t^\prime) &=e^{-\tau}  |\overline A_{+ -}|^2  |A_{\K0bar X} |^2    \,\bigg\{\cr
&e^{-\Gamma_S t^\prime }\bigg[c^-
+\, c_f^{- } \,\tau +  \,c^{\prime }\,\tau^2 \bigg]\,+\cr
&\,e^{-\Gamma_K t^\prime}  \bigg[\, \bigg(b^- +  \,b_f^- \tau  \,\bigg) \, \cos(\Delta M_K t^\prime)\cr
&+ \,\bigg(d^- + \,d_f^- \tau  \,\bigg) \, \sin(\Delta M_K t^\prime) \,\bigg]\bigg\}\,,
\end{split}\eeq
with coefficients,
\beq\label{KScitwobody}
\begin{split}
\mkern-18mu c^\pm &=1\,\pm\,2 \epsilon_R\,,~~~c_f^\pm = - \eta_f^{CP}\,(y_{12} \pm x_{12} \sin\phi_f^M )\,,\cr
\mkern-18mu c^{\prime}\,&=  \frac12 y_{12}^2\,,~~~b^\pm =\mp \,2 \,\epsilon_R\,,~~~~b_f^\pm  =    \pm 2\, \eta_f^{CP}x_{12}\,\epsilon_I  \,,\cr
\mkern-18mu d^\pm &= \mp\, 2 \,\epsilon_I \,,~~~~d_f^\pm  =   \mp 2  \, \eta_f^{CP}x_{12}  \,\epsilon_R  \,.\cr
\end{split}
\eeq
The corresponding CP asymmetries, as defined in \eqref{K0asymms1}, \eqref{K0asymms2}, are given by
\beq\label{K0asymmstwobody}  \begin{split}
\delta c&= 2\epsilon_R\,,~~~~~~~\,\delta c_f = - \, \eta_f^{CP}x_{12} \sin\phi_f^M\,,\cr
\delta b & = - 2 \epsilon_R\,,~~~~\delta b_f = 2\, \eta_f^{CP}x_{12}  \,\epsilon_I\,,\cr
\delta d  &= - 2\epsilon_I\,,~~~~\delta d_f =- 2\, \eta_f^{CP} x_{12}  \,\epsilon_R\,.
\end{split}
\eeq
Note that $\delta c_f$ is purely dispersive, similarly to $\Delta Y_f$ for SCS decays to CP eigenstates, cf. \eqref{DYfresult} (again,
the only CP even phase available for charm CPVINT is the dispersive mixing phase $\pi/2$).

Finally, the CP conserving observable, $y^f_{CP}$, for SCS decays to CP eigenstates, cf. \eqref{eq:yCPdef}, \eqref{eq:yCPdefExp},
can be carried over to the case of two body and quasi two body intermediate states discussed above. It is analogously defined as
\beq y^f_{CP} \equiv -{c_f^+ + c_f^-  \over 2}  \,.\label{yCPK0def} \,\eeq
However, the $K_S$ decay time dependence, $e^{-\Gamma_S t^\prime}$, in \eqref{timedepratesK0twobody1},\eqref{timedepratesK0twobody2}, must be accounted for in order to avoid additional systematic errors in its extraction.
Employing \eqref{KScitwobody} yields
\beq y^f_{CP} = \eta_f^{CP} y_{12} =   \eta_f^{CP} |y|\,,\label{yCPK0} \,\eeq
up to negligible corrections quadratic in CPV parameters.
For example, we expect $y^f_{CP} = - y_{12}$ for $X=\omega ,\pi^0$ (opposite in sign to $y^f_{CP}$ for $K^+ K^-$, $\pi^+ \pi^-$), and $y^f_{CP} = + y_{12}$ for $X= f_0$.

\section{Approximate Universality}
\label{sec:appuniv}
In the previous section, all indirect CPV effects were parametrized in full generality, in terms of final state dependent pairs of dispersive and absorptive weak phases ($\phi^M_f $, $\phi^\Gamma_f$).   In order to understand how best to parametrize indirect CPV effects in the upcoming precision era, we need to estimate the final state dependence.
We accomplish this via a $U$-spin flavor symmetry decomposition of the SM $\DDbar$ mixing amplitudes.
Crucially, this also yields estimates of indirect CPV effects in the SM.

\subsection{U-spin decomposition}
\label{sec:Uspindecomp}
The SM $\DDbar$ mixing amplitudes $\Gamma_{12}$ and  $M_{12}$ have flavor transitions $\Delta C= - \Delta U =2$ and $\Delta S = \Delta D = 0$.
We can write them as
\beq \Gamma^{\rm SM}_{12} = - \sum_{i,j=d,s} \lambda_i \lambda_j \Gamma_{ij},~~M^{\rm SM}_{12} = - \sum_{i,j=d,s,b} \lambda_i \lambda_j M_{ij}\,,\eeq
where $\lambda_i \equiv V_{ci} V_{ui}^*$.
At the quark level, the transition amplitudes $\Gamma_{ij}$ and $M_{ij}$ are identified with box diagrams containing, respectively, on-shell and off-shell internal $i$ and $j$ quarks.
Thus, they possess the flavor structures (Dirac structure is unimportant for our discussion)
$\Gamma_{ij}, M_{ij}  \sim  (\bar u c )^2 (\bar i  i ) (\bar j j )\sim  (\bar u c )^2 (\bar i j) (\bar j i) $,
or
\beq\begin{split}
 \Gamma_{ss} &\sim (\bar s s )^2 \,,~\Gamma_{dd} \sim (\bar d d)^2\,,~ \Gamma_{sd} \sim (\bar s s) (\bar d d)  \,,\end{split}\eeq
and similarly for the $M_{ij}$.
Employing CKM unitarity ($\lambda_d + \lambda_s+\lambda_b =0  $),
the $U$-spin decomposition of $\Gamma^{\rm SM}_{12}$ is given by
\beq
 \Gamma^{\rm SM}_{12 } = {(\lambda_s - \lambda_d )^2 \over 4 }\, \Gamma_2 + {(\lambda_s - \lambda_d ){\lambda_b} \over 2 }\, \Gamma_1 +
{{ \lambda_b^2} \over 4} \, \Gamma_0 \,, \label{Uspindecomp}\eeq
where the $U$-spin amplitudes $\Gamma_{2,1,0}$ are the $\Delta U_3 =0$ elements of the $\Delta U$= 2, 1, 0 multiplets, respectively.
This can be seen from their quark flavor structures,
\beq \begin{split} \label{Uspinamps}
\Gamma_2  &= \Gamma_{ss} + \Gamma_{dd} - 2 \Gamma_{sd} ~\sim~ (\bar s s - \bar d d )^2 =O(\epsilon^2)  \,,\cr
\Gamma_1 & = \Gamma_{ss} - \Gamma_{dd}~ \sim~ (\bar s s - \bar d d ) (\bar s s + \bar d d) =O(\epsilon)\,,\cr
\Gamma_0 & = \Gamma_{ss} +\Gamma_{dd}+2 \Gamma_{sd} ~ \sim~ (\bar s s+ \bar d d )^2 =O(1)  \,.
\end{split}
\eeq
The orders in the $U$-spin breaking parameter $\epsilon$
at which they enter are also included, corresponding to the power of the
$U$-spin breaking spurion $M_\epsilon \sim \epsilon\, (\bar s s - \bar d d)$ required to construct each $\Gamma_i$.
The $U$-spin decomposition of $M_{12}$ is analogous to \eqref{Uspindecomp},
with the exception of additional contributions to $M_1$ and $M_0$, given by $(M_{sb} - M_{db})$ and $(M_{sb} + M_{db} + M_{bb})$, respectively, and corresponding to box diagrams with internal $b$ quarks at the quark level.  %However, they do not have any qualitative impact on our conclusions
The small value of $\lambda_b$ implies that we can neglect the $\Delta U= 1, 0$ contributions to the mass and width differences, even though the $\Delta U=2$ piece is of higher order in $\epsilon$.
Thus, $x_{12}$ and $y_{12}$ are due to
$\Gamma_2$ and $M_2$, respectively, and arise at $O(\epsilon^2)$ \cite{Falk:2001hx,Falk:2004wg,Gronau:2012kq}.
Similarly, CPV in mixing arises at $O(\epsilon)$ due to $\Gamma_1$ and $M_1$, while
the contributions of $\Gamma_0 $ and $M_0$ are negligible.

The $U$-spin amplitudes $\Gamma_i$, $M_i$ are of the form,
\beq  M_i = \eta_i^M |M_i | e^{ 2 i \xi},~~~\Gamma_i = \eta_i^\Gamma |\Gamma_i | e^{ 2 i \xi}\,,~~~\eta_i^M ,\eta_i^\Gamma =\pm\,. \label{MiGammaiForms}\eeq
The exponential factors originate from the choice of meson phase convention, and trivially cancel in physical observables.  However, the $\eta_i $ in \eqref{MiGammaiForms}
are physical, can a priori be of either sign, and can be determined from experiment.  For example,
since $\phi_{12} \approx 0$, we already know that
\beq\label{M2ovGamma2} {\rm arg} [M_2 /\Gamma_2 ] =0\,,\eeq
or that $\eta^M_2 = \eta_2^\Gamma $.
Moreover, as we shall see shortly, cf. \eqref{SCSCPphiMfphiM2}, existing measurements also imply that
\beq \label{eta2eq1} \eta^M_2 = \eta_2^\Gamma = +\,.\eeq

 %are positive, modulo the meson phase factors

The inclusive \cite{Georgi:1992as,Ohl:1992sr,Bigi:2000wn,Bobrowski:2010xg,Carrasco:2014uya,Carrasco:2015pra,Bazavov:2017weg,Kirk:2017juj}
and exclusive \cite{Falk:2001hx,Falk:2004wg,Cheng:2010rv,Gronau:2012kq,Jiang:2017zwr} approaches to
estimating $\Delta \Gamma_D$ yield several observations of relevance to our discussion of CPV below.  In the inclusive OPE based approach, the flavor amplitudes satisfy $\Gamma_{ij} \sim \Gamma_D$. % (the $D^0$ lifetime),
This is reflected  in the ability of this approach to accommodate the charm meson lifetimes
\cite{Lenz:2013aua,Kirk:2017juj}.
The individual $\Gamma_{ij}$ contributions to $y_{12}$ are, therefore, about five times larger than the experimental value \cite{Lenz:2016fcv}, suggesting
that $U$-spin violation is large, e.g. ${\mathcal O}(\epsilon^2 ) \sim 20\% $
for $\Gamma_2$, cf. \eqref{Uspinamps}, \eqref{e2bound}.\footnote{Inclusive OPE based GIM-cancelations between the $\Gamma_{ij} $ yield $y$ four orders of magnitude below experiment. Evidently, $m_c$ and $(m_s - m_d )/\Lambda_{\rm QCD}$ are not sufficiently large and small, respectively, for this approach to properly account for $U$-spin breaking in $y_{12}$.}
The exclusive approach %addresses the origin of $U$-spin violation by
estimates sums over exclusive decay modes.
Unfortunately, the charm quark mass is not sufficiently light
for $D^0$ meson decays to be dominated by a few final states.  Moreover, the strong phase differences entering $y_{12}$, and the off-shell decay amplitudes
in $x_{12}$
are not calculable from first principles.   However, there is consensus in the literature
that accounting for $y_{12}$ near $1\%$ requires significant contributions from high multiplicity final states ($n \ge 4$), due to the large $SU(3)_F$ breaking near threshold.  This observation is consistent with the large $U$-spin breaking required
(potentially from duality violations)
in the OPE/HQE approach.

\subsection{CPV phases intrinsic to mixing}
\label{sec:theorphases}
We introduce three intrinsic CPV mixing phases, defined with respect to the direction of the dominant $\Delta U=2$ dispersive and absorptive mixing amplitudes in the complex plane,
\beq \label{theorphasesdef} \begin{split}
{ \phi_2^\Gamma } &\equiv  {\rm arg}\left[\frac{ \Gamma_{12}}{\frac14  (\lambda_s - \lambda_d )^2 \,\Gamma_2 } \right],~~{ \phi_2^M }\equiv  {\rm arg}\left[ \frac{M_{12} }{\frac14 (\lambda_s - \lambda_d )^2 \, M_2  }\right]\,,\cr
&~~~~~~~~~~~~~~ \phi_2  \equiv  {\rm arg}\left[\frac{q} { p}\,{(\lambda_s - \lambda_d )^2\,  \Gamma_2 \over 4}  \right]\,,\end{split}
\eeq
where $\Gamma_{12}$, $M_{12}$, and $q/p$ can contain NP contributions.
These phases can be viewed as the pure mixing analogs of the final state dependent phases $\phi_f^M$, $\phi_f^\Gamma$, and $\phi_{\lambda_f}$, respectively.
Note that they are quark and meson phase convention independent, like the final state dependent ones, as required for physical phases.
For later use we give the expressions for the (phase convention dependent) arguments of $M_{12}$ and $\Gamma_{12}$ in terms of $\phi_2^M$ and $\phi_2^\Gamma$, respectively, cf. \eqref{MiGammaiForms},
\beq \begin{split}\label{phiMphiM2}
\phi^M& = 2 \,{\rm arg}\big[\lambda_s - \lambda_d \big] + 2\, i \,\xi  +  \pi  (1 - \eta^M_2 )/2 + \phi_2^M\,,\cr
\phi^\Gamma& =2 \, {\rm arg}\big[\lambda_s - \lambda_d \big] + 2\, i\, \xi  +  \pi  (1 - \eta^\Gamma_2 )/2 + \phi_2^\Gamma\,.\end{split}\eeq

Employing \eqref{M2ovGamma2}, the theoretical or intrinsic mixing phases are seen to satisfy the relations
\beq \phi_{12} = \phi_2^M  - \phi_2^\Gamma \, , \label{phi12phiMphiGamma}\eeq
and the analog of \eqref{tan2phifnew},
\beq
\tan 2 \phi_2  = - \left(\,{x_{12}^2 \sin 2 \phi_{2}^M + y_{12}^2 \sin 2 \phi_{2}^\Gamma \over
x_{12}^2 \cos 2 \phi_{2}^M + y_{12}^2 \cos 2 \phi_{2}^\Gamma }\,\right).\label{tan2phitheory}\eeq
Combining the two relations, $\phi_2$ can be related to $\phi_{12}$, and $\phi_2^\Gamma$ or $\phi_2^M$, to first order in CPV quantities, as
\beq \begin{split}
 \tan 2 (\phi_2 + { \phi_{2}^{\Gamma}})  &\approx -\,{x_{12}^2 \over x_{12}^2 + y_{12}^2  }   \sin 2{ \phi_{12}} \cr
  \tan 2 (\phi_2 + { \phi_{2}^{M}})  &\approx {y_{12}^2 \over x_{12}^2 + y_{12}^2  }   \sin 2{ \phi_{12}} \,.
 \label{tan2phitheory2} \end{split}
 \eeq
Together with \eqref{qovpphi12}, the above relations allow translation between $\phi_2$ and $|q/p|$, and any two out of the three phases
$\phi_{2}^M$, $ \phi_{2}^\Gamma$, and $\phi_{12}$.

We estimate the magnitudes of the theoretical phases in the SM ($\Gamma_{12} = \Gamma_{12}^{\rm SM}$, $M_{12} = M_{12}^{\rm SM}$), as well as their deviations from the corresponding final state dependent phases $\phi_f^\Gamma$, $\phi_f^M$, and $\phi_{\lambda_f} $, using $U$-spin based arguments and experimental input.
To very good approximation, the CKM hierarchy $|\lambda_b /(\lambda_s - \lambda_d )|\ll 1$ yields,
\beq \begin{split} \label{theorphases}
{ \phi_2^\Gamma }& = {\rm Im}\left(  \frac{2 \lambda_b }{ \lambda_s - \lambda_d}  \frac{\Gamma_1 }{\Gamma_2} \right)
= \left| \frac{\lambda_b }{ \theta_{C} } \right| \sin \gamma  \,\times  \frac{\Gamma_1 }{\Gamma_2} \,.
\end{split}
\eeq
Taking into account the $U$-spin breaking hierarchy $\Gamma_1/ \Gamma_2 = {\mathcal O}(1/\epsilon)$, cf. \eqref{Uspinamps}, yields
the rough SM estimates\footnote{We thank Yuval Grossman for this estimate.}
\beq \begin{split} \label{theorphases2}
{ \phi_2^\Gamma }&  \sim \left| \frac{\lambda_b }{ \theta_{C} } \right| \sin \gamma  \times \frac{1} { \epsilon},
\end{split}
\eeq
and similarly for $\phi_2^M$.
In terms of %the $U$-spin breaking parameter $\epsilon$, %Taking the nominal value $\epsilon \sim 0.2$ for $U$-spin breaking,
the most recent CKM fits \cite{CKMfitter,UTfittriangle}, we obtain
\beq  \phi_{12} \sim \phi_2^\Gamma \sim \phi_2^M \sim (2.2 \times 10^{-3}) \times\left[ \frac{0.3}{\epsilon}\right]
\,.\label{phiGphiMnaive}\eeq
The third phase, $\phi_2$, is seen to be of same order, barring
large cancelations, cf. \eqref{tan2phitheory}.

An alternative expression for $\phi_2^\Gamma$ in the SM follows from \eqref{theorphases}, via the relation $|\Gamma_2 |  \cong |y| \,\Gamma_D /\lambda_s^2$,
 \beq \begin{split}|\phi_2^\Gamma| &=  \left|\frac{ \lambda_b \,\lambda_s \,\sin\gamma} {y }\right| \,\frac{|\Gamma_1 |} { \Gamma_D }\cr&= 0.005 \,\left( {0.66\% \over |y|}  \right)\, \frac{|\Gamma_1 |} { \Gamma_D }  \sim 0.005\, \epsilon \end{split} \,,\label{refined}\eeq
where in the second relation we have incorporated the current central value of $|y|$ \cite{Amhis:2016xyh}, and in the last relation we have taken $\Gamma_1 \sim \epsilon \,\Gamma_D $ (recall that the inclusive approach yields $\Gamma_{ij} \sim \Gamma_D$).
The estimates for $\phi_2^\Gamma$ in \eqref{phiGphiMnaive} and \eqref{refined} are consistent with each other (for illustrative purposes, if we identify their respective $\epsilon $ factors, the two estimates would coincide for $\epsilon \approx 0.36$).

The $\epsilon$ dependence in \eqref{refined} has been shifted to the numerator, compared to  \eqref{phiGphiMnaive} [note that $y = O(\epsilon^2 )$].   This allows us to obtain an approximate upper bound on $\phi_2^\Gamma$, which we briefly describe here.  A detailed discussion will be given elsewhere \cite{prlDmix}.  We rewrite
the ratio of widths in \eqref{refined} as
\beq\begin{split}\label{Gamma1ratio}    {|\Gamma_1 | \over \Gamma_D } & ={ |\Gamma_{sd} |\over \Gamma_D  }\, \epsilon_{1}\,,\end{split}\eeq
where, cf.  \eqref {Uspinamps},
\beq \label{eps1def} \epsilon_{1} \equiv{ |\Gamma_{dd} -\Gamma_{ss} |\over  |\Gamma_{sd}| }= O(\epsilon) \,.\eeq
%i.e. it enters at first order in $SU(3)_F$ breaking.
Moreover, $SU(3)_F$ flavor symmetry arguments yield the bound
\beq { |\Gamma_{sd} |\over \Gamma_D  } < 1 + O(\epsilon) \,.\label{Gammasdratio}\eeq
The $O(\epsilon)$ correction in \eqref{Gammasdratio} originates from
differences between the $D^0$ decay matrix elements for $U$-spin related DCS and CF final states, modulo the CKM factors.  It is expected to be small since it does not depend on $U$-spin breaking from phase space differences.\footnote{Phase space differences enter the RHS of \eqref{Gammasdratio} at $O(\epsilon^2)$ \cite{prlDmix}.}
%{\bf*** AK: checking - not sure about the size of phase space breaking***}
(It is interesting to note that $ |\Gamma_{sd} |/ \Gamma_D \approx 0.6 - 0.75$ has been obtained in the OPE  based approach
 \cite{Bobrowski:2010xg}.)
%Given that the differences between the $\Gamma_{ij}$ enter at first order in $U$-spin breaking,
%(and that they are expected to respect the hierarchy $|\Gamma_{dd}| > |\Gamma_{sd} |> |\Gamma_{ss}|$),
%the quantity $\epsilon_{sd}$ should, conservatively, satisfy the upper bound,
% \beq \epsilon_{sd} <1\,\label{epssdbnd} \,.\eeq
%%Given that the widths satisfy $|\Gamma_{dd}| > |\Gamma_{sd} |> |\Gamma_{ss}|$, and it is very conservative
%Moreover, as discussed in Appendix~\ref{sec:AppendixB}, $SU(3)_F$ flavor symmetry arguments yield the bound,
%\beq { |\Gamma_{sd} |\over \Gamma_D  } < 1 + O(\epsilon) \,.\label{Gammasdratio}\eeq
%The $O(\epsilon)$ correction is expected to be small, as it is originates from differences between the $D^0$ decay rates to $U$-spin related DCS and CF final states, modulo the CKM factors.  Thus, there are no phase space contributions or potentially large QCD penguin effects.
%Finally, combining the first relation in \eqref{refined} with \eqref{Gamma1ratio}--\eqref{Gammasdratio},
Thus, we obtain the absorptive CPV upper bound,
 \beq |\phi_2^\Gamma| <  0.005 \,\left( {0.66\% \over |y|}  \right) \epsilon_{1}  \left[1 + O(\epsilon)\,\right],\label{phi2Gbound1}\eeq
where, conservatively, $\epsilon_1 < 1$.

Combining \eqref{Gammasdratio} with the measured value of $y$ also yields the lower bound, cf. \eqref {Uspinamps},
\beq
(\epsilon_2 )^2 \equiv {| \Gamma_{ss} + \Gamma_{dd} - 2 \Gamma_{sd} | \over |\Gamma_{sd}  | } > 0.14 \left( {|y| \over 0.66\%}  \right) [1+ O(\epsilon) ]\,.\label{e2bound}\eeq
Given that $(\epsilon_2)^2 = O(\epsilon^2)$, \eqref{e2bound} confirms the existence of large $U$-spin breaking in $\DDbar$ mixing.

%
%
%Taking $y$ at its central value, and $\epsilon_{sd}\le 1$ would yield the bound
%$ |\phi_2^\Gamma | < 0.005 \,%\left( {0.66\% \over |y|}  \right)
% \left[1 + O(\epsilon)\,\right] $.
%

%where the second ratio in \eqref{phi2bound} is of $O(\epsilon)$.

%
%
%, which essentially amounts to
%imposing the inequality $\epsilon \lsim 1 $ in \eqref{refined}.
%
%%, \eqref{refined} can be employed to obtain an approximate upper bound on $\phi_2^\Gamma$.
%
%\beq \phi_2^\Gamma \lsim 0.005\,,\label{phi2Gbound} \eeq
%provided that $\epsilon \lsim 1$ therein, i.e. if the relation $|\Gamma_1 | \lsim { \Gamma_D}$ is satisfied.
%%(or $\epsilon \lsim 1$, therein)
%
%%However, the $\epsilon$ dependence in the latter has been shifted to the numerator:
%%$|y| = O(\epsilon^2 )$, while $\Gamma_1 = O(\epsilon)$.
%%Consequently, \eqref{refined} would yield the upper bound, $\phi_2^\Gamma <  0.005$,
%%if the relation $|\Gamma_1 | <  { \Gamma_D}$ %(or $\epsilon < 1$)
%%is satisfied.
%
%*****************
%if the relation
%%$\epsilon \le 1$ (corresponding to
%$|\Gamma_1 |  \le { \Gamma_D}$ (motivated by the magnitudes of the $\Gamma_{ij}$ in the OPE) were to hold.

In principle, $\Gamma_1$ can be estimated via the exclusive approach, as more data on SCS $D^0$ decay branching ratios and direct CP asymmetries become available.  It relies on the $U$-spin decomposition of exclusive contributions to $\Gamma_1$. Details can be found in  \cite{kagancharm2015}.
Unfortunately, the potentially large contributions from high multiplicity final states would complicate this program,
as in the case of $\Delta \Gamma_D$.

\subsection{Final state dependence}
\label{sec:finstatedep}

The misalignments between the final state dependent phases $\phi_f^M$, $\phi_f^\Gamma$, $\phi_{\lambda_f}$,
and their theoretical counterparts are equal in magnitude, satisfying
\beq \begin{split} \label{dphif}
{ \delta \phi_f } &\equiv \phi_{f}^\Gamma - \phi_{2}^\Gamma = \phi_{f}^M - \phi_{2}^M  = \phi_2 -\phi_{\lambda_f}\,.
\end{split} \eeq
Below, we discuss the size of $\delta \phi_f$ in the SM for (i) SCS decays, (ii) CF/DCS
decays to $K^\pm X$, and (iii) CF/DCS decays to $K^0 X$, $\K0bar X$.

\subsubsection{SCS decays}
\label{sec:SCSfinstatedep}
The amplitudes for the SCS decay modes $D^0 \to f$ and $\D0bar \to f$ in the SM can be written as, see e.g. \cite{Brod:2012ud},
\beq  \label{SCSampsgeneric}\begin{split}
A_f &= \frac12 \big(\lambda^*_s - \lambda^*_d \big)\, {\cal A}_{f,1}  + \lambda^*_b \,{\cal A}_{f,0} %+\delta {\cal A}_f^{\rm NP}
\,,\cr
\overline A_f &= \frac12 \big(\lambda_s - \lambda_d \big)\, \overline {\cal A}_{f,1}  + \lambda_b \,\overline{\cal A}_{f,0} %+\delta \overline {\cal A}_f^{\rm NP}
\,,\end{split}\eeq
with substitutions $f \to \bar f$ for the CP conjugate modes.
The first and second terms in each relation are the
$\Delta U=1$ and $\Delta U=0$ transition amplitudes, respectively, where the former is due to the current-current operators $Q_1 , Q_2$, and the latter is dominated by their QCD penguin contractions.
Generically, both amplitudes are $O(1)$ in $SU(3)_F$ breaking,  %, yielding ${\cal A}_{f,0} /{\cal A}_{f,1} = O(1)$.
and the $\Delta U=0$ amplitude is parametrically suppressed by $O(\lambda_b /\theta_{C} )$. (Two exceptions are mentioned below).

The amplitudes for decays to CP eigenstates are generally of the form given in \eqref{eq:twoamp}.
In the case of SCS decays, comparison with \eqref{SCSampsgeneric} yields the weak phase,
\beq\label{SCSCPargAfAfbar}
{\rm arg}\left[\eta_f^{CP} \,{A_f \over \overline A_f } \right] = - 2 \,{\rm arg}\big[\lambda_s - \lambda_d \big] - 2 i \xi + 2 r_f \cos\delta_f \sin\phi_f ,\eeq
where the sum of the first two terms on the RHS is identified with $2 \phi_f^0$ (the second term originates from the choice of meson phase convention), and in the SM,
 \beq \begin{split}\label{rfetcSCS}
\delta_f & = {\rm arg}[{\cal A}_{f,0}/{\cal A}_{f,1}] \,,~~~~~~~~~\phi_f  =- \gamma\,,\cr
&~~~~~~~~~~~~~r_f = \left|{\lambda_b \over \theta_{C}}  \,{{\cal A}_{f,0} \over {\cal A}_{f,1} }\right|\,.\end{split}\eeq
Combining \eqref{phiMphiM2} and \eqref{SCSCPargAfAfbar} yields the following expressions for the CPVINT phases $\phi^M_f$, $\phi_f^\Gamma$, cf. \eqref{lambdaMf}, \eqref{phiMfCP},
\beq \begin{split}\label{SCSCPphiMfphiM2}
\phi^M_f & =  \pi  (1 - \eta^M_2 )/2   +   \phi^M_2 -  2 r_f \cos\delta_f \sin\gamma \,,\cr
\phi^\Gamma_f & =  \pi  (1 - \eta^\Gamma_2 )/2   +   \phi^\Gamma_2 -  2 r_f \cos\delta_f \sin\gamma \,.\end{split}\eeq
Given that
$\phi_f^M$, $\phi_f^\Gamma \approx 0$ (rather than $\pi$) for $f=\pi^+\pi^-$, $K^+ K^-$, cf. \eqref{phiM0},
we learn that the first term on the RHS of each relation in \eqref{SCSCPphiMfphiM2} must vanish, i.e. $\eta^M_2 = \eta_2^\Gamma = +$, as claimed in \eqref{eta2eq1}.
In turn, the misalignment in \eqref{dphif} for a CP eigenstate final state, is given by
 \beq\label{dphifSCSCP}  \delta \phi_f =-  2 r_f \cos\delta_f \sin\gamma = - a_f^d  \cot\delta_f\,,\eeq
where the direct CP asymmetry, $a_f^d$, has been defined in \eqref{afdCP}.

It is instructive to rewrite the CPVINT asymmetry $\Delta Y_f$, cf. \eqref{DYfresult}, in terms of $\phi_2^M$, and the subleading decay amplitude parameters $r_f$, $\phi_f$, and $\delta_f$, cf. \eqref{rfetcSCS},
\beq  \label{DYfull} {\Delta Y_f \over  \eta_{CP}^f} =- x_{12} \sin\phi_2^M  - 2 r_f \sin\phi_f \big( x_{12} \cos\delta_f + y_{12} \sin\delta_f  \big) \,.\eeq
Previously, we saw that the leading amplitude contribution is purely dispersive for CP eigenstate final states, because the requisite CP-even phase difference is only present in the dispersive mixing amplitude ($\delta = \pi/2$).  Similarly, it is now clear that the strong phase dependence of the dispersive and absorptive contributions entering at first order in the subleading amplitudes, cf. \eqref{DYfull},
can be attributed to the strong phase differences $\pi/2 + \delta_f$ and $\delta_f$, between their respective interfering decay chains.

In the case of SCS decays to non-CP eigenstates,  the misalignments of the CPVINT phases, %$\phi_f^{M}$, $\phi_f^{\Gamma}$
cf. \eqref{lambdaMfnonCP}--\eqref{phiMfnonCP},
generalize as
 \beq\label{dphifSCSnonCP}  \begin{split} \delta \phi_f  & =-   (r_f \cos\delta_f +r_{\bar f} \cos\delta_{\bar f} )  \sin\gamma  \cr
 &=
  -\,(a_f^d  \cot\delta_f+a_{\bar f}^d  \cot\delta_{\bar f} )/2 \,,\end{split}\eeq
where $r_f$, $\delta_f$ are defined as in \eqref{rfetcSCS}; $r_{\bar f}$, $\delta_{\bar f}$ correspond to the substitutions
$f \to \bar f$ therein; and $\phi_f = \phi_{\bar f} = -\gamma$.
The direct CP asymmetries have been defined in \eqref{dirCPnonCP}.

The misalignments \eqref{dphifSCSCP}, \eqref{dphifSCSnonCP} for SCS decays are non-perturbative, and incalculable at present, like the direct CP asymmetries.
However, the strong phases are expected to satisfy $\delta_{f,\bar f} = O(1)$, due to large rescattering at the charm mass scale, yielding
the order of magnitude estimates $\delta \phi_{f} =   {\mathcal O}( \lambda_b \sin\gamma /\theta_{C} ) $.  In particular, the misalignments, like the direct CP asymmetries $a_f^d$  are $O(1)$ in $SU(3)_F$ breaking.  Thus, they are parametrically suppressed relative to the theoretical phases in the SM, cf.~\eqref{theorphases},
\beq \label{SCSappuniv} {\delta \phi_f  \over \phi_{2}^{M} }\,, {\delta \phi_f  \over \phi_{2}^{\Gamma} } = {\mathcal O}({ \epsilon})  \,.\eeq

For example, the recent LHCb discovery \cite{Aaij:2019kcg} of a non-vanishing difference between the $D^0 \to K^+ K^- $ and $D^0 \to \pi^+ \pi^-$ direct CP asymmetries yields the world average \cite{Amhis:2016xyh},
\beq\label{dacp} \Delta a_{CP}^{\rm dir} \equiv a^d_{K^+ K^- } - a^d_{\pi^+ \pi^-} = -0.00164 \pm 0.00028 \,.\eeq
In the $U$-spin symmetric limit, $a^d_{\pi^+ \pi^- } = - a^d_{K^+ K^-}$ \cite{Grossman:2006jg}, implying the rough estimate
$\delta \phi_f \sim 0.08\%$ for these decays.  Dividing by the SM estimates for
$\phi_2^M$ and $\phi_2^\Gamma$ in \eqref{phiGphiMnaive} or \eqref{refined} yields significant misalignments, consistent with the parametric suppression in \eqref{SCSappuniv} for sizable $\epsilon \sim 0.4$.

Fortunately, the $K^+ K^-$ and $\pi^+ \pi^-$ misalignments, like the direct CP asymmetries \cite{Grossman:2006jg}, are equal and opposite in the $U$-spin limit, i.e.
\beq\begin{split} \label{SCSUspin}( \delta \phi_{K^+ K^- }+ \delta \phi_{\pi^+ \pi^-})&= O(\epsilon\, \delta \phi_{K^+ K^-,\pi^+ \pi^-})\,,\cr
(a^d_{K^+ K^- } +a^d_{\pi^+ \pi^-} )&= O(\epsilon\, a^d_{K^+ K^- ,\pi^+ \pi^-})\,.\end{split}\eeq
Thus, the average of $\phi_f^{M,\Gamma}$ over $f= K^+ K^-, \,\pi^+ \pi^-$ satisfies,
\beq \label{Uspinmisalign}\frac12 ( \phi^{M,\Gamma}_{K^+ K^-} +  \phi^{M,\Gamma}_{\pi^+\pi^- } )= \phi_2^{M,\Gamma} [1 + O(\epsilon^2) ] \,,\eeq
and the average of the time dependent CP asymmetries in \eqref{DYfresult} satisfies,
\beq \begin{split} \label{AGammaappuniv} A_\Gamma&= - x_{12}\, \phi_2^M [ 1 + O(\epsilon^2) ]\,,
\end{split}
\eeq
where we have used the relations $x_{12}\sim y_{12}$ and $\delta\phi_f \sim a_f^d $.

As has already been noted, large $U$-spin violation
is likely to play an important role in mixing.  Moreover, the $\delta \phi_f$ for SCS decays
are inherently non-perturbative.  Therefore, while \eqref{SCSappuniv} implies that
the order of magnitude
estimates \eqref{phiGphiMnaive}, \eqref{refined} for $\phi_2^{M,\Gamma}$ apply equally well to the measured phases
$\phi_f^{M,\Gamma}$ in the SM, $O(1) $ variations can not be ruled out.
The latter possibility would correspond to the weakest form of approximate universality.
Ultimately, precision measurements of the indirect and direct CP asymmetries in a host of SCS decays will clarify the situation.

We point out that in the presence of NP in SCS decays,
the expressions for the misalignments, $\delta \phi_f$, in the second relations of \eqref{dphifSCSCP}, \eqref{dphifSCSnonCP} remain valid. In particular, the direct CP asymmetries $a_{f,\bar f}^d$ and the strong phases $\delta_{f, \bar f}$ now depend on the total subleading amplitudes, i.e the sums of the QCD penguin and NP amplitudes.  The $\delta \phi_f$ would be of same order as in the SM, %Again, $\lsim O(1)$ variations in the measured phases $\phi_f^{M,\Gamma}$ are expected,
provided that the CP-odd NP amplitudes are similar in size, or smaller than the SM QCD penguin amplitudes,
as already hinted at by the current bounds on direct CPV in $D^0 \to K^+ K^- , \pi^+ \pi^- $ decays.

Finally, we mention two SCS decay modes, $D^0 \to K^0 \K0bar$  and $D^0 \to K^{*0} \K0bar$,
which violate the $O(\epsilon)$ counting in \eqref{SCSappuniv}.  For $D^0 \to K^0 \K0bar$, the first term in \eqref{SCSampsgeneric} is suppressed by $O(\epsilon)$ (as reflected in the rate),
yielding $O(1/\epsilon)$ enhancements of
$\delta \phi_f$, the direct CP asymmetry \cite{Brod:2011re}, \cite{Nierste:2015zra},
and the misalignment, i.e. $ {\delta \phi_f  /\phi_{2}^{M,\Gamma} } = O(1)$ in the SM.
For $D^0 \to K^{*0} \K0bar$, the first term in \eqref{SCSampsgeneric} is not formally suppressed by $O(\epsilon)$. However, a large accidental cancelation between contributions related by $K^{*0} \leftrightarrow \K0bar$ interchange (again reflected in the measured decay rate), again enhances $\delta \phi_f$, and the direct CP asymmetry \cite{Nierste:2017cua}.  Thus, in effect, the misalignment could be $O(1)$, as for $K^0 \K0bar$.

\subsubsection{CF/DCS decays to $K^\pm X$ }
\label{sec:finstatedepKpm}

The CPVINT observables in this class
are given in \eqref{lambdaMfnonCP}, \eqref{lambdaMfbarnonCP}, with the modified sign convention of \eqref{lambdaCFDCS}.
The CKM factors enter the CF/DCS amplitudes as $A_{f} \propto V_{cs}^* V_{ud} $ (CF) and $\bar A_f  \propto V_{cd} V_{us}^* $ (DCS).  Thus, in the SM and, more generally, in models with negligible new weak phases in CF/DCS decays, Eqs. \eqref{phiMfnonCPCFDCS} and \eqref{phiMphiM2}  %(with $\eta_2^{M,\Gamma}=1$)
yield the absorptive and dispersive phases,
\beq \label{phiMGCFDCS}
\phi^{M \,(\Gamma)}_f = \phi_2^{M\,(\Gamma)} + {\rm arg}\left[- {V_{cs}^* V_{ud} \over V_{cd} V_{us}^* } \,(\lambda_s - \lambda_d)^2 \right]\,.\eeq
Employing CKM unitarity, the misalignments, given by the second term on the RHS, are seen to satisfy
\beq \delta \phi_f =
O\left({\lambda_b^2 \over  \lambda_d^2 }\right). \label{dphifCFDCDnonCP}
\eeq
Thus, for CF/DCS decays to $K^\pm X$, the misalignments vanish    % $\delta \phi_f =0$,
up to a negligible (and precisely known) final-state independent correction of $O(10^{-6})$.
This represents the strongest form of approximate universality, i.e. the universal limit $\phi^{M \,(\Gamma)}_f = \phi_2^{M\,(\Gamma)}$. 
In particular, CPVINT measurements in these decays directly determine the theoretical phases.

\subsubsection{CF/DCS decays to $K^0 X$, $\K0bar X$}
\label{finstatedepK0}

We begin with a discussion of the misalignments %$\delta \phi_f$
in this class of decays in the limit that the DCS decays are neglected.
Expressions for the CPVINT observables and time-dependent decay widths in this approximation are given in \eqref{lamKSvsKL}--\eqref{phiMfCFK0} and Section \ref{sec:K0Xtimedep}, respectively.
The misalignments follow from  \eqref{phiMfCFK0}.
One ingredient is the phase of $q_K /p_K$.  To excellent approximation \cite{Nir:1992uv}, this ratio satisfies the relation
\beq\label{qovpeps} {q_K\over p_K}= {A_0 \over \overline A_0} \,( 1- 2\,\epsilon_K) \,,\eeq
where $A_{0\,,2}$ denote the $K^0\to (\pi\pi)_{I=0\,,2}$ amplitudes, respectively, i.e. they are
$\Delta I =1/2\,, 3/2$ transitions.
Keeping track of the CKM factors, these amplitudes can be written as
\beq \begin{split}\label{A0A2CKM}
A_{0\,(2) } &= V_{ud} V_{us}^*  \,{\cal A}_{0\,(2)}  +  V_{td} V_{ts}^* \, {\cal B}_{0\,(2)}\cr
&=V_{ud} V_{us}^* \, {\cal A}_{0\,(2)}\, \big[1  + r_{0\,(2)} \big]\,,\end{split}\eeq
yielding
\beq \label{argqovp} {\rm arg}\!\left[{q_K\over p_K}\right] = 2\, {\rm arg} \big[V_{ud} V_{us}^* \big]  - 2 \epsilon_I + 2 {\rm Im} [r_0] \,.\eeq
A second ingredient is the $CP$-odd phase in the ratio of CF amplitudes, ${A_{\K0bar X}/\overline A_{K^0 X }}$,
\beq \label{phiK0bar} \begin{split} 2 \phi_{\K0bar X}^0 &=  2\, {\rm arg}\big[V^*_{cs} V_{ud}\big] - 2 i \xi\cr
&= 2\, {\rm arg}\big[V^*_{us} V_{ud}\big]+ 2\, {\rm arg}\big[\lambda^*_s \big]- 2 i \xi \,.\end{split}\eeq
Finally, combining \eqref{phiMphiM2},\eqref{argqovp}, and \eqref{phiK0bar} yields the final state independent absorptive and dispersive phases,
\beq \label{phiMf2K0}\begin{split} \phi^{M\,(\Gamma)}_f &= \phi_2^{M\,(\Gamma)} +2 \,\epsilon_I+ \left|\lambda_b\over\lambda_s \right|\sin\gamma - 2\, {\rm Im}[r_0] \,. \end{split}\eeq

The last term in \eqref{phiMf2K0} %, ${{\rm Im}[r_0]}$,
is non-perturbative in origin.  However, it enters the kaon CPV observable, $\epsilon_K^\prime /\epsilon_K$, as\footnote{In a phase convention commonly employed for discussions of ${\epsilon_K^\prime / \epsilon_K}$,
${\rm Im}[r_{0\,(2)}] = {\rm Im}[A_{0\,(2)}]/{\rm Re}[A_{0\,(2)}]$.}
 \beq \label{epspepsanatomy} \begin{split}  {\rm Re}\!\left[{\epsilon_K^\prime \over \epsilon_K }\right] &=(1.66 \pm 0.23)\times 10^{-3}~~~ \text{\cite{PDG}}\cr
 &= -{\omega \over \sqrt{2} |\epsilon|} \left(  {{\rm Im}[r_0]} -  {{\rm Im}[r_2]}\right) \,,\end{split}\eeq
where $\omega \equiv ({\cal A}_{2}/{\cal A}_{0})\approx 1/22  $.
Equating the measured value of $ {\rm Re}[{\epsilon_K^\prime / \epsilon_K }]$ with the first term on the RHS of the second relation in \eqref{epspepsanatomy}, i.e. assuming modest cancelation with $A_2$ \cite{Gisbert:2018tuf},
yields the estimate
\beq \label{ImA0ReA0}  {{\rm Im}[r_0]} \approx 1.2\times 10^{-4}\,.\eeq
Similarly, the dominant chirally enhanced penguin operator ($Q_6$) contribution to $A_0$ yields
 \cite{Gisbert:2018tuf},
\beq \label{ImA0ReA0Q6}  {{\rm Im}[r_0]} \approx 1.5\times 10^{-4} B_6^{(1/2)} \,,\eeq
where the matrix element parameter $B_6^{(1/2)} =1$ in the large $N_C $ limit.
(A recent study \cite{Aebischer:2018rrz} claiming that the SM prediction for $\epsilon^\prime/\epsilon$ could be significantly smaller than the measured value obtains $ {{\rm Im}[r_0]}< 10^{-4}$).

Thus, in the limit that the DCS amplitudes are neglected, the misalignments satisfy
\beq \delta \phi_f = 2 \,\epsilon_I+ \left|\lambda_b\over\lambda_s \right|\sin\gamma =  3.7 \times 10^{-3}, \label{phiepsapp} \eeq
up to a small CP-odd ratio of $K\to \pi\pi$ amplitudes, given by $-2  {{\rm Im}[r_0]}= O(10^{-4})$. The latter lies an order of magnitude below our SM estimates for the theoretical phases $\phi_2^M$, $\phi_2^\Gamma$ in \eqref{phiGphiMnaive}, \eqref{refined} and can be neglected.

Finally, we address the impact of the DCS amplitudes. % on $\phi^M_f$, $\phi^\Gamma_f$.
Expanding the CPVINT observables in \eqref{genlambdaK0} to first order in the DCS amplitudes,
the weak and strong phases in $\lambda^{M,\Gamma}_{K_{S/L} X}$ are seen to be related to those in $\lambda_f^{M,\Gamma}$ (cf. \eqref{genlambdaK0form} and \eqref{lambdaMfCFK0}, respectively), as
\beq
\begin{split}\label{DCScorrnsphases}
\phi^{M} [K_{S/L} X] &= \phi_f^{M} \pm (r_f \cos\delta_f + r_{\bar f} \cos\delta_{\bar f} ) \,\delta\phi_f\,,\cr
\phi^{\Gamma} [K_{S/L} X] &= \phi_f^{\Gamma} \pm (r_f \cos\delta_f + r_{\bar f} \cos\delta_{\bar f} ) \,\delta\phi_f\,,\cr
\Delta [K_{S/L} X]&=\Delta_f \pm (r_f \sin\delta_f - r_{\bar f} \sin\delta_{\bar f}) \,,\cr
\end{split}
\eeq
where $\delta \phi_f$ is given in \eqref{phiepsapp}.
We recall that $\phi_f^{M,\Gamma}$ are the CPV phases in the absence of the DCS amplitudes, $r_f$ and $r_{\bar f}$ are the magnitudes of DCS to CF amplitude ratios,
\beq
r_f  = \left| {A_{K^0 X}\over A_{\K0bar X}}\right|,~~~r_{\bar f}  = \left| {\overline A_{\K0bar X}\over \overline A_{K^0 X}}\right|\,,\eeq
and $\delta_f$, $\delta_{\bar f}$ are the strong phase differences of the corresponding amplitude ratios.
Finally, their magnitudes are related as
\beq\begin{split}\label{DCScorrnsmags}
\left|\lambda^{M}_{K_{S/L} X} \right|&=\left|\lambda_f^{M} \right|\,\big(1- [r_f  \cos\delta_f  - r_{\bar f} \cos\delta_{\bar f } \,]\big)\,,\cr
\left|\lambda^{M}_{\overline{K_{S/L} X}} \right|&=\left|\lambda_{\bar f}^{M} \right|\,\big(1+ [r_f  \cos\delta_f  - r_{\bar f} \cos\delta_{\bar f } \,]\big)\,,\end{split}\eeq
and similarly for $M \to \Gamma$.

Expressions for the time dependent decay widths, including the DCS amplitudes, are obtained via insertion of
the CPVINT observables \eqref{genlambdaK0form} %(including the expansions in \eqref{DCScorrnsphases}, \eqref{DCScorrnsmags} for the phases and magnitudes),
and the full expressions for the decay amplitudes \eqref{DtoKaXamps} into the general formulae \eqref{twotimeamps2} for the time-dependent amplitudes.
The result can be brought into the same general form as \eqref{timedepratesK0f}, \eqref{timedepratesK0fbar}.
Effectively, the
prefactors in Eqs. \eqref{timedepratesK0f}, \eqref{timedepratesK0fbar}, the ratios $\sqrt{R_f}$, and the expressions \eqref{KSci}, \eqref{KSKLintbidi} for the coefficients are modified at $O(r_f  \,,r_{\bar f}  )$, i.e. $O(\theta_{C}^2)$.   For example, the coefficients contain new CP-even terms of $O(r_{f,\bar f} )$,
and new CP-odd terms of $O(\epsilon_K\,r_{f,\bar f})$.
These corrections produce relative shifts in the CP averaged decay rates, as well as the indirect CP asymmetries
listed in \eqref{K0asymms1}, \eqref{K0asymms2}, \eqref{K0asymmstwobody}, of $O(\theta_{C}^2)$.

Our primary focus here is on the absorptive and dispersive CPVINT phases. As previously noted, they only reside in the pure $K_S$ contributions to the time dependent widths (to first order in CPV).
In particular, $\phi_f^{M,\Gamma}$ are replaced by $\phi^{M,\Gamma} [K_{S} X]$ in the coefficients $c_{f}^\pm$, $c_{\bar f}^\pm$, cf.   \eqref{DCScorrnsphases}, \eqref{KSci}.
Consequently, the misalignments \eqref{phiepsapp} are modified as
\beq \begin{split}  \label{dphifK0wDCS}
\delta{\phi_f }&\equiv \phi^{M\,(\Gamma)} [K_{S} X] - \phi_2^{M\,(\Gamma)} \cr
&=  \left(2 \,\epsilon_I+ \left|\lambda_b\over\lambda_s \right|\sin\gamma\right)  (1 +  r_f \cos\delta_f + r_{\bar f} \cos\delta_{\bar f} )\,\cr
&= \left(2 \,\epsilon_I+ \left|\lambda_b\over\lambda_s \right|\sin\gamma\right)\, \big( 1+ O\big[\theta_{C}^2\big ] \big)\,.
\end{split}\eeq
Thus, while the DCS corrections to the CPVINT phases are final state dependent, they are of $O( 2 \theta_{C}^2 \,\epsilon_I )$, or $O(0.1\, \phi_2^{M,\Gamma})$ in the SM.  This represents a more generic form of approximate universality than what we found in the previous two classes of decays, i.e. an $O(10\%)$ variation among the $\phi_f^M$ and $\phi_f^\Gamma$, corresponding to a similar variation in the CPVINT asymmetries.
The shifts in the asymmetries remain at this order when taking all of the DCS corrections to the widths into account.
We therefore conclude that their inclusion
in \eqref{timedepratesK0f}, \eqref{timedepratesK0fbar} is not warranted for the interpretation of CPVINT data at SM sensitivity.

\section{Implementation of approximate universality}
\label{sec:implement}

In this section, we discuss how to convert the general expressions for the time dependent decay widths and indirect CP asymmetries obtained in Section~\ref{sec:hadronicdecays} to the approximate universality parametrization, in the three classes of decays.  For CF/DCS decays to $K^0 X$, $\K0bar X$,  we pay special attention to $\epsilon_K$ induced effects at LHCb and Belle-II.

\subsection{SCS decays}
\label{sec:SCSimplement}
For SCS decays, the theoretical absorptive and dispersive CPV phases replace the final state dependent ones via the substitutions,
\beq\label{SCSsubs} \phi^{M}_f \to  \phi_2^M,~~~ \phi^{\Gamma}_f \to  \phi_2^\Gamma\,,\eeq
in the expressions for the time dependent decay widths and CP asymmetries. % in Secs. \ref{sec:SCSCPwidths}, \ref{sec:SCSnonCPwidths}.
For decays to CP eigenstates, they enter the expressions for the decay widths \eqref{timedepSCSCP} (via Eq.~\eqref{cpmSCSCP} for $c_f^\pm$) %\eqref{eq:yCPdefExp}, \eqref{yCPapp},
and the CP asymmetry $\Delta Y_f $ \eqref{DYfresult}.
For decays to non-CP eigenstates, they enter
the expressions for the decay widths \eqref{timedepSCSCPf}, \eqref{timedepSCSCPfbar} (via Eq. \eqref{eq:coeffsSCSnonCP} for $c_f^\pm$) and the indirect CP asymmetries $\Delta Y_f$, $\Delta Y_{\bar f}$ \eqref{DYfnonCPdetail}.
Note that the misalignments $\delta \phi_f$ are dropped on the RHS of \eqref{SCSsubs}, as they are not calculable from first principles QCD.
Moreover, while formally of $O(\epsilon)$ in $U$-spin breaking relative to $\phi_2^{M,\Gamma}$, they could, in principle, yield $O(1)$ variations in $\phi_f^M$ and $\phi_f^\Gamma$ in the SM.
In Section~\ref{sec:appunivfits} we discuss  %the limitations of this assumption, and
a strategy for fits carried out once SM sensitivity is achieved, and
final state dependent effects in $\phi_f^M$, $\phi_f^\Gamma$ become accessible to experiment.

The direct CPV ($a_f^d$)
and misalignment ($\delta\phi_f $) contributions to
the CPVINT asymmetries in \eqref{DYfresult}, \eqref{DYfnonCPdetail} are of same order, cf. \eqref{dphifSCSCP}.
Therefore, consistency requires us to drop the $a_f^d , a_{\bar f}^d$ terms
in the CPVINT asymmetries, if we neglect $\delta \phi_f$ in \eqref{SCSsubs}.
For example, for CP eigenstate final states,
and in the approximate universality parametrization, \eqref{DYfresult} reduces to,
\beq \label{dYfappuniv} \Delta Y_f  = -\eta_f x_{12} \sin\phi_2^M \,,\eeq
and similarly for the non-CP eigenstates (the first line of each asymmetry in \eqref{DYfnonCPdetail} is kept, with $\phi_f^{M,\Gamma}\to \phi_2^{M,\Gamma}$).
However, we recall that in the average of $\Delta Y_f$ over $f=K^+ K^- , \pi^+ \pi^-$, i.e. $A_\Gamma$, the
error incurred by dropping $\delta \phi_f$ and $a_f^d$ %in \eqref{DYfresult}
is of $O(\epsilon^2)$, cf. \eqref{Uspinmisalign}
\eqref{AGammaappuniv}.

\subsection{CF/DCS decays to $K^\pm X$}
\label{sec:CFDCSKpmimplement}

For CF/DCS decays to $K^\pm X$, substitute
\beq\label{DCSsubs} \phi^{M}_f \to  \phi_2^M,~~~ \phi^{\Gamma}_f \to  \phi_2^\Gamma\,,\eeq
in the expressions for the decay widths \eqref{timedepWS} (via Eq. \eqref{cpm} for the coefficients $c^\pm$), and the indirect CP asymmetries $\delta c_f$ \eqref{CFDCSCP}. However, in contrast to the SCS decays, the misalignments are entirely negligible, cf. \eqref{dphifCFDCDnonCP}. %and

\subsection{CF/DCS decays to $K^0 X$, $\K0bar X$}
\label{sec:CFDCSK0implement}

In CF/DCS decays to $K^0 X$, $\K0bar X$, the final state dependent phases for $f = \pi^+ \pi^- X$
are replaced by the theoretical phases via the substitutions,
 \beq\label{phi2phiepssum} \phi^{M,\Gamma}_f \to  \phi_2^{M,\Gamma} +2 \,\epsilon_I+ \left|\lambda_b\over\lambda_s \right|\sin\gamma\,,\eeq
in the widths \eqref{timedepratesK0f}, \eqref{timedepratesK0fbar} (via Eq. \eqref{KSci} for the coefficients $c_f^{\pm}$, $c_{\bar f}^\pm$), and in the indirect CP asymmetries $\delta c_f$, $\delta c_{\bar f}$ \eqref{K0asymms1}.
The sum of the last two terms in  \eqref{phi2phiepssum} equals the misalignment $\delta \phi_f $ \eqref{phiepsapp},
up to negligible corrections lying an order of magnitude below our SM estimates of $\phi_2^{M,\Gamma}$, cf. \eqref{ImA0ReA0}, \eqref{ImA0ReA0Q6},\eqref{dphifK0wDCS}.

At LHCb, the bulk of observed $K^0 / \K0bar \to \pi^+ \pi^-$ decays take place within
a time interval\footnote{We thank Marco Gersabek for correspondence on this point.} $t^\prime \lsim \tau_S /3$,
while at Belle-II they can be detected over far longer time intervals\footnote{We thank David Cinabro for correspondence on this point}, e.g. $t^\prime \lsim O(10 \,\tau_S )$.  This has important consequences for the impact of $\epsilon_K$ on the CP asymmetries,
e.g. in $D^0 \to K_S \pi^+ \pi^-$ decays, which we discuss below.

The total time dependent CP asymmetries, following from
 \eqref{timedepratesK0f}, \eqref{timedepratesK0fbar}, \eqref{K0asymms1}, \eqref{K0asymms2},
can be expressed (up to an overall normalization factor) as
\beq \label{timedepDGammaf}
\begin{split}
\Gamma_f - \overline \Gamma_{\bar f}  &= - 2  \,e^{-\tau}   |\overline A_{+ -}|^2  |A_{\K0bar X} |^2   \,\bigg\{2 \epsilon_R \, F_0 (t^\prime )
\cr
+ & \sqrt{R_f} \, \tau\,\bigg[ 2  \epsilon_I\,( x_{12} \cos\Delta_f +  y_{12} \sin\Delta_f  )\, F_1 (t^\prime) \cr
+&\big( x_{12}\cos\Delta_f  \sin {\tilde\phi}_2^M + y_{12} \sin\Delta_f  \sin{\tilde\phi}_2^\Gamma\, \big)e^{-\Gamma_S t^\prime } \bigg]    \bigg\}  \,,\end{split}\eeq
and
\beq \label{timedepDGammafbar}
\begin{split}\Gamma_{\bar f} - \overline \Gamma_{ f}  &= - 2  \,e^{-\tau}   |\overline A_{+ -}|^2  |A_{\K0bar X} |^2   \,\bigg\{2 \epsilon_R \, F_0 (t^\prime )
\cr
+ & \sqrt{R_f} \, \tau\,\bigg[ 2  \epsilon_I\, ( x_{12} \cos\Delta_f-  y_{12} \sin\Delta_f  )\, F_1 (t^\prime) \cr
+&\big( x_{12}\cos\Delta_f  \sin {\tilde\phi}_2^M - y_{12} \sin\Delta_f  \sin{\tilde\phi}_2^\Gamma\, \big)e^{-\Gamma_S t^\prime } \bigg]    \bigg\}\,, \end{split}\eeq
where, for convenience, we have introduced the phase
\beq \label{phi2tilde}   {\tilde\phi}_2^{M,\Gamma} \equiv \phi_2^{M,\Gamma} +  \left|\lambda_b/\lambda_s \right|\sin\gamma\,.\eeq
The CKM  term in \eqref{phi2tilde} is $\approx 6.6 \times 10^{-4}$. %, a factor of a few smaller than our estimates of $\phi_2^{M,\Gamma}$.
The functions $F_0$, $F_1$ satisfy, %entering the $\epsilon_K$ contributions
\beq\begin{split} \label{F0F1exp} F_0 (t ) &= -e^{-\Gamma_S t} +e^{-\Gamma_K t}\! \left( \cos \Delta m_K t +
 {\epsilon_I \over \epsilon_R } \sin\Delta m_K t \right),\cr
 F_1 (t) &=  e^{-\Gamma_S t }-e^{-\Gamma_K t} \! \left( \cos \Delta m_K t -
 {\epsilon_R \over \epsilon_I } \sin\Delta m_K t \right)\,.
 \end{split}\eeq
Note that the ratio $\epsilon_I /\epsilon_R  =1$, up to a small $\approx 5\%$ correction, cf.~\eqref{epsKexp}.
Negligible CP asymmetries entering at $O(\tau^2)$ have not been included in  \eqref{timedepDGammaf}, \eqref{timedepDGammafbar}.
Dividing by the sums over the CP conjugate decay widths yields the normalized time dependent CP asymmetries,
\beq \label{timedepDGammafNorm}
\begin{split}
{\Gamma_f - \overline \Gamma_{\bar f} \over \Gamma_f +  \overline \Gamma_{\bar f} } &= -  \,\bigg\{2 \epsilon_R \, e^{\Gamma_S t^\prime } F_0 (t^\prime )
\cr
+&  \sqrt{R_f} \, \tau\,\bigg[ 2  \epsilon_I\,( x_{12} \cos\Delta_f +  y_{12} \sin\Delta_f  )\, e^{\Gamma_S t^\prime } F_1 (t^\prime) \cr
+&\big( x_{12}\cos\Delta_f  \sin {\tilde\phi}_2^M + y_{12} \sin\Delta_f  \sin{\tilde\phi}_2^\Gamma\, \big)\bigg]    \bigg\}  \,,\end{split}\eeq
and
\beq \label{timedepDGammafbarNorm}
\begin{split}{\Gamma_{\bar f} - \overline \Gamma_{ f}  \over \Gamma_{\bar f} + \overline \Gamma_{ f} } &= - \,\bigg\{2 \epsilon_R \, e^{\Gamma_S t^\prime } F_0 (t^\prime )
\cr
+ & \sqrt{R_f} \, \tau\,\bigg[ 2  \epsilon_I\, ( x_{12} \cos\Delta_f-  y_{12} \sin\Delta_f  )\, e^{\Gamma_S t^\prime } F_1 (t^\prime) \cr
+&\big( x_{12}\cos\Delta_f  \sin {\tilde\phi}_2^M - y_{12} \sin\Delta_f  \sin{\tilde\phi}_2^\Gamma\, \big) \bigg]    \bigg\}\,.\end{split}\eeq

\begin{figure*}[htb]
  \centering
  \includegraphics[width=0.4\textwidth]{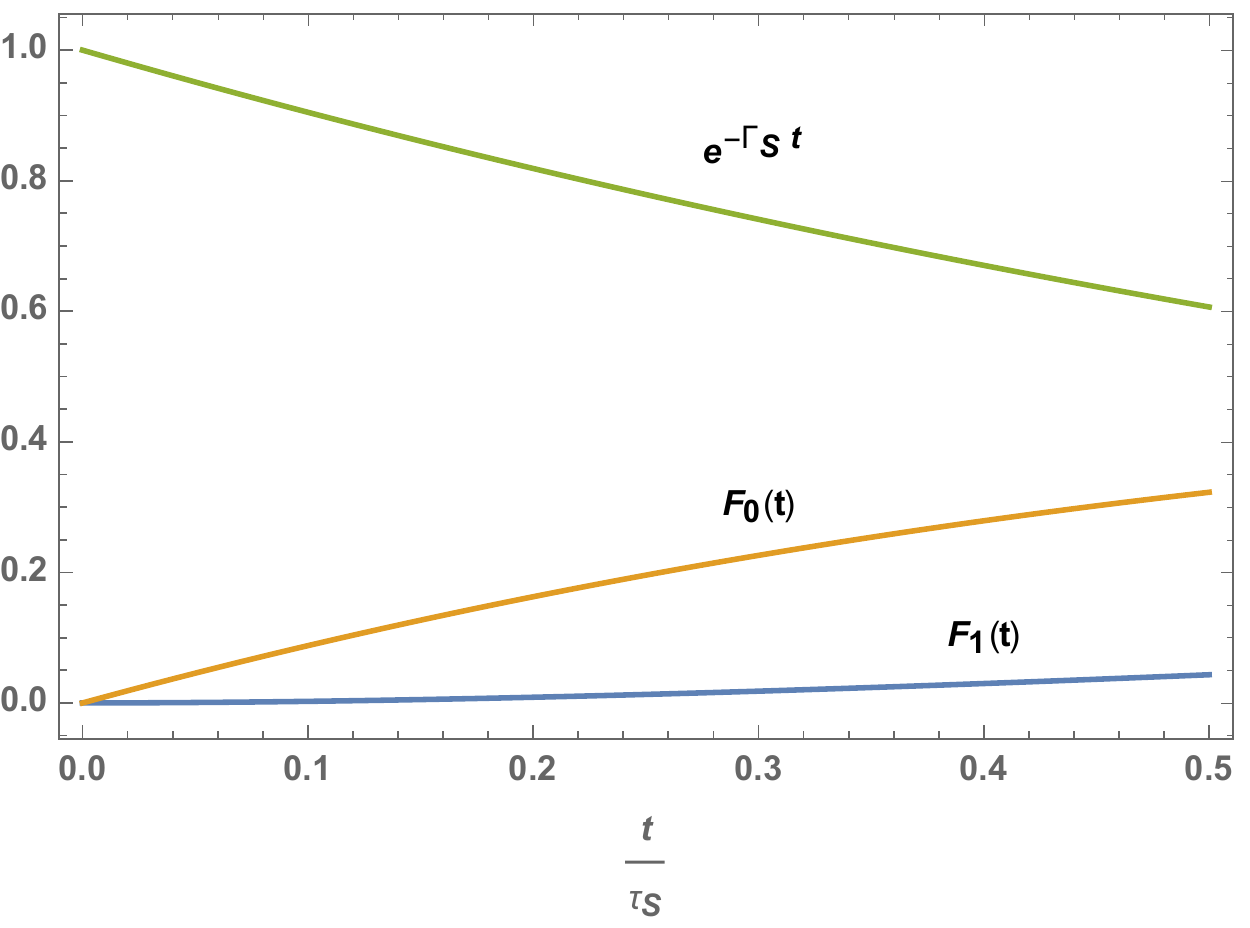}
  \includegraphics[width=0.4\textwidth]{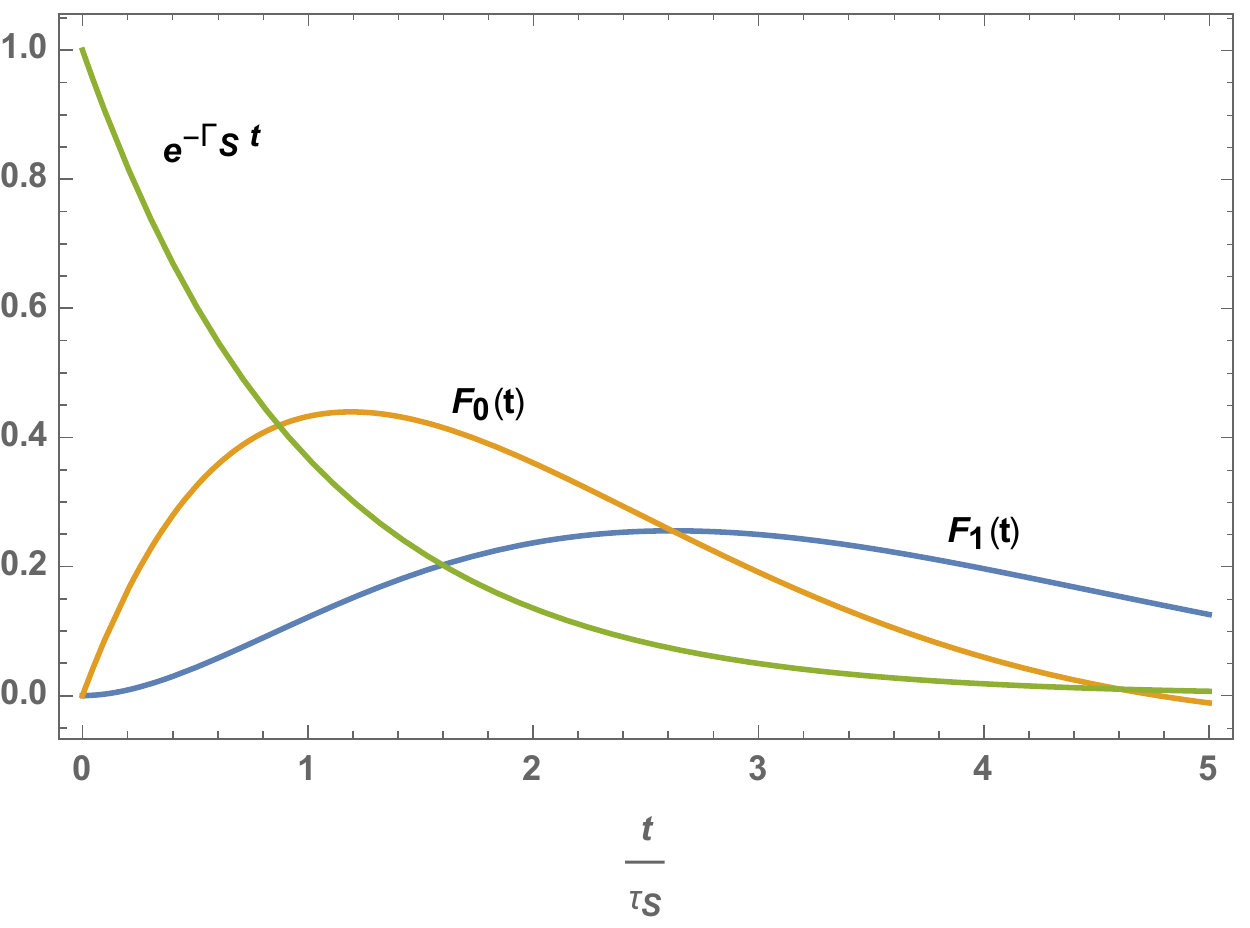} \\  \caption{The functions $F_0 (t)$, $F_1 (t)$, and $\exp[-{\Gamma_S t}]$, plotted over a short time interval of relevance to LHCb (left), and a longer time interval of relevance to Belle-II (right), cf. \eqref{timedepDGammaf}--
\eqref{F0F1exp}.  }
  \label{fig:F0F1plots}
\end{figure*}

The function $F_0$ is associated with direct CPV  %(entirely due to $\epsilon_K$ in the SM)
via integration over $\tau$, and agrees with the expression obtained in \cite{Grossman:2011zk}.
The functions $F_1$ and $e^{-\Gamma_S t^\prime } $
are associated with the contributions of $\epsilon_K$ and $\phi_2^{M,\Gamma}$ to the CPVINT asymmetries, respectively.
In Fig.~\ref{fig:F0F1plots}, we plot the three functions over a short time interval of relevance to LHCb,
and a longer time interval of relevance to Belle-II.  Over the entire time scale for observed $K^0$'s at LHCb, e.g.
$t^\prime \lsim 0.5 \tau_S$, the function $F_1  $ undergoes a remarkable cancelation down to the few percent level, while $e^{-\Gamma_S t^\prime } = O(1)$.  Thus, at LHCb, the contributions of $\epsilon_K$ to the CPVINT asymmetries are highly suppressed
compared to those of $\phi_2^{M,\Gamma}$  (recall that $\phi_2^{M,\Gamma}\sim \epsilon_{I,R}$ in the SM).

The cancelation in $F_1$ at short times takes place between the contributions to CPVINT from $K_L - K_S$ interference  [$\delta b_{f,\bar f}$,
$\delta d_{f,\bar f}$ in \eqref{K0asymms2}], and from the $\epsilon_I$ term in $\phi_f^{M,\Gamma}$ \eqref{phiepsapp} [via $\delta c_{f,\bar f}\,$ in \eqref{K0asymms1}]. %\eqref{phieps}, \eqref{phi2phiepssum}.
Thus, for simplicity, analyses of CPVINT in $D^0 \to K_{S,L} \pi^+\pi^-$ decays at LHCb
could omit a  fit to the interference terms [$\propto e^{-\Gamma_K t^\prime} \tau $ in \eqref{timedepratesK0f}, \eqref{timedepratesK0fbar}], if they substitute
\beq \label{LHCbsub}\phi^{M,\Gamma}_f \to  \phi_2^{M,\Gamma} + \left|\lambda_b/\lambda_s \right|\sin\gamma \,,\eeq
rather than \eqref{phi2phiepssum}.   
In contrast, over the longer $K^0$ decay time scales that can be explored at Belle-II,
the cancelation in $F_1$ subsides, and
$\epsilon_K$ ultimately dominates the CPVINT asymmetries in the SM, cf. Fig.~\ref{fig:F0F1plots} (right).  Thus, Belle-II CPVINT analyses must fit for
$K_L - K_S$ interference % in \eqref{timedepratesK0f}, \eqref{timedepratesK0fbar},
and employ the substitutions in \eqref{phi2phiepssum}, in order to extract $\phi_2^{M,\Gamma}$.
Finally, the function $F_0$ undergoes some cancelation at small time intervals, e.g. $t^\prime \lsim \tau_S/ 3$, leading to moderate suppression of direct CPV at LHCb.

\section{Current status and projections}

\label{sec:fits}

We perform two global analyses of the current experimental data, collected in Table \ref{tab:currentexp},
in order to assess the
current sensitivity to the phases $\phi_{2}^M$ and $\phi_{2}^\Gamma$.  (The  $x_{\rm CP}$, $y_{\rm CP}$, $\Delta x$, $\Delta y$
entries in Tables \ref{tab:currentexp}, \ref{tab:future_exp} correspond to $K_S \pi^+ \pi^-$). We also report on future projections.

\subsection{Superweak limit}

Until recently, fits to measurements of indirect CPV were sensitive to values of $\phi_{12}$ down to the
100 mrad level.  This level of precision probed for large short-distance NP effects.
In particular,
the effects of weak phases in the subleading decay amplitudes could be safely neglected in the indirect CPV observables.
In this limit, referred to as the superweak limit,
a non-vanishing $\phi_{12}$ would be entirely due to short-distance NP in $M_{12}$, with the CPVINT phases satisfying
 \beq \phi_{f}^M =\phi_2^M =  \phi_{12},~~~ \phi_{f}^\Gamma = 0\,,~~~\phi_{\lambda_f} = \phi_2 \,.\eeq
For example, the expression for the SCS time dependent CP asymmetry in \eqref{DYfresult} would reduce to\footnote{In the superweak limit, the effects of weak phases in the SCS decay amplitudes are neglected in time dependent CP asymmetries, but they are kept in time integrated ones, where they are not suppressed by $x_{12}$, $y_{12}$.}
\beq \label{dYfsuperweak} \Delta Y_f = -  \eta_{CP}^f x_{12} \sin\phi^M_{2}\,.\eeq
Thus, the phase $\phi^M_{2}$ (or $\phi_{12}$) would be the only source of indirect CPV.  Consequently, CPVMIX and CPVINT would be related as~\cite{Ciuchini:2007cw,Grossman:2009mn,Kagan:2009gb},
\beq \label{2phivsphi12}
\tan 2\phi_2  \approx  -{x_{12}^2 \over x_{12}^2 + y_{12}^2  } \sin 2 \phi^M_{2}\, ,\eeq
or, equivalently, as
\beq \label{2phivsqovp}
\tan \phi_2 \approx \left(1-\left|{q \over p}\right|\right) \, { x\over y}\,,
\eeq
where \eqref{2phivsphi12} is the superweak limit of \eqref{tan2phifnew}. %, with $\phi_2 $ replaced by the phenomenological observable $\phi$.

\begin{figure*}[htb]
  \centering
  \includegraphics[width=0.3\textwidth]{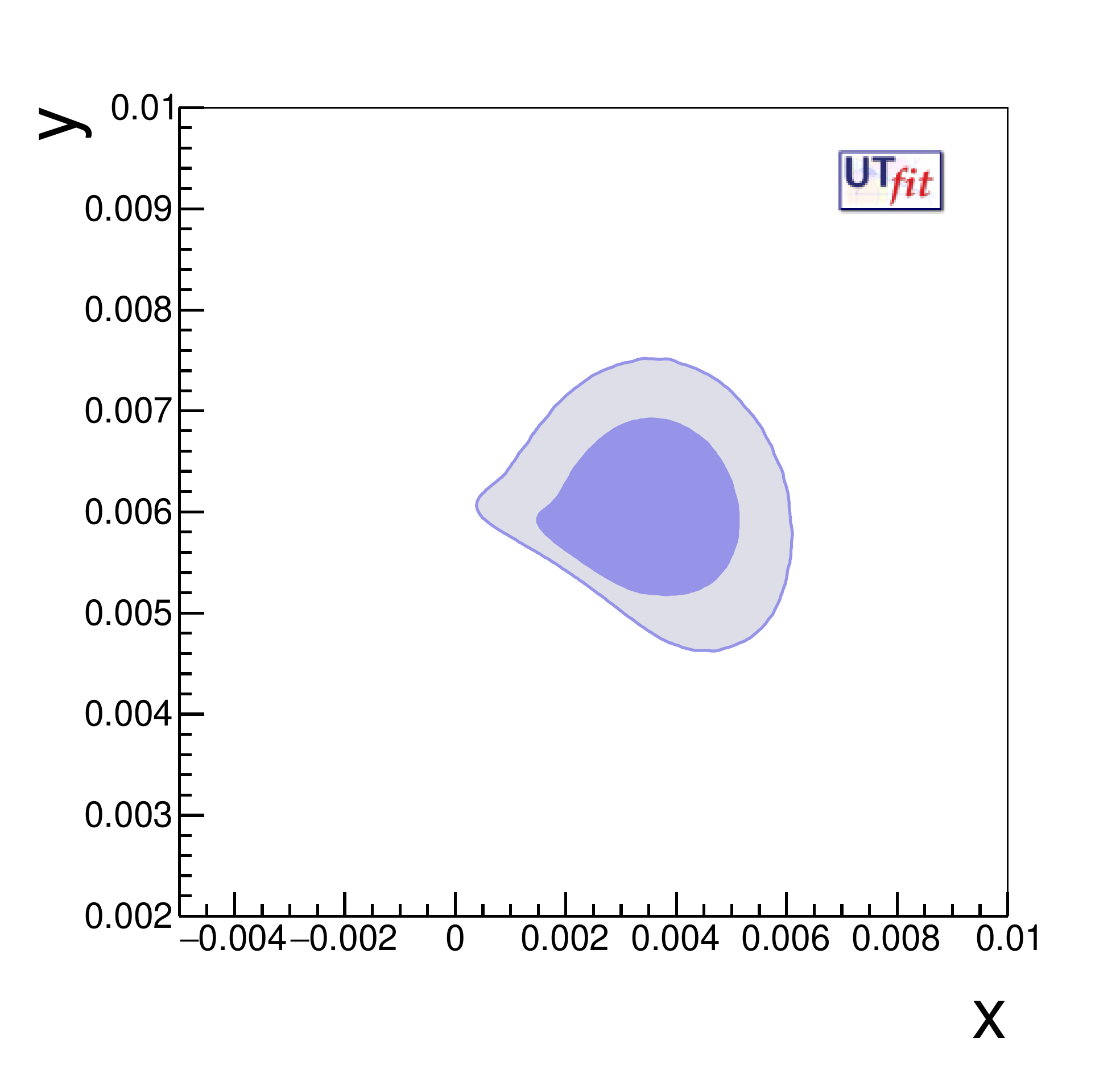}
  \includegraphics[width=0.3\textwidth]{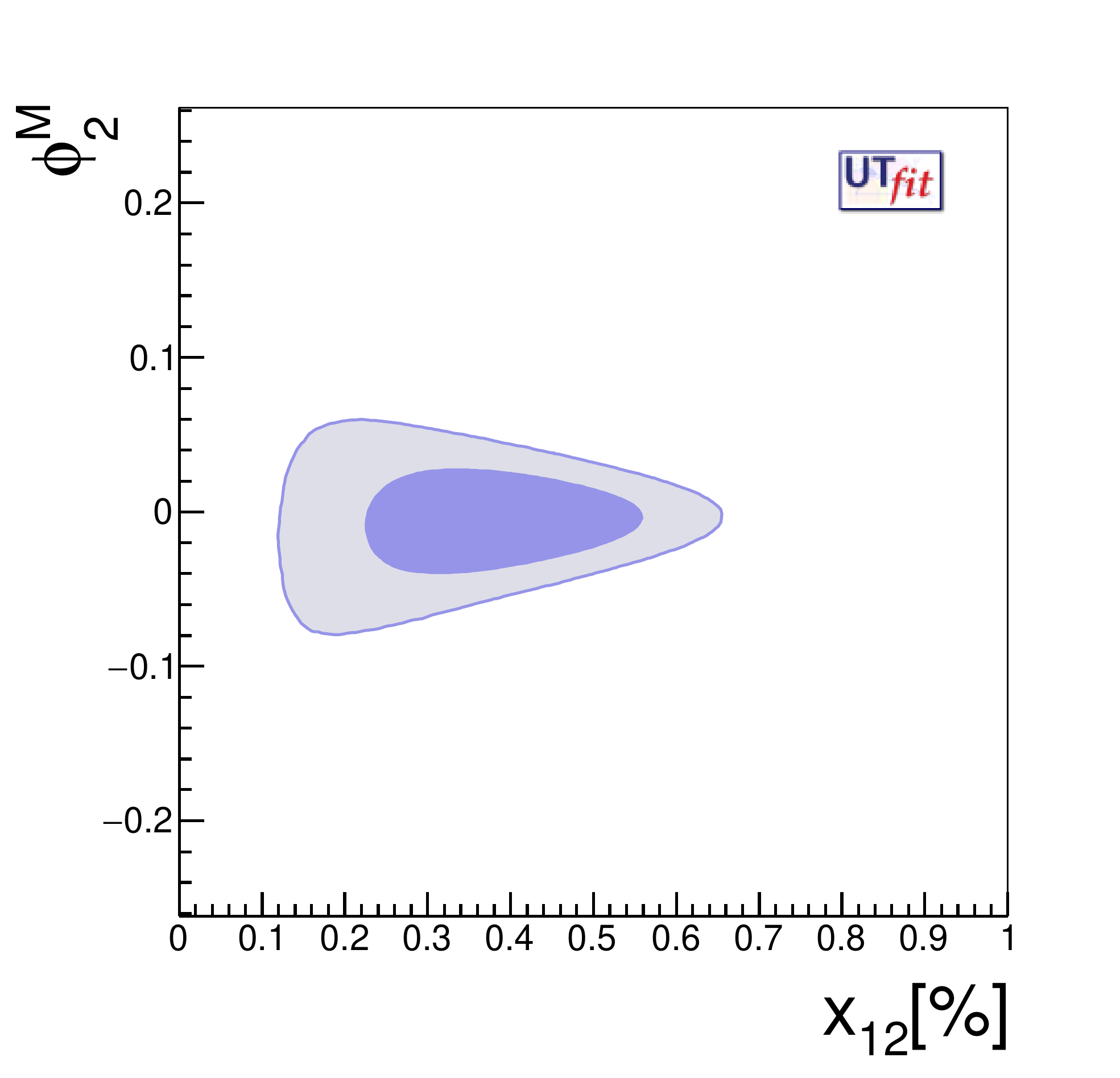}
  \includegraphics[width=0.3\textwidth]{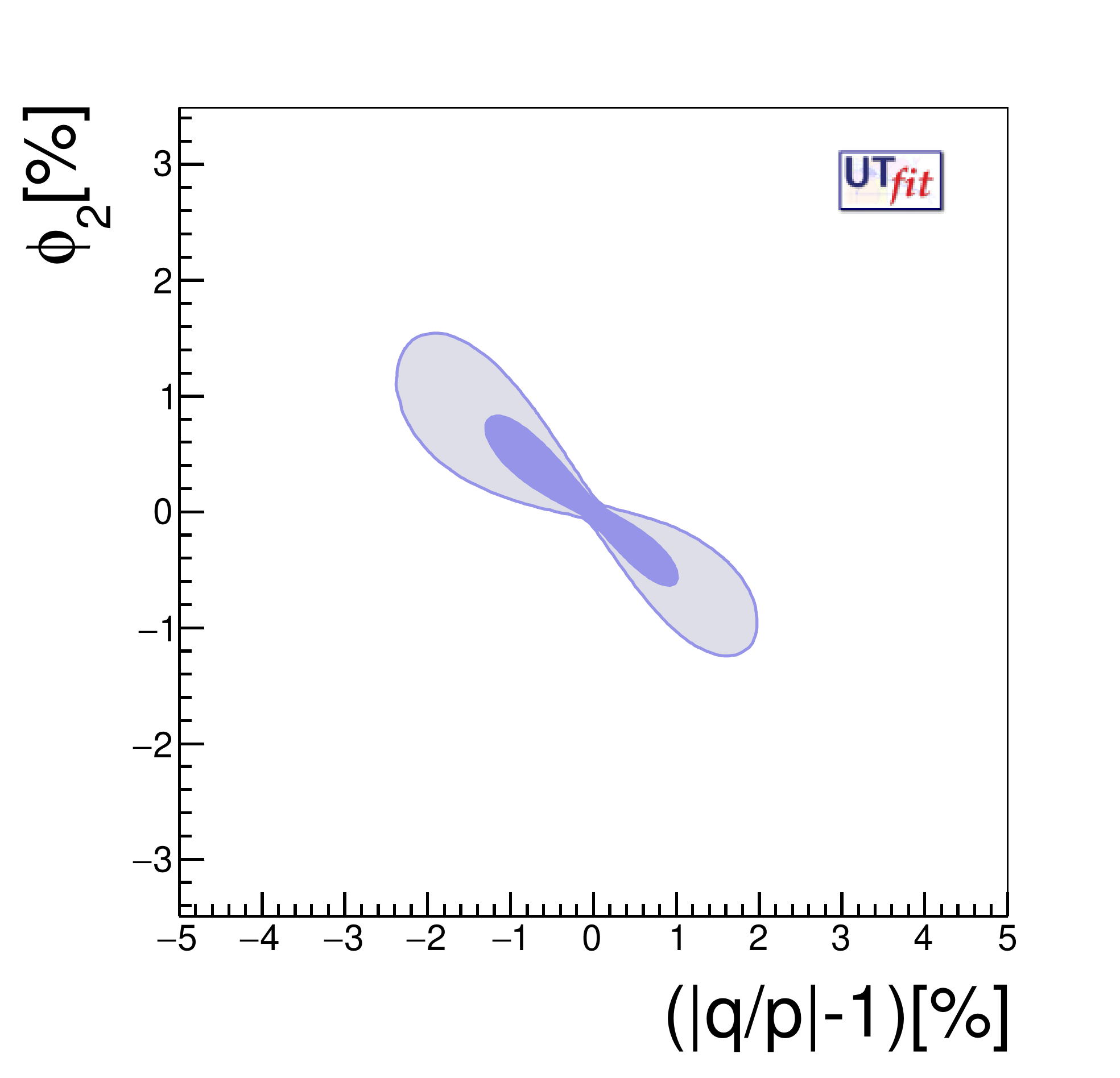} \\
  \includegraphics[width=0.24\textwidth]{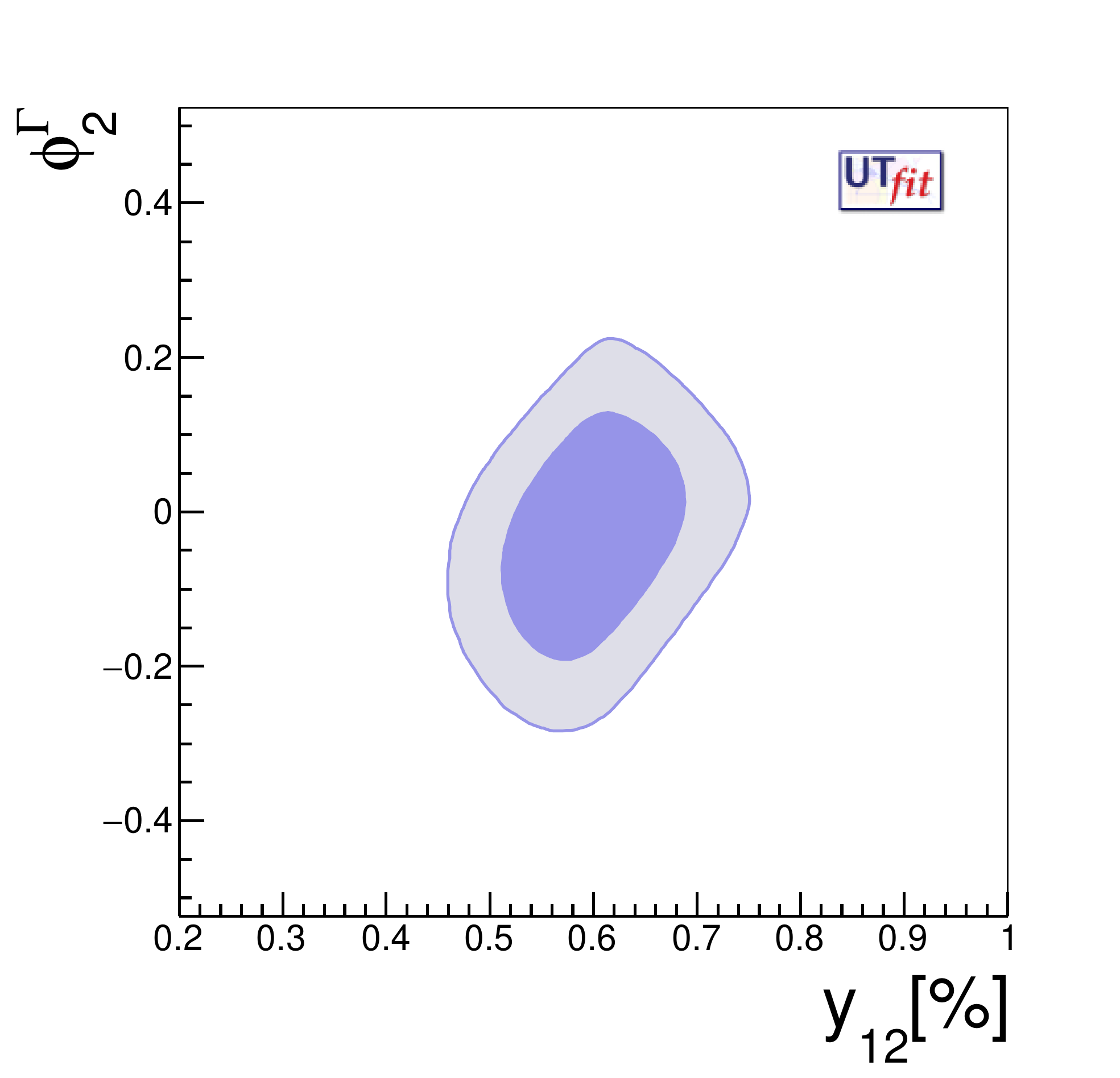}
  \includegraphics[width=0.24\textwidth]{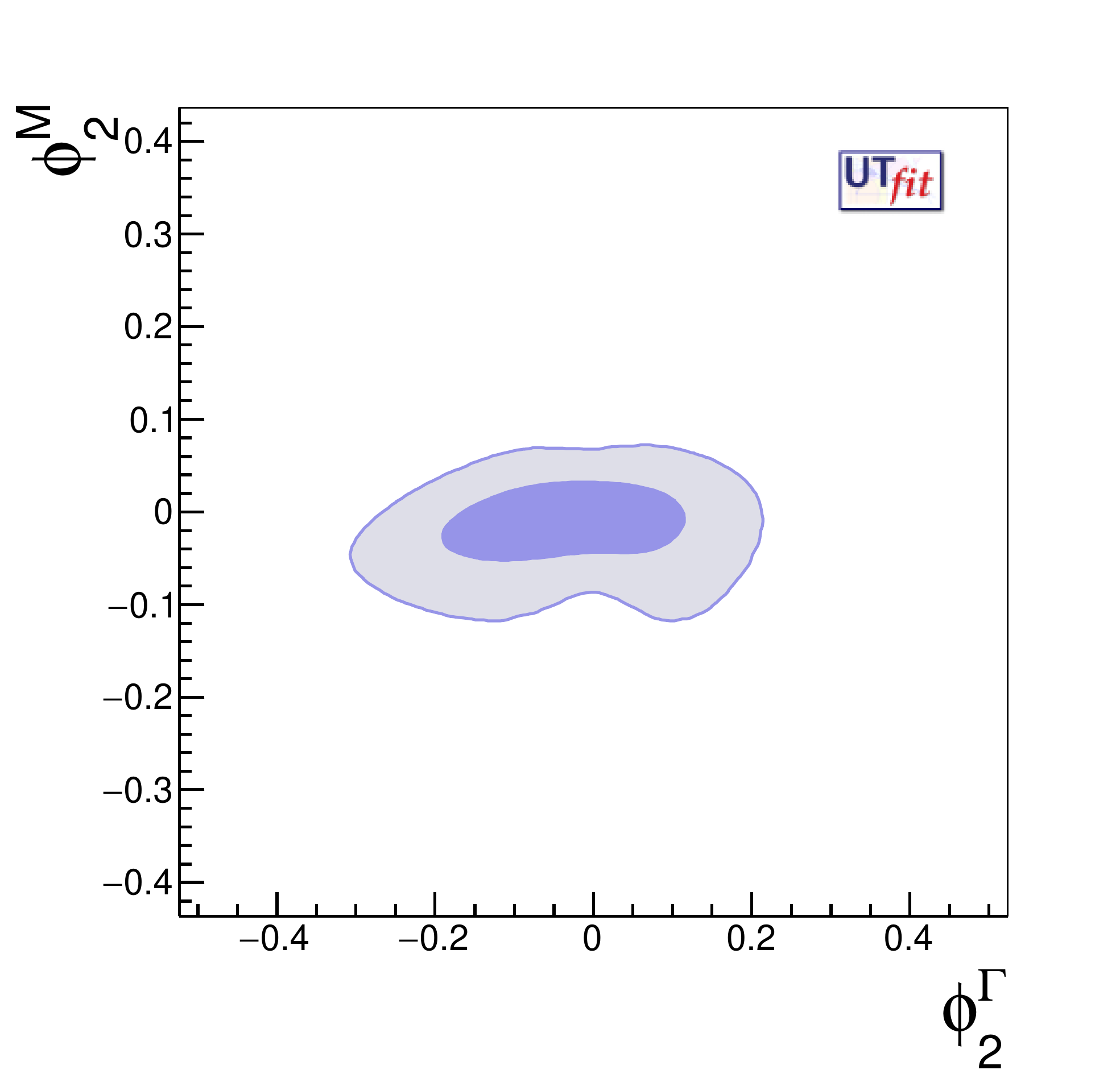}
  \includegraphics[width=0.24\textwidth]{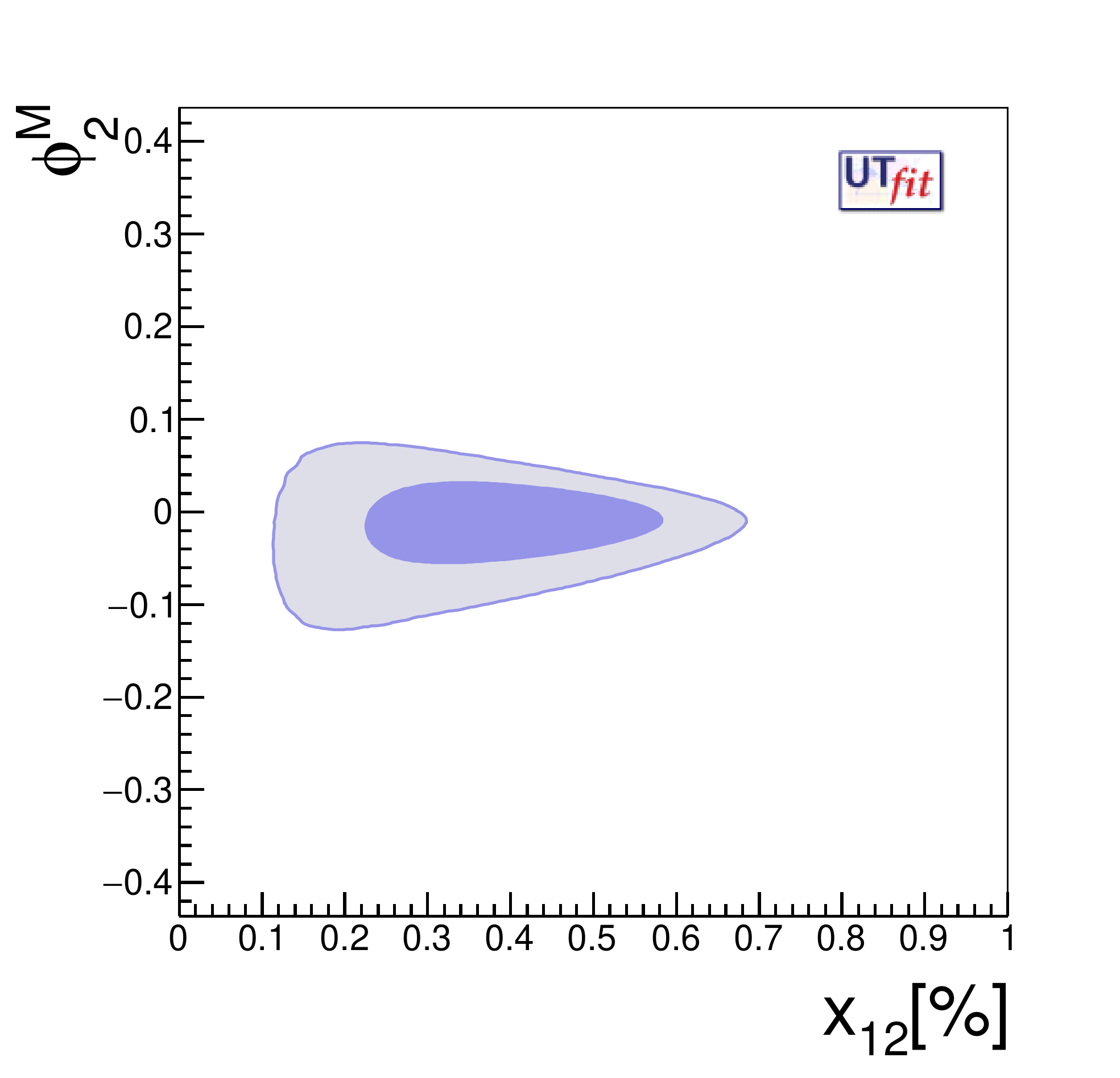}
  \includegraphics[width=0.24\textwidth]{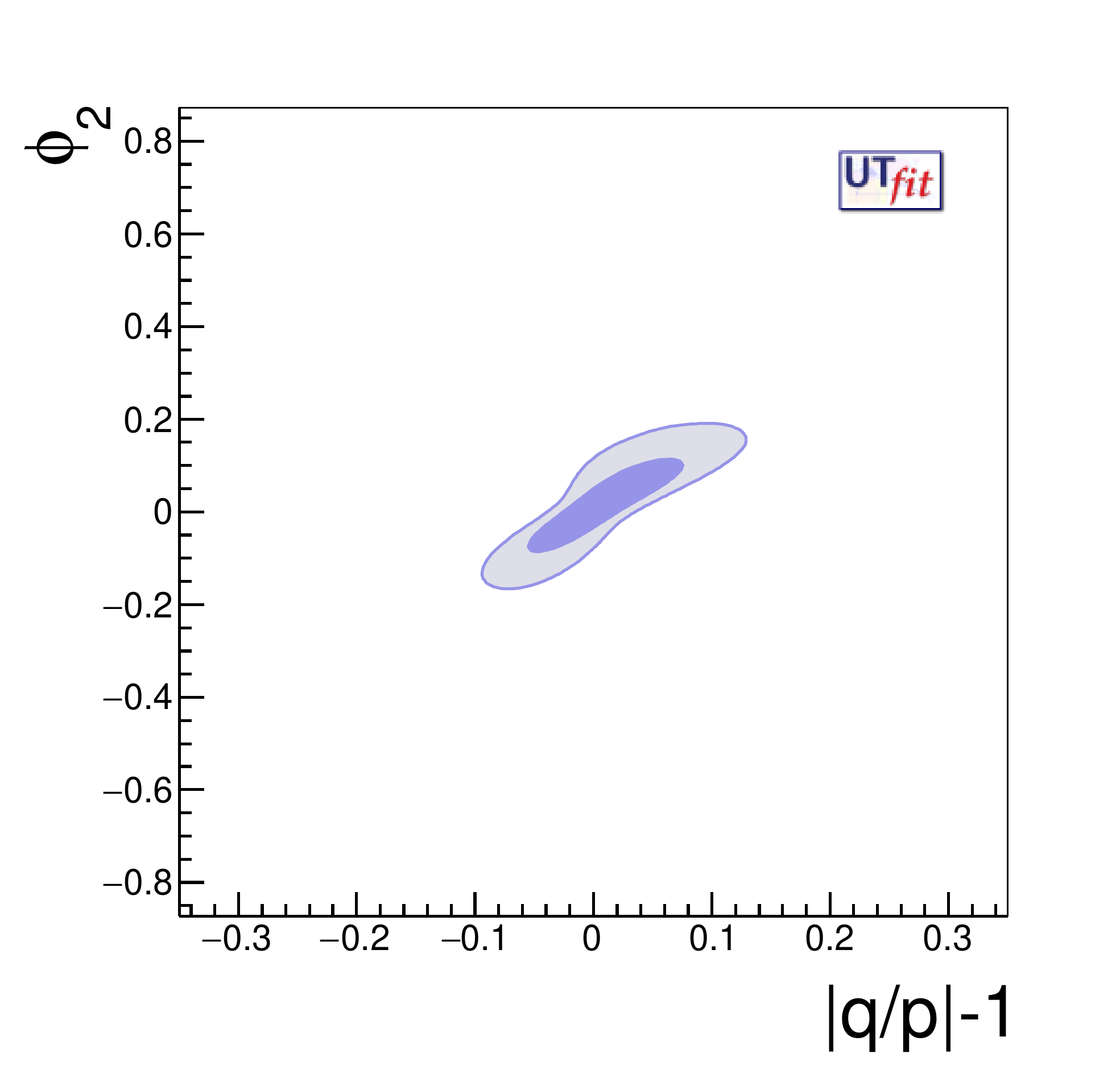}
  \caption{P.d.f.'s for mixing parameters in the superweak (first row) and
    approximate universality scenarios, see text. Darker (lighter)
    regions correspond to $68\%$ ($95\%$) probability. Notice the
    order-of-magnitude difference in the scale of the rightmost plots.}
  \label{fig:current}
\end{figure*}

Superweak
fits to the data are highly constrained, given that
there is only one CPV parameter controlling all of indirect CPV.
The second column in Table \ref{tab:res_current} contains the results of our fit to the
mixing parameters with current data in the superweak framework.
We see that sensitivity to $\phi^M_{2}$ is $\approx 22$ mrad at
$1\sigma$, and $\approx 54$ mrad  at 95\% probability,
while sensitivity to $\phi_2$ is $\approx 5$ mrad at $1\sigma$, and
$\approx 11$ mrad at 95\% probability.\footnote{Smaller errors for $\phi_2$ than
$\phi^M_{2}$ in the superweak fit can be traced to the small central
value of the prefactor in  \eqref{2phivsphi12},  $x_{12}^2 /(x_{12}^2
+ y_{12}^2) \approx  0.26$.}
Some superweak correlation plots are also shown in
the first row of Fig.~\ref{fig:current}.
The Heavy Flavor Averaging Group (HFLAV)~\cite{Amhis:2016xyh}
has obtained similar results,
\beq \begin{split}
\phi_2^M &=-0.004 \pm 0.016~(1\sigma),~~~\phi_2 = 0.001 \pm 0.005~(1\sigma)\,.\cr
\end{split}\label{superweakfit}\eeq
Comparison with the SM ranges \eqref{phiGphiMnaive}
implies that an order of magnitude window for NP remains, at 95\% probability, in the CPVINT phases.

\begin{table*}[htbp!]
  \centering
  \begin{tabular}{|cccccccc|}
    \hline
    Observable & Value & \multicolumn{5}{c}{Correlation Coeff.} &
    Reference\\ \hline
    $y_{CP}$ & $(0.72 \pm 0.11)\%$ & & & & & &
    \cite{Link:2000cu,Csorna:2001ww,Zupanc:2009sy,Lees:2012qh,Ablikim:2015hih,Staric:2015sta,Aaij:2018qiw,Nayak:2019byo}
    \\ \hline
    $A_\Gamma $ & $(-0.031 \pm 0.020)\%$ & & & & & &
    \cite{Aitala:1999dt,Lees:2012qh,Aaltonen:2014efa,Staric:2015sta,Aaij:2017idz,Aaij:2019yas}
    \\ \hline
    $x$ & $(0.53 \pm 0.19 \pm 0.06 \pm 0.07)\%$ & 1 & 0.054 & -0.074 &
    -0.031 & & \cite{Peng:2014oda}\\
    $y$ & $(0.28 \pm 0.15 \pm 0.05 \pm 0.05)\%$ & & 1 & 0.034 &
    -0.019 & & \cite{Peng:2014oda}\\
    $\vert q/p \vert$ & $(0.91 \pm 0.16 \pm 0.05 \pm 0.06)$ &  & & 1 &
                                                                       0.044
                                       & & \cite{Peng:2014oda}\\
    $\phi$ & $(-6 \pm 11 \pm 3 \pm 4)^\circ$ &
    & &
     & 1 & & \cite{Peng:2014oda}\\\hline
    $x_{\mathrm{CP}}$ & $(0.27 \pm 0.16 \pm 0.04)\%$ & 1 & -0.17 & 0.04 &
    -0.02 & & \cite{Amhis:2016xyh}\\
    $y_{\mathrm{CP}}$ & $(0.74 \pm 0.36 \pm 0.11)\%$ & & 1 & -0.03 &
    0.01 & & \cite{Amhis:2016xyh}\\
    $\Delta x$ & $(-0.053 \pm 0.07 \pm 0.022)\%$ &  & & 1 &
    -0.13 & & \cite{Amhis:2016xyh}\\
    $\Delta y$ & $(0.06 \pm 0.16 \pm 0.03)\%$ &
    & &
     & 1 & & \cite{Amhis:2016xyh}\\\hline
    $x$ & $(0.16 \pm 0.23 \pm 0.12 \pm 0.08)\%$ & 1 & 0.0615 & & & &
    \cite{delAmoSanchez:2010xz}\\
    $y$ & $(0.57 \pm 0.20 \pm 0.13 \pm 0.07)\%$ & 0.0615 & 1 & & & &
    \cite{delAmoSanchez:2010xz}\\
    \hline
    $R_M$ & $(0.0130 \pm 0.0269)\%$ & & & & & &
    \cite{Aitala:1996vz,Cawlfield:2005ze,Aubert:2004bn,Aubert:2007aa,Bitenc:2008bk}
    \\ \hline
    $(x^{2}+y^{2})/4$ & $(0.0048 \pm 0.0018)\%$ & & & & & &
    \cite{Aaij:2016rhq}
    \\\hline
    $(x^\prime_+)_{K\pi\pi}$ & $(2.48 \pm 0.59 \pm 0.39)\%$ & 1 & -0.69 &  &   &
    & \cite{Aubert:2008zh}\\
    $(y^\prime_+)_{K\pi\pi}$ & $(-0.07 \pm 0.65 \pm 0.50)\%$ & -0.69 &
    1 &  &   &
    & \cite{Aubert:2008zh}\\
    $(x^\prime_-)_{K\pi\pi}$ & $(3.50 \pm 0.78 \pm 0.65)\%$ & 1 & -0.66 &  &   &
    & \cite{Aubert:2008zh}\\
    $(y^\prime_-)_{K\pi\pi}$ & $(-0.82 \pm 0.68 \pm 0.41)\%$ & -0.66 &
    1 &  &   &
    & \cite{Aubert:2008zh}\\ \hline
    $R_D$ & $(0.533 \pm 0.107 \pm 0.045)\%$ & 1 & 0 & 0 & -0.42 & 0.01
    & \cite{Asner:2012xb}\\
    $x^2$ & $(0.06 \pm 0.23 \pm 0.11)\%$ & 0 & 1 & -0.73 & 0.39
    & 0.02
    & \cite{Asner:2012xb}\\
    $y$ & $(4.2 \pm 2 \pm 1)\%$ & 0. & -0.73 & 1 & -0.53  & -0.03
    & \cite{Asner:2012xb}\\
    $\cos\delta_{K\pi}$ & $(0.84 \pm 0.2 \pm 0.06)$ & -0.42 &
    0.39 & -0.53 & 1  & 0.04
    & \cite{Asner:2012xb}\\
    $\sin\delta_{K\pi}$ & $(-0.01 \pm 0.41 \pm 0.04)$ & 0.01
    & 0.02 & -0.03 & 0.04  & 1
    & \cite{Asner:2012xb}\\
   \hline
    $R_D$ & $(0.3030 \pm 0.0189)\%$ & 1 & 0.77 & -0.87 &   &
    & \cite{Aubert:2007wf}\\
    $(x^\prime_+)^2_{K\pi}$ & $(-0.024 \pm 0.052)\%$ & 0.77 & 1 & -0.94 &   &
    & \cite{Aubert:2007wf}\\
    $(y^\prime_+)_{K\pi}$ & $(0.98 \pm 0.78)\%$ & -0.87 & -0.94 & 1  & &
    & \cite{Aubert:2007wf}\\\hline
    $A_D$ & $(-2.1 \pm 5.4)\%$ & 1 & 0.77 & -0.87 &   &
    & \cite{Aubert:2007wf}\\
    $(x^\prime_-)^2_{K\pi}$ & $(-0.020 \pm 0.050)\%$ & 0.77 & 1 & -0.94 &   &
    & \cite{Aubert:2007wf}\\
    $(y^\prime_-)_{K\pi}$ & $(0.96 \pm 0.75)\%$ & -0.87 & -0.94 & 1  & &
    & \cite{Aubert:2007wf}\\\hline
    $R_D$ & $(0.364 \pm 0.018)\%$ & 1 & 0.655 & -0.834 &   &
    & \cite{Zhang:2006dp}\\
    $(x^\prime_+)^2_{K\pi}$ & $(0.032 \pm 0.037)\%$ & 0.655 & 1 & -0.909 &   &
    & \cite{Zhang:2006dp}\\
    $(y^\prime_+)_{K\pi}$ & $(-0.12 \pm 0.58)\%$ & -0.834 & -0.909 & 1  & &
    & \cite{Zhang:2006dp}\\\hline
    $A_D$ & $(2.3 \pm 4.7)\%$ & 1 & 0.655 & -0.834 &   &
    & \cite{Zhang:2006dp}\\
    $(x^\prime_-)^2_{K\pi}$ & $(0.006 \pm 0.034)\%$ & 0.655 & 1 & -0.909 &   &
    & \cite{Zhang:2006dp}\\
    $(y^\prime_-)_{K\pi}$ & $(0.20 \pm 0.54)\%$ & -0.834 & -0.909 & 1  & &
    & \cite{Zhang:2006dp}\\\hline
    $R_D$ & $(0.351 \pm 0.035)\%$ & 1 & -0.967 & 0.900 &   &
    & \cite{Aaltonen:2013pja}\\
    $(y^\prime_\mathrm{CPA})_{K\pi}$ & $(0.43 \pm 0.43)\%$ & -0.967 & 1 & -0.975 &   &
    & \cite{Aaltonen:2013pja}\\
    $(x^\prime_\mathrm{CPA})^2_{K\pi}$ & $(0.008 \pm 0.018)\%$ & 0.900 & -0.975 & 1  & &
    & \cite{Aaltonen:2013pja}\\\hline
    $R_D$ & $(0.3454 \pm 0.0028 \pm 0.0014)\%$ & 1 & -0.883 & 0.745 & -0.883  &
    0.749 & \cite{Aaij:2017urz}\\
    $(y^\prime_+)_{K\pi}$ & $(0.501 \pm 0.048 \pm 0.029)\%$ &  & 1
    & -0.944  & 0.758 & -0.644
    & \cite{Aaij:2017urz}\\
    $(x^\prime_+)^2_{K\pi}$ & $(6.1 \pm 2.6 \pm 1.6)10^{-5}$ &  &
     & 1 & -0.642  & 0.545
    & \cite{Aaij:2017urz}\\
    $(y^\prime_-)_{K\pi}$ & $(0.554 \pm 0.048 \pm 0.029)\%$ &  &
     & & 1  & -0.946
    & \cite{Aaij:2017urz}\\
    $(x^\prime_-)^2_{K\pi}$ & $(1.6 \pm 2.6 \pm 1.6)10^{-5}$ &  &
     &   &  & 1
    & \cite{Aaij:2017urz}\\\hline
  \end{tabular}
  \caption{Experimental data used in the analysis, mostly from
    ref.~\cite{Amhis:2016xyh}. Asymmetric
    errors have been symmetrized. }
  \label{tab:currentexp}
\end{table*}

\begin{table*}[htb!]
  \centering
  \begin{tabular}{|c|cc|cc|c|}
    \hline
    parameter & \multicolumn{2}{c|}{superweak -- current} &
                \multicolumn{2}{c|}{approx. univ. --  current} &
                approx. univ. --  future\\
    \hline
    & $68\%$ prob. & $95\%$ prob.  & $68\%$ prob. & $95\%$ prob.
    & estimated $68\%$ prob.  \\
    \hline
    $10^{3} x_{12}$ & $3.6 \pm 1.1$ & $[1.3,5.7]$ & $3.7 \pm 1.2 $ & $[1.3, 5.9]$
         & $\pm 0.017$ \\
    $10^{4} y_{12}$ & $60.3 \pm 5.7 $ & $[49, 73]$ & $59.6 \pm 5.6$ & $[49, 71]$
         & $\pm 0.19$ \\
    $10^{2} {\phi}_2^{M}$ [rad] & $-0.5 \pm 2.2$ & $[-6.1,4.7]$ & $-1.0 \pm 2.9$ & $[-10.0,5.7]$
         & $\pm 0.12$ \\
    $10^{2}\phi_2^{\Gamma}$ [rad] & $0$ & $0$ & $-3.2 \pm 9.9$ & $[-23,16]$
         & $\pm 0.17$ \\
    $10^{2}\phi_{12}$ [rad] & $-0.5 \pm 2.2$ & $[-6.1,4.7]$ & $2.6 \pm 9.7$ & $[-20,22]$
         & $\pm 0.21$ \\
    \hline
    $10^{3}x$ & $3.6\pm 1.1$ & $[1.3, 5.8] $ & $3.7\pm 1.2$ & $[1.3, 6.0]$
         &  $\pm 0.017$\\
    $10^{4}y$ & $60.3\pm 5.7$ & $[49, 73]$ & $59.5\pm 5.6$ & $[48, 71]$
         & $\pm 0.19$ \\
    $10^{3}\left(\vert q/p\vert-1\right)$ & $-2.3\pm 9.0$ & $[-21,16]$
                       & $8\pm 41$ & $[-73,99]$
         & $\pm 0.92$ \\
    $10^{2} \phi_2$ [rad] & $0.12\pm 0.51$ & $[-0.96,1.26]$ & $2.5\pm 7.2$ & $[-13,17]$
         & $\pm 0.13$  \\
  \hline
  \end{tabular}
  \caption{Results of fits to the current and future $D$ mixing data
    within the superweak
    %(second column)
    and approximate universality
   % (third column)
    frameworks, where the phases are defined in Eq. \eqref{theorphasesdef}.}
    %$\Phi^{M} = \phi_{12}$, $\Phi= \phi$ (superweak),
    %and $\Phi^{M,\Gamma}=\phi^{M,\Gamma}_2$, $\Phi=\phi_2$ (approximate universality)
  \label{tab:res_current}
\end{table*}

\subsection{Approximate universality fits}

\label{sec:appunivfits}

It is encouraging that the $1\sigma$ error on $\phi_2$ in the superweak fit ($5$ mrad), and the $U$-spin based SM estimates for
$\phi_2^{M,\Gamma}$,
$\phi_{12}$
in \eqref{phiGphiMnaive}, \eqref{refined} are only about a factor of two apart.  However, this means that
the approximate universality parametrization is advisable moving forward. Inspection
of the relations between $\phi_2$ and $\phi_2^{M,\Gamma}$ in \eqref{tan2phitheory}, \eqref{tan2phitheory2}, reinforces this conclusion.
Approximate universality fits are less constrained, given that they employ two CPV parameters rather than a single one to describe indirect CPV.
Hopefully, this will be overcome in the high statistics LHCb and Belle-II precision era, and SM sensitivity in $\phi_2^{M,\Gamma}$ will be achieved.
This possibility is assessed below.

We remark that an approximate universality fit for any two of the phases $\phi_2^M$, $\phi_2^\Gamma$, and $\phi_{12}$ is equivalent to a (traditional) two-parameter fit for $\phi_2$ and $|q/p|$, with translations provided by  \eqref{qovpphi12}, \eqref{phi12phiMphiGamma}--\eqref{tan2phitheory2}.  General formulae for the decay widths, given in terms of $\phi_{\lambda_f}$ and $|q/p|$, can be converted to approximate universality formulae which depend on $\phi_2$ and $|q/p|$,
via the substitutions
$\phi_{\lambda_f} \to \phi_2$ (SCS), $\phi_{\lambda_f} \to \phi_2$ (CF/DCS $K^\pm X$),  $\phi_{\lambda_f} \to \phi_2 -2 \,\epsilon_I - \left|\lambda_b /\lambda_s \right|\sin\gamma$ (CF/DCS $K^0 X$, general), and $\phi_{\lambda_f} \to \phi_2 - \left|\lambda_b /\lambda_s \right|\sin\gamma$ (CF/DCS $K^0 X$, LHCb). These are analogous to the substitutions for $\phi_f^{M,\Gamma}$ in \eqref{SCSsubs}, \eqref{DCSsubs}, \eqref{phi2phiepssum} , and \eqref{LHCbsub}, respectively.

We begin with a fit to the current data, cf.~Table \ref{tab:currentexp}, for the phases $\phi_2^M$ and $\phi_2^\Gamma$.
We implement the substitutions for $\phi_f^{M,\Gamma} $ given in \eqref{SCSsubs}, \eqref{DCSsubs}, \eqref{LHCbsub}, and employ the expression for $\Delta Y_f$ in \eqref{dYfappuniv}.  The $K_L-K_S$ interference terms in the $D\to K_{S,L}  \,\pi^+\pi^-$ decay widths \eqref{timedepratesK0f}, \eqref{timedepratesK0fbar} are ignored, as in the experimental analyses.  As explained in Section~\ref{sec:CFDCSK0implement}, this does not affect the determination of $\phi_2^{M,\Gamma}$ at LHCb, provided that the substitution in \eqref{LHCbsub} is employed. %, cf. Sec. \ref{sec:CFDCSK0implement}.
For the Belle $D^0 \to K_{S,L} \, \pi^+ \pi^-$ analysis \cite{Peng:2014oda}, omission of $K_L -K_S$ interference is not an issue, given its experimental precision.

  % {\bf ***}add more details on the fit - refer to eqs for substitutions, etc.{\bf ***}
The results of the approximate universality fit appear in the third column of Table \ref{tab:res_current}, and in the second row of correlation plots in
Fig.~\ref{fig:current}.
It is interesting to notice that the
error on $\phi^{M}_{2}$ is about a factor of three smaller than the error on $\phi_2^\Gamma$, and is similar to the corresponding superweak error.
This can be traced, in part, to the observable $A_\Gamma= -\Delta Y_f$, for $f=\pi^+\pi^-$, $K^+ K^-$.  It has a relatively small experimental error, and it only depends on the product $x_{12}\,\sin \phi_2^M$ in the fit [compare \eqref{dYfappuniv}, \eqref{dYfsuperweak}].
However, both
$\phi_2$ and $|{q/p}|-1$ are determined with order of magnitude
larger uncertainties in the approximate universality framework, due to
their dependence on both $\phi_2^M$ and $\phi^{\Gamma}_{2}$. %, the second CPV parameter in the fit.

\begin{table}[htb!]
  \centering
  \begin{tabular}{|c|c|c|c|c|}
    \hline
    $\delta(x_{\mathrm{CP}})$ & $\delta(y_{\mathrm{CP}})$ & $\delta(\Delta x)$ & $\delta(\Delta y)$ &
            \cite{Aaij:2019jot} scaled \\
    $3.8 \cdot 10^{-5}$ & $8.6 \cdot 10^{-5}$ & $1.7 \cdot 10^{-5}$ & $3.8 \cdot 10^{-5}$ &
            by luminosity \\
    \hline
    $\delta(y^\prime_+)_{K\pi}$ & $\delta(y^\prime_-)_{K\pi}$ & $\delta(x^\prime_+)^{2}_{K\pi}$ & $\delta(x^\prime_-)^{2}_{K\pi}$&
            \cite{Aaij:2017urz} scaled \\
    $3.2 \cdot 10^{-5}$ & $3.2 \cdot 10^{-5}$ & $1.7 \cdot 10^{-6}$ & $1.7 \cdot 10^{-6}$ &
            by luminosity \\
    \hline
   $\delta(x_{K\pi\pi\pi})$ & $\delta(y_{K\pi\pi\pi})$ & $\delta(\vert q/p\vert_{K\pi\pi\pi})$ & $\delta(\phi_{K\pi\pi\pi})$ &
            \cite{Cerri:2018ypt} \\
    $2 \cdot 10^{-5}$ & $2 \cdot 10^{-5}$ & $2 \cdot 10^{-3}$ & $0.1^{\circ}$ &
             \\  \hline
  \end{tabular}
  \caption{Estimated uncertainties on mixing parameters from CF/DCS
    decays in the LHCb Phase II Upgrade. Correlations from current
    results have been used where available.}
  \label{tab:future_exp}
\end{table}

\begin{figure*}[htb]
  \centering
  \includegraphics[width=0.24\textwidth]{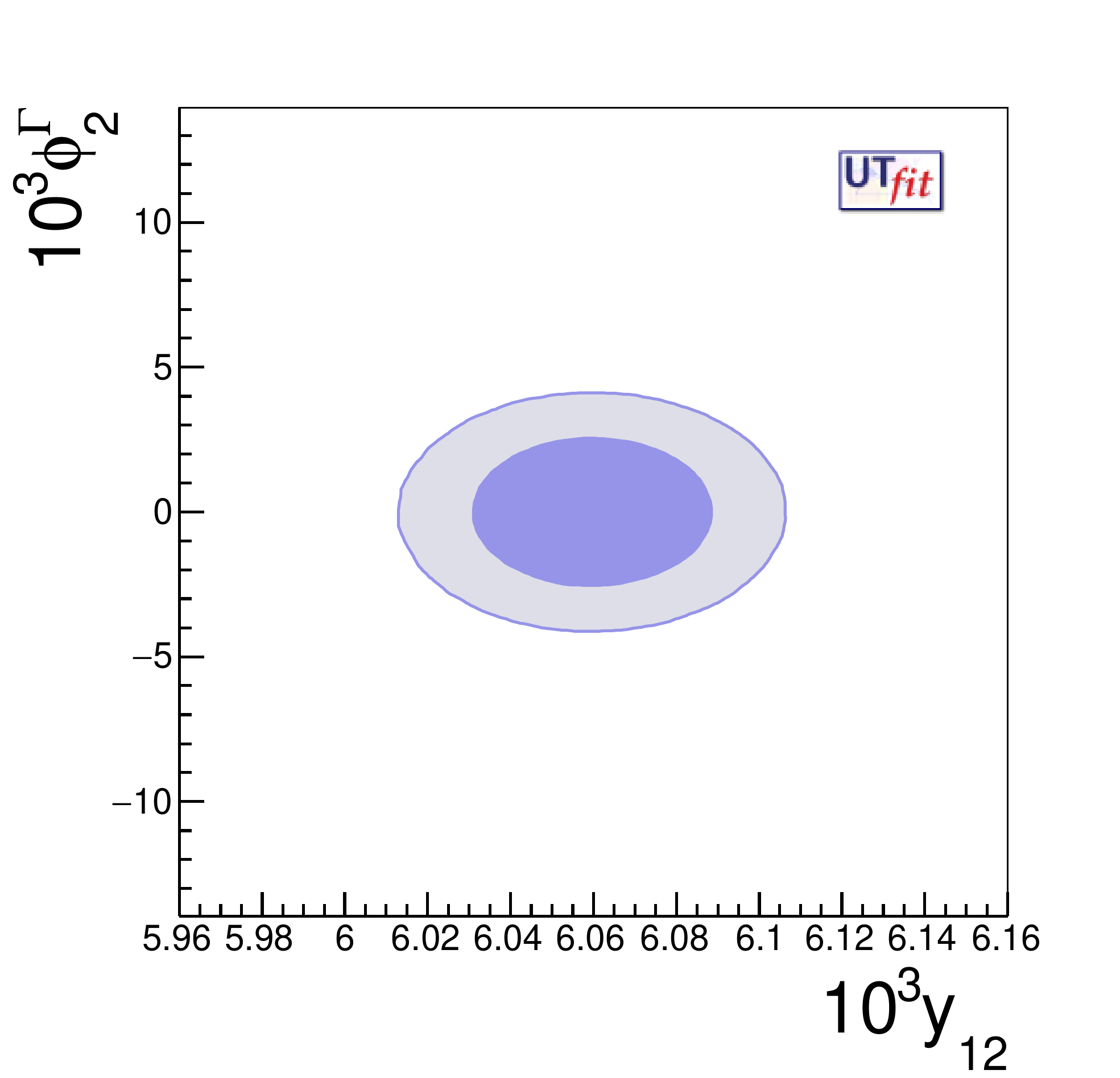}
  \includegraphics[width=0.24\textwidth]{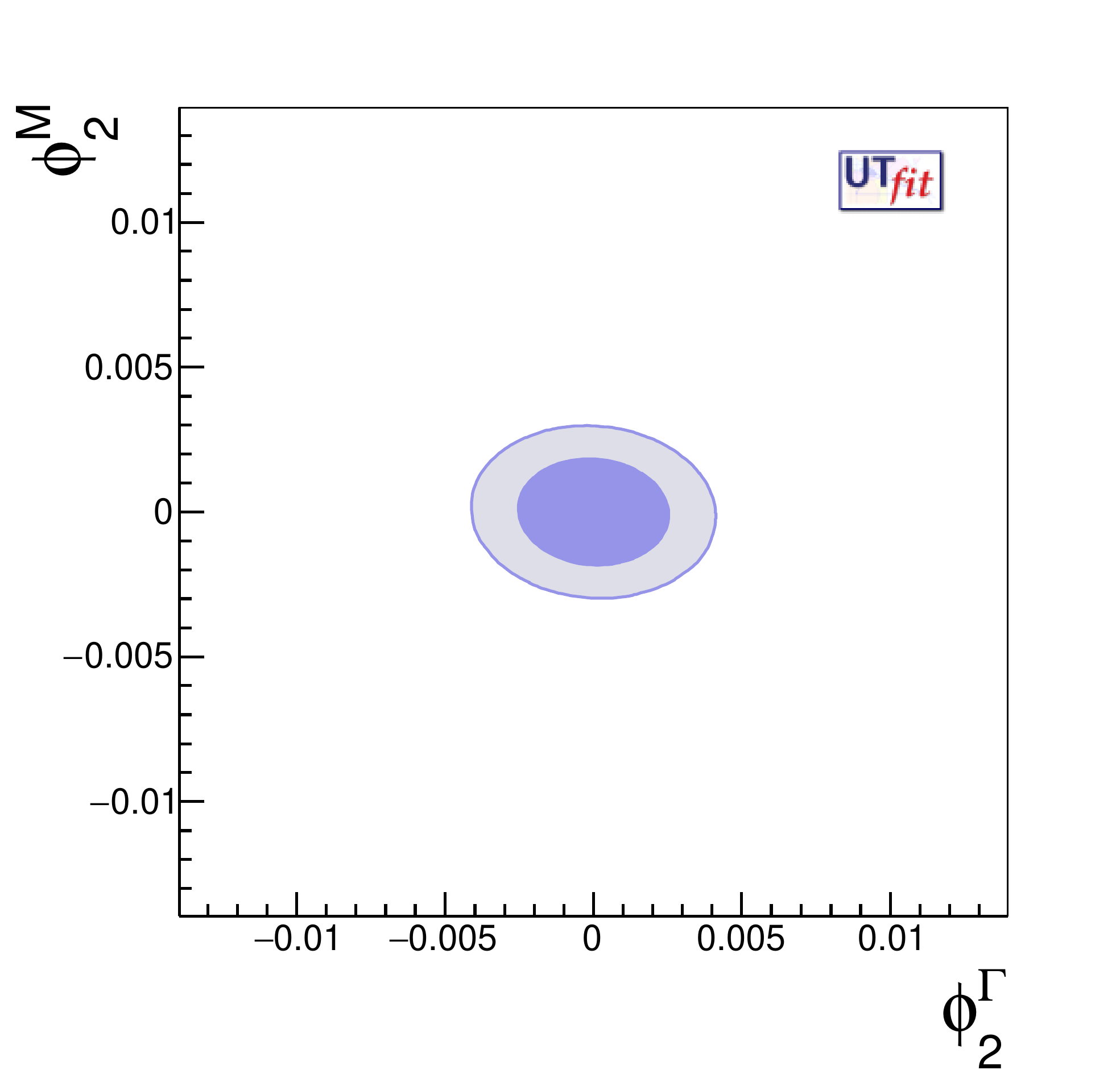}
  \includegraphics[width=0.24\textwidth]{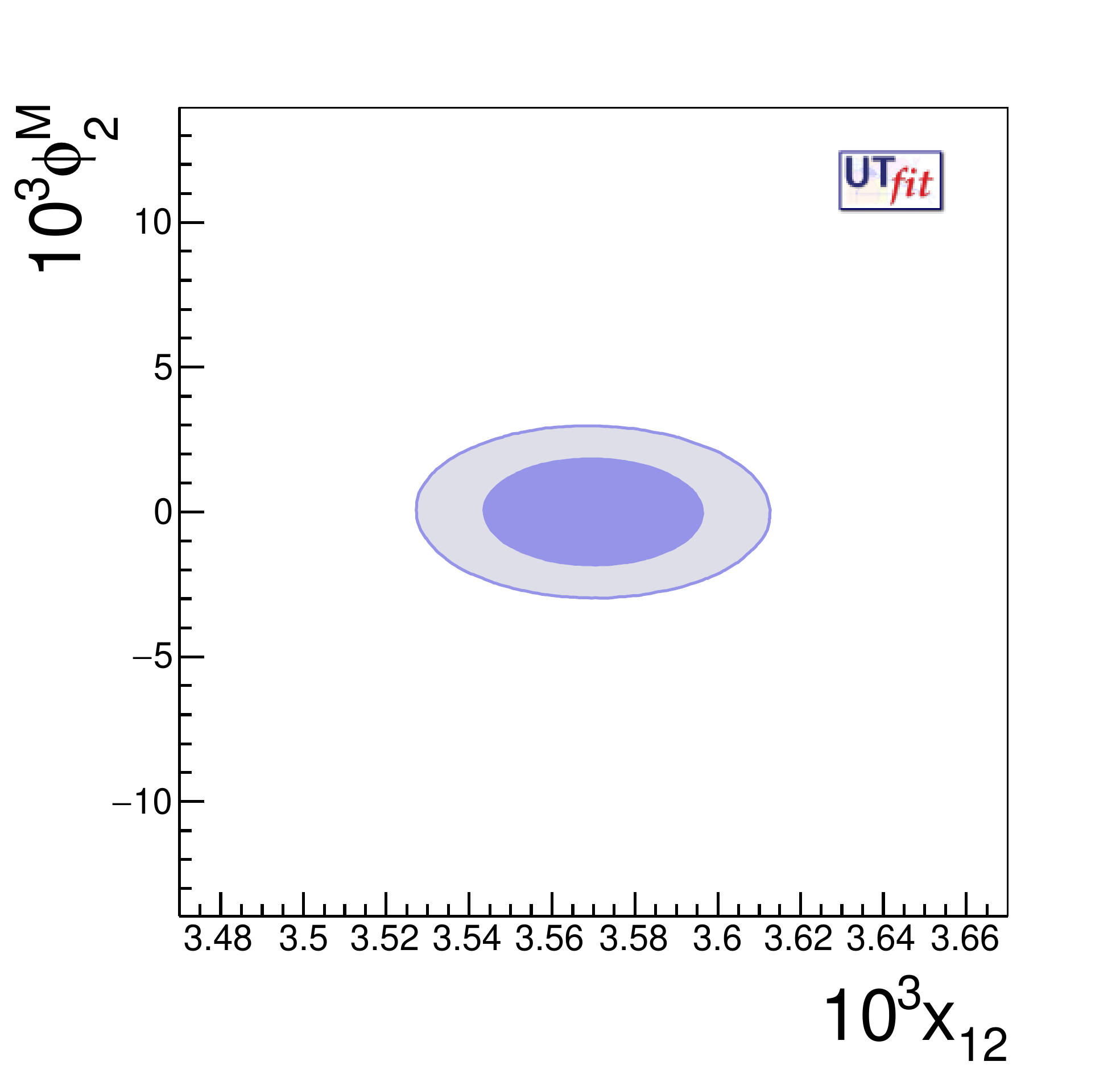}
  \includegraphics[width=0.24\textwidth]{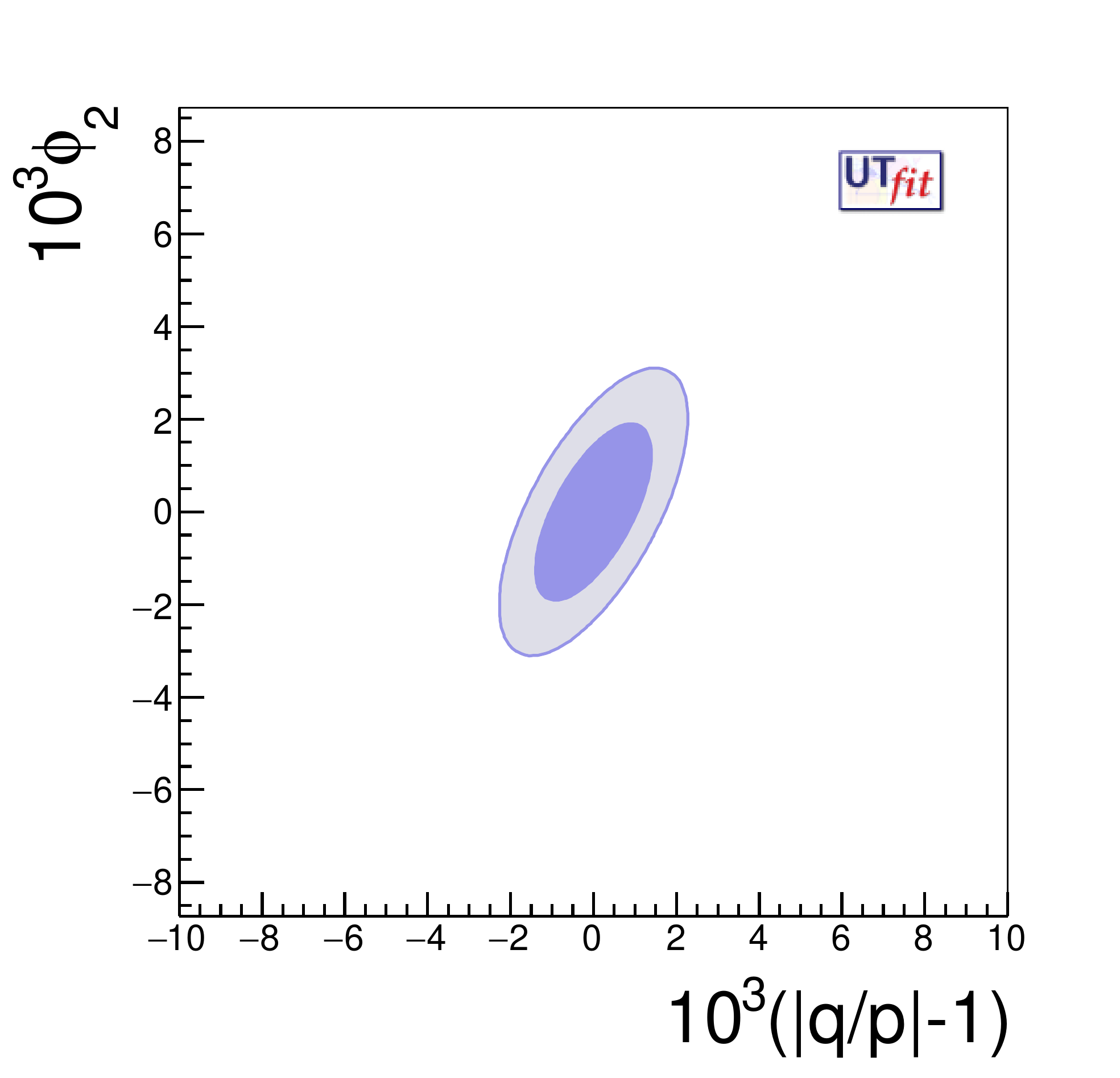}
  \caption{P.d.f.'s for mixing parameters in the
    approximate universality future scenario, see text. Darker (lighter)
    regions correspond to $68\%$ ($95\%$) probability.}
  \label{fig:future}
\end{figure*}

In the future, as SM sensitivity in CPVINT is approached, a modified strategy will be appropriate.
As discussed in Section~\ref{sec:SCSfinstatedep}, significant and non-universal misalignment ratios $\delta \phi_f /\phi_2^{M,\Gamma}$ could manifest themselves in the SCS measurements, even though they are formally $O(\epsilon)$ in $U$-spin breaking.  In contrast, the misalignments in CF/DCS decays are either negligible ($K^\pm X$), or known to very good approximation ($K^0 X ,\, \K0bar X$), cf. Secs. \ref{sec:finstatedepKpm}, \ref{finstatedepK0}.
Thus, at that this point one could simply drop the SCS observables from the global fits to $\phi_2^{M}$, $\phi_2^\Gamma$.
Alternatively, one could only include the SCS final states
$\pi^+ \pi^-$ and $K^+ K^-$ in the global fits, via their averaged time dependent CP asymmetry $A_\Gamma$, thus taking advantage of the $O(\epsilon^2)$ suppression of the averaged QCD penguin pollution, cf. \eqref{AGammaappuniv}.

It is interesting to point out that simultaneous knowledge of $\phi_2^{M,\Gamma}$ from CF/DCS decays, and of the direct CP asymmetries in the SCS decays could be used to determine the relative magnitudes and strong phases of the corresponding subleading SCS decay amplitudes in the SM, i.e. $r_f$ and $\delta_f$.
This can be seen for CP eigenstate final states via \eqref{afdCP} with $\phi_f = \gamma$, \eqref{DYfresult} with $\phi_f^M = \phi_2^M + \delta \phi_f$, and \eqref{dphifSCSCP}, and similarly for non-CP eigenstate final states.  Thus, important information on the QCD anatomy of these decays could be obtained.

To illustrate the potential for probing the SM in the precision era, we
use the (na{\"\i}vely) estimated experimental sensitivities reported in Table
\ref{tab:future_exp} for the LHCb Phase II Upgrade era, for three decay modes: $D^0 \to K_{S,L} \pi^+ \pi^- $, $K^+\pi^-$, and
$K^+ \pi^- \pi^+ \pi^-$.  We caution that scaling the errors on the individual measurements purely based on the expected statistics may be optimistic.
The results of the fit are presented in the
rightmost columns in Table \ref{tab:res_current} and in Figure
\ref{fig:future} (including the SCS observable $A_\Gamma$ leads to marginal improvement in the sensitivity to $\phi_2^M$ in Phase II).
They suggest that SM sensitivity to $\phi_2^{M,\Gamma}$ may be achievable, particularly if these phases lie on the high end of
our $U$-spin based estimates.
Moreover, additional input from Belle-II indirect CPV measurements at 50 ab$^{-1}$ \cite{Kou:2018nap}, e.g. for the decays
$D^0 \to K_{S,L}\pi^+ \pi^-$, $K^+ \pi^- $, $K^+ \pi^- \pi^0$, and $A_\Gamma$, may improve the sensitivity.

 \section{Discussion}
\label{sec:conclusion}

In this paper we have developed the description of CP violation in $\DDbar$ mixing in terms of the final state dependent
dispersive and absorptive weak phases $\phi_f^M$ and $\phi_f^\Gamma$.  They govern CP violation in the interference between decays with and without dispersive mixing, and with and without absorptive mixing, respectively.   The expressions for the time dependent decay widths and CP asymmetries undergo extensive simplifications compared to the familiar parametrization in terms of $|q/p|$ and $\phi_{\lambda_f}$ (translations are provided), and become physically transparent.  For instance, their dependence on the strong phases in the decay amplitudes, and the CP-even dispersive mixing phase $\pi/2$, are easily understood.  This understanding extends to the strong phases of the subleading decay amplitudes, e.g. those responsible for direct CP violation in $D^0 \to K^+ K^- , \pi^+\pi^-$.
 An important consequence is that the time dependent CP asymmetries for decays to CP eigenstate final states, e.g. $f=  K^+ K^- , \pi^+ \pi^- $, depend on $\phi_f^M$ (dispersive CP violation), but not on $\phi_f^\Gamma$ (absorptive CP violation).  Conversely, the $\phi_f^\Gamma$ can only be probed in decays to non-CP eigenstate final states, e.g. the CF/DCS final states $f=K^+ \pi^- , \,K_{S,L} \, \pi^+ \pi^-$.

We have applied the dispersive/absorptive formalism to the three classes of decays which contribute to $\DDbar$ mixing,  (i) CF/DCS decays to $K^\pm X$, (ii) CF/DCS decays to
$K^0 X$, $\K0bar X$, and (iii)
SCS decays (both CP eigenstate and non-CP eigenstate final states).
Derivations and expressions have been provided for the time dependent decay widths and asymmetries in all three cases.
The CF/DCS decays to $K^0 X$, $\K0bar X$ require special care due to the effects of CPV in $K^0 - \K0bar$ mixing. Moreover,
their widths depend on two elapsed time intervals, the $D$ and $K$ decay times, following their respective production.
Appendix~\ref{sec:AppendixA} contains expressions for a selection of time-integrated CP asymmetries, demonstrating that they can also be used to separately measure $\phi_f^{M}$ and $\phi_f^{\Gamma}$.

Measurements of the final state dependent phases $\phi_f^M$ and $\phi_f^\Gamma$ ultimately determine a pair of
intrinsic mixing phases $\phi_2^M$ and $\phi_2^\Gamma$, respectively, cf. \eqref{theorphasesdef}.  The latter are the arguments, in the complex mixing plane, of the total dispersive and absorptive mixing amplitudes $M_{12}$ and $\Gamma_{12}$, relative to their dominant $\Delta U=2$ ($U$-spin) components. The latter are responsible for the neutral $D$ meson mass and width differences.  The intrinsic mixing analog ($\phi_2$) of the final state dependent phenomenological phases $\phi_{\lambda_f}$, is similarly defined as the argument of $q/p$ relative to the $\Delta U=2$ mixing amplitude.
The $U$-spin decomposition of the dispersive and absorptive mixing amplitudes yields the SM estimates $\phi_2^M, \phi_2^\Gamma = O(0.2\%)$, cf. \eqref{theorphases}--\eqref{refined}, \eqref{phi2Gbound1}, with $\phi_2$ of same order.
We also obtain an upper bound on the absorptive phase in the SM, $|\phi_2^\Gamma | < 0.005 $ \cite{prlDmix}, when taking $\Delta \Gamma_D$ equal to its measured central value, and conservatively assuming that a certain $U$-spin breaking parameter satisfies $\epsilon_1 <1 $, cf. \eqref{eps1def}, \eqref{phi2Gbound1}.

The intrinsic mixing phases are experimentally accessible due to {\it approximate universality}.  In particular, we have shown that there is minimal uncontrolled final-state dependent pollution from the decay amplitudes in the measured phases $\phi_f^M$, $\phi_f^\Gamma$:\\

\noindent $\bullet$  For the CF/DCS $K^\pm X$ final states, e.g. $K^+ \pi^-$, in the SM and in extensions with negligible new weak phases in these decays, the difference $\delta \phi_f $ between $ \phi_2^{M, \Gamma}$ and $ \phi_f^{M, \Gamma}$
is known, final state independent, and entirely negligible, i.e. it is $O(\lambda_b^2 / \lambda_d^2)\sim 10^{-6}$, cf. \eqref{phiMGCFDCS},\eqref{dphifCFDCDnonCP}.\\

\noindent $\bullet$   For the CF/DCS $K^0 X$ final states, e.g. $K_{S,L} \,\pi^+ \pi^-$, in the SM and under the same NP assumptions,
there are two contributions to the misalignments, $\delta\phi_f$:  a small incalculable final state dependent one of $O( 2\, \theta_{C}^2\, {\rm Im}[\epsilon_K ]\, )\sim 0.1\, \phi_2^{M,\Gamma}$, due to the subleading DCS amplitudes, and a precisely known one of $O(2 \,{\rm Im}[\epsilon_K ]\, )\sim \phi_2^{M,\Gamma}$ which can be subtracted from the measured values of $\phi_f^{M,\Gamma}$, cf. \eqref{dphifK0wDCS}.\\

 \noindent $\bullet$  For the SCS decays, e.g. $f=K^{+} K^- $, $\pi^+ \pi^- $, there is uncontrolled final state dependent QCD penguin pollution. In the SM, and for extensions with CP-odd QCD penguins of same order, the misalignments satisfy $\delta \phi_f / \phi_2^{M,\Gamma}  = O(\epsilon)$ in $U$-spin breaking.  This could be sizable for certain decays.  A $U$-spin based estimate, taking into account $\Delta A_{CP}$, yields
the representative value $ \epsilon \sim 0.4$, or
$\delta\phi_{K^{+} K^-}, \delta\phi_{\pi^+ \pi^- } =O(0.4) \,\phi_{2}^{M,\Gamma}$, cf. \eqref{SCSappuniv}--\eqref{Uspinmisalign}.
 %$ \epsilon \sim 0.4$
Fortunately, the
average over $\phi_{K^{+} K^- }^{M,\Gamma}$ and $\phi_{\pi^{+} \pi^- }^{M,\Gamma}$ differs from $\phi_2^{M,\Gamma}$ by $O(\epsilon^2 )$.\\

Expressions for the time dependent decay widths in the approximate universality parametrization, i.e. in terms of $\phi_2^M$, $\phi_2^\Gamma$, have been discussed in detail for the three classes of decays, cf. Section \ref{sec:implement}.  Our results for the $K^0 X$ final states are particularly noteworthy.  On the time scale of sequential $K^0$ decays at LHCb ($t \lsim 0.5\, \tau_S$), the effect of kaon CP violation on the
time dependent CP asymmetries (due to $K_L X - K_S X$ interference, and an ${\rm Im}[\epsilon_K]$ component in $\phi_f^{M,\Gamma}$)
undergoes a cancelation at the few percent level.  Thus, to very good approximation, LHCb analyses of these modes can neglect the effects of
kaon CP violation in measurements of $\phi_2^{M,\Gamma}$ from the time dependent CP asymmetries.
In contrast, over the longer $K^0$ decay time scales that can be explored at Belle-II,
the cancelation subsides, and
$\epsilon_K$ ultimately dominates the time dependent CP asymmetries.  Thus, Belle-II analyses must fit for
$K_L - K_S$ interference effects, and account for ${\rm Im}[\epsilon_K]$ in the extraction of $\phi_2^{M,\Gamma}$.

In the future, the values of $\phi_2^{M,\Gamma}$ obtained from the CF/DCS decays will allow a determination of the misalignments, $\delta\phi_f$, in the SCS decays.  In combination with measurements of the SCS direct CP asymmetries, $a_f^d$, it will be possible to determine
the anatomy of the QCD penguins in the SM, e.g. for $f=K^+ K^-$, $\pi^+ \pi^-$.  In particular, taking the SM value $\gamma$ for the  weak phases of the penguin amplitudes relative to the dominant ``tree" amplitudes, it will be possible to measure their relative magnitudes and strong phases.  This would provide an important test of QCD dynamics, if lattice measurements of these quantities become available.

Past fits to the mixing data were sensitive to values of $\phi_{12}= {\rm arg}[M_{12} /\Gamma_{12} ] = \phi_2^M-\phi_2^\Gamma$ down to the
100 mrad level.  This level of precision probed for large short-distance new physics contributions.  Thus,
the effects of weak phases in the subleading decay amplitudes could be safely neglected in the indirect CPV observables.  In this limit, referred to as superweak, the mixing phases satisfy $\phi_{12} = \phi_2^M$,
and $\phi_2^\Gamma=0$.
We have carried out a fit to the current data set in this limit, yielding $\phi_2^M = (-0.5 \pm 2.2)\%$ at $1\sigma$, consistent with the HFLAV fit result,  and corresponding to an $O(10)$ window for New Physics at $2\sigma$.

The approximate universality fit is less constrained, given the description of indirect CP violation in terms of two phases, $\phi_2^{M}$ and $\phi_2^{\Gamma}$, rather than just one.  Interestingly, in this case, our errors for $\phi_2^{M}$ ($\approx \,29$ mrad) are similar to the superweak fit result, and about a factor of three smaller than the errors for $\phi_2^\Gamma$ ($\approx\,99$ mrad).  This is due, in part, to the observable $A_\Gamma= -\Delta Y_f$ ($f=\pi^+\pi^-$, $K^+ K^-$), which depends on $\phi_2^M$ but not on $\phi_2^\Gamma$, and has a relatively small experimental error.
The phenomenologically motivated phase $\phi_2$ is a weighted sum over $\phi_2^M$ and $\phi_2^\Gamma$, where the weights are equal to the leading CP averaged dispersive %($\propto x_{12}^2$)
and absorptive
%($\propto y_{12}^2$)
mixing probabilities, respectively, cf. \eqref{tan2phitheory}.  This explains why the error on $\phi_2$ ($\approx \,72$ mrad) is similar to the error on $\phi_2^\Gamma$.

The $U$-spin based estimates of $\phi_2^{M}$ and $\phi_2^{\Gamma}$ imply that probing the SM will require
a precision of a few mrad or better for both phases.  Given the large theoretical uncertainties, a null result
as this sensitivity is approached would effectively close the window for new physics in charm indirect CP violation.
Alternatively, the most likely origin for a significantly enhanced signal would be CP violating short distance new physics, yielding
$\phi_2^M \gg \phi_2^\Gamma $, with the latter given by its SM value.  A second possibility, light CP violating new physics, would enter both the dispersive and absorptive mixing amplitudes via new $D^0$ decay modes,
likely enhancing both $\phi_2^M$ and $\phi_2^\Gamma $.
This appears unlikely, given the upper bounds on exotic $D^0$ decay rates.
For instance, for invisible $D^0$ decays, the upper bound on the branching ratio, ${\rm Br}_{\rm inv} < 9.4 \times 10^{-5}$ (90\% CL) \cite{PDG},
constrains
the invisible contribution to $\phi_2^\Gamma$ as $\delta \phi_2^\Gamma \lsim {\rm Br}_{\rm inv}/ \theta_{C}^2 \sim 0.2 \%$, i.e. the upper bound lies at the SM level (before taking into account additional suppression due to the relative magnitudes of the interfering invisible decay amplitudes, and their weak and strong phase differences).
Moreover, the upper bound on contributions from $D^0 \to  K^0 +\,$invisibles is about a factor of 30 smaller.\footnote{An upper bound on $D^+\to K^+ + \,$invisibles, ${\rm Br}_{K^+ +{\rm\, inv} }< 8 \times 10^{-6}$ \cite{jurenewpaper}, yields $\delta \phi_2^\Gamma \lsim  ({\rm Br}_{K^+ +{\rm inv}  }\,/\,\theta_{C}^2\,)(\Gamma_{D^+}/\Gamma_{D^0}) \sim 6.5 \times 10^{-5}$, well below the SM estimates, where we have assumed similar widths for the semi-invisible $D^+$ and $D^0$ decays.}

Finally, based on available LHCb Phase II projections for the decays $D^0 \to K_{S,L} \pi^+ \pi^-$, $K^+ \pi^- $, $K^+ \pi^- \pi^+ \pi^-$, and $A_\Gamma$, we have estimated the
precision that could be reached for $\phi_2^{M,\Gamma}$ in the upcoming high statistics charm era, using an approximate universality fit.
Note that our results are intended to be illustrative, given that the LHCb phase II projections do not include systematic errors.
The resulting $1\sigma$ errors for $\phi_2^M$ ($\approx 1.2$ mrad) and $\phi_2^\Gamma$ ($\approx 1.7$ mrad) suggest that sensitivity
to $\phi_2^{M,\Gamma}$ in the SM may be achievable, particularly if these phases lie on the high end of the $U$-spin based estimates.
Measurements of $\phi_2^{M,\Gamma}$ could one day become available on the lattice.  Comparison with their measured values would provide   the ultimate precision test for the SM origin of CP violation in charm mixing.

\vspace{0.5cm}
\mysection{Acknowledgements}
We are indebted to Yuval Grossman, Zoltan Ligeti, Alexey Petrov, and Gilad Perez for their collaboration during earlier stages of this work.
We thank Marco Gersabek, Bostjan Golob, Uli Nierste, Alan Schwartz, Mike Sokoloff and Jure Zupan for discussions.  We are especially grateful to 
Tommaso Pajero for a detailed reading of the manuscript and for pointing out typos in several equations.
This project has received funding from DOE grant DE-SC0011784, and
from the European Research Council (ERC) under the European
Union's Horizon 2020 research and innovation program (grant agreement
n$^o$ 772369). AK thanks the Aspen Center of Physics,
supported by the NSF grant PHY-1607611, where parts of this work were
carried out.
 \appendix

\section{CPVINT phases $\phi_2^M$, $\phi_2^\Gamma$ from time-integrated CP asymmetries}
\label{sec:AppendixA}
We give expressions for a few time integrated CP asymmetries, illustrating the possibility of determining
the theoretical CPVINT phases purely from time-integrated decays.
We begin with the tagged and untagged CP asymmetries for the CF/DCS final states $f= K^+ \pi^- $,  $\bar f = K^- \pi^+ $ ($A_{\bar f}$, $\bar A_f$ are the DCS amplitudes):
\beq
\begin{split}
&A_{\rm CP}^{\rm tag,\,DCS\, (CF)}  \equiv  {\int dt \left( \Gamma_{D^0 (t) \to \bar f (f )} -  \Gamma _{\bar D^0 (t) \to f (\bar f) } \right)\over
\int dt \left( \Gamma_{D^0 (t) \to \bar f  (f)} +  \Gamma_{\bar D^0 (t) \to
  f (\bar f )} \right)}\,, \nonumber \cr
&A_{\rm CP}^{\rm untag} \equiv \nonumber \cr
& {\int dt \left(\Gamma_{D^0 (t) \to \bar f }+  \Gamma_{\bar D^0 (t) \to \bar f } - \Gamma_{D^0 (t) \to  f }-  \Gamma_{\bar D^0 (t) \to  f } \right)\over
\int dt \left(\Gamma_{D^0 (t) \to \bar f }+  \Gamma_{\bar D^0 (t) \to \bar f } + \Gamma_{D^0 (t) \to  f }+  \Gamma_{\bar D^0 (t) \to  f }  \right)}\, .\nonumber
\end{split}
\eeq
To obtain their dependence on the CPVINT phases, we must keep the subleading DCS amplitudes in 
 \eqref{timedepRS}, in analogy to the CF contributions in \eqref{timedepWS}. 
Assuming no new weak phases in the CF/DCS decays as in the SM, hence no direct CPV, the amplitude ratios simplify as
$R_f =1/R_{\bar f}=R_f^\pm  $, cf. \eqref{Rfpm}.   Thus, Eqs.  \eqref{timedepRS}, \eqref{timedepWS} yield
\beq \begin{split}\sqrt{R_f} A_{\rm CP}^{\rm tag, DCS} &=  x_{12} \sin\phi_f^M \cos\Delta_f - y_{12} \sin\phi_f^\Gamma \sin \Delta_f \,,\cr
{A_{\rm CP}^{\rm tag, CF}\over \sqrt{R_f} } &=  x_{12} \sin\phi_f^M \cos\Delta_f + y_{12} \sin\phi_f^\Gamma \sin \Delta_f\,.\end{split}
\eeq
The absorptive and dispersive CPV phases are then readily separated as %(we use $R_f =1/R_{\bar f}=R_f^\pm  $, cf. \eqref{Rfpm}, in the absence of direct CPV),
\beq \begin{split}
{A_{\rm CP}^{\rm tag, CF}\over \sqrt{R_f} }&- \sqrt{R_f} A_{\rm CP}^{\rm tag, DCS}
 =- {{(1+R_f)  A_{\rm CP}^{\rm untag} } \over  \sqrt{R_f} }\cr&~~~~~~~~~~~~~~~~~~~~~~= 2 y_{12} \sin\phi_2^\Gamma \sin\Delta_f \, \cr
 {A_{\rm CP}^{\rm tag, CF}\over \sqrt{R_f} }&+ \sqrt{R_f} A_{\rm CP}^{\rm tag, DCS}=
2 x_{12} \sin\phi_2^M \cos\Delta_f \,,
\end{split}
\eeq
where $\Delta_f $ is the $K^+\pi^-$ strong phase, cf. \eqref{lambdaCFDCS}.
We have taken $\phi_f^{M,\Gamma} = \phi_2^{M,\Gamma}$, cf. \eqref{phiMGCFDCS}, \eqref{dphifCFDCDnonCP}.
Note that the untagged CP asymmetry is purely absorptive.

We end with the time integrated CP asymmetries for the SCS final states $f= \pi^+ \pi^- , K^+ K^-$:
\beq A_{\rm CP, f}^{SCS}  \equiv  {\int dt ( \Gamma_{D^0 (t) \to f} -  \Gamma _{\bar D^0 (t) \to  f) } )\over
\int dt ( \Gamma_{D^0 (t) \to f} +  \Gamma_{\bar D^0 (t) \to f } )} .\eeq
We obtain the expression
\beq A_{\rm CP, f}^{SCS}= a_f^d  +  {\langle t \rangle \over \tau_D } \Delta Y_f =  a_f^d  + {\langle t \rangle \over \tau_D } (-x_{12} \sin\phi_f^M + y_{12} a_f^d) \label{ACPSCSTI}\,,\eeq
where $\langle t \rangle$ is the average (acceptance dependent) decay time of the $D^0$ mesons in the experimental sample.
The ratio $\langle t \rangle / \tau_D$ is very close to 1 at the B factories; at LHCb, it exceeds 1 by about $5\%-10\% $ for the muon-tagged sample \cite{Aaij:2019kcg}, while
it is in the $1.7-1.8$ range for the $D^{*+}$-tagged sample \cite{Betti:2669175}.\footnote{We thank T. Pajero for pointing this out to us.}
Recall that in the SM, for SCS decays,
\beq \phi_f^{M}
= \phi_2^{M} - a_f^d \cot \delta_f
 = \phi_2^{M} [1 + O(\epsilon)], \eeq
where $\delta_f$ is the strong phase difference between the leading and subleading $D^0 \to f$ decay amplitudes, and
$a^d_{f}$ is the direct CP asymmetry, cf. \eqref{dphifSCSCP}.  
However, the average of $\phi_f^{M}$ over
$f=K^+K^-, \,\pi^+ \pi^-$ differs from $\phi_2^{M}$ by $O(\epsilon^2)$ in $U$-spin breaking, cf. \eqref{dphifSCSCP}, \eqref{SCSappuniv},  \eqref{Uspinmisalign}. 

The time integrated CP asymmetry difference $\Delta A_{CP} = A_{CP,K^+ K^-} - A_{CP,\pi^+ \pi^-} $ \cite{Aaij:2019kcg} can be expressed in terms of $\phi_2^M$ and the direct CP asymmetries as
%\beq
%\begin{split} \label{dacp2} \Delta A_{\rm CP}&= a_{K}^d  - a_{\pi}^d -  {\langle t_K \rangle - \langle t_\pi \rangle  \over \tau_D } x_{12} \sin\phi_2^M \cr
%&-  {x_{12} \over \tau_D} \,\left(\langle t_K \rangle  a_{K}^d \cot\delta_{K} - \langle t_\pi \rangle  a_{\pi}^d \cot\delta_{\pi}  \right)\,,
%\end{split} \eeq
\beq
\begin{split} \label{dacp2} \Delta A_{\rm CP}&= a_{K}^d  - a_{\pi}^d \cr
% - \langle t_\pi \rangle  \over \tau_D } x_{12}\, \sin\phi_2^M \cr
+& {\langle t_K \rangle + \langle t_\pi \rangle  \over 2\, \tau_D } \Big(x_{12} \,\large[  a_{K}^d \cot\delta_{K} -  a_{\pi}^d \cot\delta_{\pi}  \large] \cr
&~~~~~~~~~~~~~~~~~~~~~~~~~~~~~~~~~~~~~+y_{12} \large[a_K^d - a_\pi^d \large] \Big)\cr
-&  {\langle t_K \rangle - \langle t_\pi \rangle  \over 2\, \tau_D }\Big( x_{12} \,\large[  2\sin\phi_2^M - a_{K}^d \cot\delta_{K} - a_{\pi}^d \cot\delta_{\pi} \large]\cr
&~~~~~~~~~~~~~~~~~~~~~~~~~~~~~~~~~~~~~~- y_{12} \large[a_K^d+ a_\pi^d \large] \Big)\,,
%&-  {\langle t_K \rangle + \langle t_\pi \rangle  \over 2\, \tau_D } x_{12} \,\left(  a_{K}^d \cot\delta_{K} -  a_{\pi}^d \cot\delta_{\pi}  \right)\,,
\end{split} \eeq
where the subscripts $K$ and $\pi$ refer to the $K^+  K^-$ and $\pi^+\pi^-$ final states, respectively.
%where $\delta_{K,\pi}$ are the strong phase differences between the leading and subleading $K^+ K^-$ and $\pi^+\pi^-$ decay amplitudes, respectively,
%$a^d_{K,\pi}$ are the two direct CP asymmetries. 
At LHCb the difference of the two average decay times satisfies $\langle t_{K}\rangle -\langle t_{\pi}\rangle\approx 0.12 \,\tau_D$.
The corrections to the first line in \eqref{dacp2} are negligible, as is well known.  In particular, we find that the contribution proportional to the sum of the average decay times is of $O(x_{12}\,a_{f}^d ,\, y_{12}\,a_{f}^d )$.  The contribution proportional to the difference of decay times is of $O(0.1\,x_{12} \,\phi_2^M )$, given that $(a_{K}^d  +  a_{\pi}^d )$ and $(a_{K}^d \cot\delta_K +  a_{\pi}^d \cot\delta_{\pi})$ are formally of $O(\epsilon^2  \cdot\,\phi_2^M )$.

%\section{The ratio $\Gamma_{sd} /\Gamma_D$}
%\label{sec:AppendixB}

\end{document}